\documentclass[reprint, twocolumn, amsfonts, amsmath,amssymb, superscriptaddress, prx]{revtex4-2}

\usepackage{graphicx}
\usepackage{dcolumn}

\usepackage{bm}
\usepackage{physics}
\usepackage{layouts}
\usepackage[normalem]{ulem}

\usepackage[dvipsnames]{xcolor}
\usepackage{xfrac}
\usepackage[linkcolor=Blue,citecolor=Blue,urlcolor=Blue,colorlinks=true]{hyperref}
\usepackage{notes2bib}
\bibnotesetup{ note-name = , use-sort-key = false}

\newcommand{\updownarrows}{\mathbin\uparrow\downarrow}
\renewcommand{\upuparrows}{\mathbin\uparrow\uparrow}
\newcommand{\updownarrowsnarrow}{\mathbin\uparrow\hspace{-.3em}\downarrow}
\newcommand{\upuparrowsnarrow}{\mathbin\uparrow\hspace{-.3em}\uparrow}

\newcommand{\vc}[1]{\mathbf{#1}}
\newcommand{\vcu}[1]{\mathbf{\hat{#1}}}
\newcommand{\delt}{\partial_t}
\newcommand{\dpar}{\nabla_\parallel}
\newcommand{\dperp}{\nabla_\perp}
\newcommand{\rv}{\vc{r}}
\newcommand{\kr}{k_r}
\newcommand{\kn}{n}
\newcommand{\kv}{\vc{k}}
\newcommand{\rhoc}{\rho_c}
\newcommand{\rhocn}{\rho_c^{(n)}}
\newcommand{\rhocp}{\rho_c^{(p)}}
\newcommand{\polv}{\vc{p}}
\newcommand{\pol}{p}
\newcommand{\nv}{\vcu{n}}
\newcommand{\npv}{\vcu{n}_\perp}
\newcommand{\xiv}{\bar{\vc{r}}}
\newcommand{\xic}{\bar{r}}
\newcommand{\xicu}{\bar{u}}
\newcommand{\xicv}{\bar{v}}
\newcommand{\TS}{^{\mathsf{TS}}}
\newcommand{\Deff}{D_{\mathrm{eff}}}
\newcommand{\ellBL}{\ell^{}_{\mathrm{BL}}}

\newcommand{\LBL}{\Lambda_{\mathrm{BL}}}
\newcommand{\LMIC}{\Lambda_{\mathrm{M}}}
\newcommand{\DrMIC}{\Delta\rho^{}_{\mathrm{M}}}
\newcommand{\DrBL}{\Delta\rho^{}_{\mathrm{BL}}}

\newcommand{\wcrit}{\omega_c}
\newcommand{\wcritpol}{\omega_c^{(p)}}
\newcommand{\wcritnem}{\omega_c^{(n)}}
\newcommand{\wcritqp}{\omega_c^{*}}

\DeclareMathOperator{\sign}{sign}

\usepackage[inline]{enumitem}
\usepackage{xcolor}
\usepackage{lipsum}
\usepackage{ulem}
\definecolor{myGreen}{RGB}{89,132,63}
\definecolor{cadmiumorange}{rgb}{0.93, 0.53, 0.18}
\definecolor{olivegreen}{rgb}{0.43, 0.72, 0.48}
\definecolor{wildstrawberry}{rgb}{1., 0.26, 0.64}
\definecolor{mypurple}{rgb}{50.2, 0., 50.2}
\definecolor{darkpastelgreen}{rgb}{0.1, 0.75, 0.24}

\begin{document}

\title{Supramolecular assemblies in active motor-filament systems:\\micelles, bilayers, and foams}

\author{Filippo De Luca}
\affiliation{Arnold Sommerfeld Center for Theoretical Physics and Center for NanoScience, Department of Physics, Ludwig-Maximilians-Universit\"at M\"unchen, Theresienstra\ss e 37, D-80333 Munich, Germany}
\author{Ivan Maryshev}
\affiliation{Arnold Sommerfeld Center for Theoretical Physics and Center for NanoScience, Department of Physics, Ludwig-Maximilians-Universit\"at M\"unchen, Theresienstra\ss e 37, D-80333 Munich, Germany}
\author{Erwin Frey}
\email{frey@lmu.de}
\affiliation{Arnold Sommerfeld Center for Theoretical Physics and Center for NanoScience, Department of Physics, Ludwig-Maximilians-Universit\"at M\"unchen, Theresienstra\ss e 37, D-80333 Munich, Germany}
\affiliation{Max Planck School Matter to Life, Hofgartenstraße 8, D-80539 Munich, Germany}

\date{January 10, 2024}

\begin{abstract}
Active matter systems evade the constraints of thermal equilibrium, leading to the emergence of intriguing collective behavior. A paradigmatic example is given by motor-filament mixtures, where the motion of motor proteins drives alignment and sliding interactions between filaments and their self-organization into macroscopic structures. After defining a microscopic model for these systems, we derive continuum equations, exhibiting the formation of active supramolecular assemblies such as micelles, bilayers and foams. The transition between these structures is driven by a branching instability, which destabilizes the orientational order within the micelles, leading to the growth of bilayers at high microtubule densities. Additionally, we identify a fingering instability, modulating the shape of the micelle interface at high motor densities. We study the role of various mechanisms in these two instabilities, such as contractility, active splay, and anchoring, allowing for generalization beyond the system considered here.
\end{abstract}

\maketitle

\section{\label{sec:introduction}Introduction}

Active matter exhibits complex collective behavior absent in thermal equilibrium systems due to its constituents' ability to consume energy locally, thereby breaking detailed balance~\cite{Ramaswamy2010, Saintillan2013, Marchetti2013, Needleman2017, Doostmohammadi2018,  Baer2019, Chate2020, Han2021}.
A fascinating class of paradigmatic model systems for active matter are mixtures of cytoskeletal filaments and molecular motors.
These include motility assays, where filaments are self-propelled by motors attached to a substrate \cite{Schaller2010,Huber2018,Sumino2012}. In contrast, in bulk motor-filament mixtures, the activity is manifested in the relative motion rather than self-propulsion of the filaments. 
A prime example for such a system  are microtubule-motor mixtures. 
Here, the relative motion is induced by molecular motors such as kinesin or dynein, which cross-link microtubule filaments and exert torques and forces as they walk along them.
The resulting motor-mediated interaction between different microtubule filaments drives the formation of a variety of large-scale structures \cite{Dogterom2013, Shelley2014}, including asters and vortices \cite{Nedelec1997, Surrey2001, Hentrich2010, Tan2018, Dogic2021}, extensile bundles \cite{Sanchez2011, Sanchez2012, Dogic2021}, and foam-like patterns \cite{Dogic2021}.

Understanding the self-organization of microtubules, driven by molecular motors, into complex large-scale structures holds significant importance in a cell biology context. 
For example, it sheds light on essential processes such as the formation of the mitotic spindle~\cite{Sawin1992, Burbank2007, Mogilner2010, Brugues2014, Cross2014} and of cell-like structures observed in \textit{Xenopus} egg extracts~\cite{Cheng2019}. 
More generally, unraveling the mechanisms driving this self-organization can offer profound insights into the physics of systems operating far from thermal equilibrium, transcending the fraction of phase space typically observed in living systems~\cite{Needleman2017}.

The present theoretical study is motivated by recent experimental work on mixtures of microtubules and kinesin-4 motors, revealing a novel non-equilibrium phase termed \emph{active foam}, which consists of a foam-like network of microtubule bilayers~\cite{Dogic2021}.
Each bilayer within the active foam displays microtubules pointing in opposite directions on either side, with their plus-ends directed towards the bilayer midplane, where kinesin motors accumulate.
The interconnected bilayer network is observed to undergo sustained rearrangements and does not coarsen over time, highlighting the inherently non-equilibrium nature of this active foam. 
Notably, in contrast to equilibrium foams, the cells within the active foam exhibit diverse shapes, including non-convex ones, and present loose bilayer edges extending into the cell bulk.
While previous theoretical studies have discussed active foam states in a phenomenological, top-down approach \cite{MaryshevSM2020,Fausti2021}, there is currently no comprehensive bottom-up theory for the emergence of active foams from microscopic interactions with reference to a specific physical system.

In this study, we address this critical gap in understanding, introducing a novel non-equilibrium field theory for motor-filament mixtures. 
We derive this field theory from a microscopic model for the motor-mediated interaction between microtubules, employing the Boltzmann-Ginzburg-Landau (BGL) approach \cite{Bertin2006, Bertin2009, Peshkov2014}.
This approach enables us to bridge the gap between the microscopic and macroscopic scales, linking the properties of filament-filament interactions with the collective states that emerge macroscopically due to these interactions. 
It allows us to gain critical insights into the mechanisms driving the self-organization of active filament systems into complex structures.

In modelling the microscopic scale, our primary focus lies in the role of motor proteins as facilitators of alignment and sliding interactions between microtubules.
Although the molecular interaction between filaments and motors is complex in detail, it exhibits several generic features, which are all crucial in the cell environment \cite{Cross2014}:
Molecular motors drive the \emph{sliding} of microtubules relative to each other by applying forces on filament pairs \cite{Kapitein2005,Wijeratne2018,Tan2018,Fink2009}, an essential process in mitotic spindle formation.
They induce relative \textit{alignment} by exerting torques on the microtubules \cite{Kapitein2005, Wijeratne2018, Tan2018}, a vital process for organizing microtubule arrays within cells.
Finally, microtubules serve as molecular tracks for motor proteins as they walk from one end to the other, facilitating intracellular transport processes \cite{Cross2014}. 
In microtubule-motor mixtures, the procession of molecular motors along microtubules leads to to a spatially and temporally \emph{inhomogeneous motor density} across the system.
Previous work has focused on various individual aspects, such as the significance of parallel alignment in the presence and absence of sliding~\cite{AransonTsimring2006, Ziebert2007}, as well as the role of the interaction kernel \cite{MaryshevPRE2018} or parallel versus antiparallel alignment~\cite{MaryshevSM2019}. 
However, to date, the interplay between the above general features of motor-mediated filament interactions has not been explored, which, as we show here, leads to the formation of novel supramolecular structures.

Most importantly, earlier studies have neglected the possibility of an asymmetry between parallel and antiparallel alignment. We refer to this essential property as {\emph{parity symmetry breaking} of the alignment interaction}. When parity symmetry is not broken, the ensuing patterns can only exhibit nematic order~\cite{MaryshevSM2019}, excluding the polar structures observed in experimental studies~\cite{Dogic2021}.
In our microscopic model, we account for the {parity symmetry breaking} by introducing a critical angle ${\wcrit \neq 90^\circ}$ that separates the ranges of crossing angles where parallel or antiparallel alignment interactions occur.
We show that the broken parity symmetry in the alignment interaction plays a pivotal role in the selection of the relevant orientational order, which can be polar or nematic.

In the polar regime, we observe the emergence of a diverse array of patterns, involving the self-organization of microtubules into micelles, bilayers, and active foams. The active micelles are characterized by a radially symmetric arrangement of microtubules, reminiscent of lipid molecules in passive micelles, with the microtubule ends pointing towards the center of the assembled structure. These structures are stable at small microtubule and motor densities. 
However, as the densities are increased, they are subject to two distinct instabilities that eventually break their radial symmetry. 
The first instability, termed the \emph{fingering} instability, emerges at high motor concentrations. 
It leads to the modulation of the micelle away from a circular shape, bending its interface into lobes.
The second instability, referred to as the \emph{branching} instability, emerges at high microtubule concentrations and causes the micelle perimeter to fragment into bilayer-like branches.
Remarkably, for even higher microtubule concentrations, we observe the formation of active networks of bilayers.
These networks closely resemble the active foams reported in recent experimental studies~\cite{Dogic2021}, exhibiting non-convex cells and loose ends. 

Consistent with these experiments, we find that microtubule density is the control parameter that determines whether active micelles or foams are formed.
Our theory reveals three critical features of the microscopic interaction required for the formation of polar bilayers and the assembly of these bilayers into foams: 
First, the \emph{breaking of parity symmetry} in the alignment interaction is crucial for the emergence of polar order. 
Intermediate values of the critical angle $\wcrit$ ensure that the opposing polar order on the two sides of the bilayer is stable.
Second, \emph{antiparallel sliding} provides polarity sorting, which is essential for the formation of well-defined bilayers. 
Third, the \emph{inhomogeneous motor field} gives rise to spatial modulation of the interaction strength across the system. This leads to the emergence of regions where polar order can form locally, enabling the assembly of bilayers and micelles.

These three features play a crucial role in forming and maintaining the microtubule bilayers as the elementary mesoscopic structure composing the active foam network. Interestingly, the microtubule bilayers have a structure reminiscent of lipid bilayers, just as the active micelles parallel lipid micelles. However, these active supramolecular assemblies differ fundamentally in their formation and maintenance mechanisms from their lipid counterparts.
In lipid bilayers, the amphiphilic nature of lipid molecules is the critical molecular feature driving their formation \cite{Terentjev2015}. Conversely, the microtubule bilayers in our theory emerge through active processes, namely the motor-mediated interactions described by our theory. This difference highlights the intriguing role of molecular motors in the dynamic behavior of microtubules, leading to the self-assembly of these unique bilayer structures and of the active foams.

The paper is structured as follows. In sec.~\ref{sec:derivedmodel}, we formulate our microscopic model for microtubule-motor mixtures and discuss the derivation of the continuum equations, highlighting the mechanisms encoded into those equations that will prove crucial for the phenomena we observe. In sec.~\ref{sec:PD}, we present numerical simulations of the continuum model and describe the structure of its phase diagram, involving micellar structures and active foams. Then, in sec.~\ref{sec:phen}, we perform a phenomenological generalization of our derived model, which we use to validate the analytical predictions made in the following sections. In sec.~\ref{sec:profiles}, we discuss the stationary profiles of the bilayer and micelle solutions, discussing how they differ from their equilibrium counterparts. The stability of the homogeneous ordered state and the micelle solutions is studied in sec.~\ref{sec:instab}, where we characterize the instabilities that drive micelle branching and fingering and we connect these instabilities with the phase diagram of the derived model. Finally, sec.~\ref{sec:discussion} contains a discussion of our results.

\section{
\label{sec:derivedmodel}
Boltzmann-Ginzburg-Landau theory}

\subsection{
\label{sec:micro}
Microscopic interactions}

Microtubules (MT) are biopolymers consisting of tubulin dimers. 
In a cell environment, they undergo constant polymerization and depolymerization, while \emph{in vitro} they can be stabilized to ensure that their lengths stay constant. 
The tubulin dimers are anisotropic, so that the microtubule has an intrinsic polarity, with two distinct ends which are conventionally referred to as plus-ends and minus-ends \cite{Desai1997}. 
Molecular motors are proteins that can walk along microtubules in a directed fashion towards either of the two ends of a filament. Furthermore, they can crosslink pairs of filaments, thereby mediating interactions between them \cite{Cross2014}. 
Some motor proteins, like kinesin-5, have two sets of motor domains that can walk along two filaments simultaneously \cite{Kapitein2005}; others, like kinesin-14, can passively bind to one microtubule with their tail domain while their head domain walks along a second one \cite{Fink2009}. 
Regardless of the motor-specific mechanism, the interconnection of filaments combined with directed motion allows the motors to exert forces and torques on pairs of microtubules, leading to relative sliding and alignment. 
In microtubule-motor mixtures, this motor-mediated interaction drives the emergence of collective dynamics, in which microphase separation and local orientational order show complex interplay \cite{Dogic2021}.

The goal of this section is to set up a microscopic model for microtubule-motor mixtures in two dimensions, which we then proceed to coarse-grain to obtain continuum equations that describe the dynamics of the concentrations and of the orientational order. 
We treat the microtubules as perfectly rigid polar rods of fixed length $L$. 
The state of each microtubule can be described by the position of its center of mass $\rv$ and a unit vector $\nv$ indicating its orientation. 
We choose this vector to point along the direction of motion of the motors, i.e., from the minus-end to the plus-end for most kinesins. 
In two dimensions, this unit vector can be expressed in terms of a single orientation angle $\phi$ such that $\nv(\phi)=(\cos\phi,\sin\phi)$. 
Individual microtubules are subject to thermal fluctuations leading to Brownian diffusion of their center of mass and their orientation angle. 
Unlike the filaments in motility assays, which can be modeled as self-propelled rods \cite{Baskaran2008, Schaller2010, Peshkov2012, Peshkov2014, Suzuki2015, Huber2018}, the microtubules in microtubule-motor mixtures do not show persistent motion along their body axis, so we do not incorporate self-propulsion in our model.

\begin{figure}
\includegraphics[width=\linewidth]{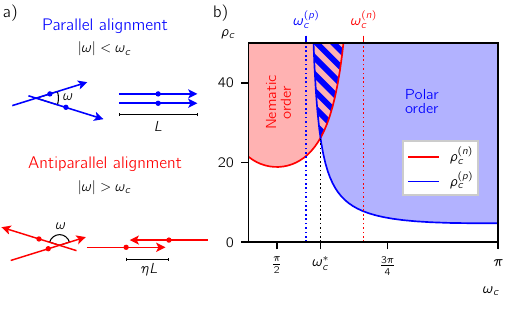}
\caption{\label{fig:micro}
Microscopic model and ordering transitions. a) Microscopic interaction rules. For intersection angles $\omega$ smaller than a critical angle $\wcrit$, an interaction results in parallel alignment, as well as relative sliding until the microtubules coincide. For intersection angles larger than $\wcrit$, the interaction is antipolar and the tubules are aligned in opposite directions. Here, relative sliding results in the separation of the filaments' centers of mass by a distance $\eta L$, with $\eta\in[0,1]$. b) Dependence of the polar critical density $\rhocp$ (in blue) and nematic critical density $\rhocn$ (in red) on the critical angle $\wcrit$. Depending on the choice of this angle, different kinds of order can emerge above a critical density, determining the stability of the isotropic state. The nematic critical density $\rhocn$ is positive for $\wcrit<\wcritnem\approx125.3^\circ$ (red dotted line), while the polar critical density $\rhocp$ is positive for critical angles larger than $\wcritpol\approx 101.9^\circ$ (blue dotted line). For $\wcrit<\pi-\wcritnem$ (not shown), no order can form at all. For densities in the hashed region, the isotropic state is unstable to the emergence of both kinds of orders, and they are expected to coexist. In this paper, we focus on the region where only polar order can emerge and the nematic order is enslaved to it, i.e., $\wcrit>\wcritqp$ (dotted black line) and $m\rho<\rhocn$ (see sec.~\ref{sec:coarsegrain}).}
\end{figure}

We introduce two types of interactions into the model. 
Firstly, we consider steric repulsive interactions. This is a passive effect resulting from the finite extension of the microtubules. Following the treatment in Ref.~\cite{MaryshevPRE2018}, we take into account the two-filament and three-filament contributions to the excluded volume. They result in a density-dependent increase of the isotropic diffusivity (see App.~\ref{app:BGL}).
Secondly, we incorporate the active interaction mediated by molecular motors. 
This interaction can be modeled explicitly on a more microscopic scale, taking into account the torques and forces exerted by the motors on the filaments \cite{Ahmadi2006,Gao2015,Striebel2020,Lamson2021,Furthauer2021}.
Here, we take a simpler approach, modeling the interaction as alignment and sliding events between pairs of microtubules (see Fig.~\ref{fig:micro}a). This approach reduces the number of parameters, allowing us to identify the key qualitative features of the interaction that lead to the emergent collective behavior.
We assume that the mixture is dilute enough, so that binary interactions constitute the dominant contribution to the dynamics. 
Furthermore, we treat the interaction events as instantaneous, which is justified in the case of a separation of time scales between the slow collective dynamics and the fast interaction events. 
We incorporate the motors into the model as a local and time-dependent concentration field $m (\rv,t)$. This field enters the model via the rate of interaction, which we take to be given by $G\cdot m$, where $G$ is a proportionality constant \footnote{Other functional dependences of the interaction rate on $m$ are conceivable. Here, for simplicity, we assume linearity, following Ref.~\cite{AransonTsimring2006}.}. We measure $m$ in units of its (physical) mean value ${\bar{m}_\mathrm{ph}=M/V}$, absorbing it into $G$; here $M$ is the total number of motors and $V$ the system volume \footnote{Thus, $G$ represents the rate of interaction at the concentration $\bar{m}_\mathrm{ph}$, while the mean value of the field $m$ is 1 by definition.}.

Experiments have shown that motors such as kinesin-4 and kinesin-5 can lead to both parallel and antiparallel alignment of pairs of microtubules \cite{Kapitein2005,Wijeratne2018}. We incorporate this feature into our model by allowing for both kinds of alignment, depending on the initial intersection angle $\omega = \phi_2-\phi_1$ between the two filaments. 
For small intersection angles close to $\omega=0$, parallel alignment (polar interaction) is expected. 
On the other hand, for large initial intersection angles close to $\omega=\pi$ the filaments align in an antiparallel fashion (antipolar interaction). 
We connect these two cases by introducing a tunable critical angle $\wcrit$ that determines the boundary between these two regimes: 
polar interactions occur for $\abs{\omega} < \wcrit$ and antipolar interactions for $\abs{\omega} > \wcrit$ (see Fig.~\ref{fig:micro}a). 
Previous work considered either exclusively polar interactions~\cite{AransonTsimring2006,MaryshevPRE2018}, i.e., $\wcrit=\pi$, or perfectly symmetric admixtures of polar and antipolar interactions~\cite{MaryshevSM2019}, i.e., the case $\wcrit=\pi/2$. 
A general critical angle $\wcrit$ introduces an asymmetry in the alignment rules, breaking the parity symmetry between polar and antipolar interactions.
Such a parity symmetry breaking emerged in recent numerical studies on the forces and torques involved in the motor-mediated interactions between two filaments, which demonstrates the validity of an effective mesoscopic description in terms of alignment interactions using a critical angle $\omega_c \neq \pi/2$~\cite{Lamson2021}.

In addition to aligning filament pairs, it was observed experimentally that the motors can drive relative sliding of the filaments with respect to each other \cite{Kapitein2005,Wijeratne2018,Fink2009}. 
In our model, a polar interaction slides the filaments together until their tips coincide (Fig. \ref{fig:micro}a). 
In an antipolar interaction, on the other hand, the filaments are slid apart. 
Due to stalling effects caused by the crowding of motors at the tips, this separation can come to a halt for non-zero overlaps \cite{Wijeratne2018}. 
In our model, we allow for partial antiparallel sliding by introducing a parameter $\eta$. 
For $\eta=0$, the microtubules fully overlap after the interaction, while for $\eta=1$ they separate completely until only their tips are touching. 
In both polar and antipolar interactions, we enforce the conservation of the center of mass of the microtubule pair. 
This means that the only motion allowed in our interaction model is the relative motion of the filaments in a pair. In Ref.~\cite{Ahmadi2006}, it was shown that a filament pair can experience a net translation due to the anisotropy of the viscous drag. 
Here, we neglect this effect, focusing exclusively on relative motion.

In summary, our model for microtubule-motor mixtures takes into account filament diffusion, steric interactions, and a motor-mediated alignment and sliding interaction. 
By introducing a critical angle $\wcrit$, we allow for an asymmetric interaction rule with a tunable bias towards parallel alignment, breaking the parity symmetry of the interaction. 
We also introduce the possibility of tuning the antiparallel sliding strength via the parameter $\eta$. 
The motor-mediated interaction defined above is the only active element in our microscopic model. 
Thus, in contrast to systems such as collections of active colloids or self-propelled rods, here activity manifests itself not in the self-propulsion, but in the relative motion of the microscopic constituents of the system.

\subsection{Coarse-graining}
\label{sec:coarsegrain}
Having set up the microscopic model, we can now proceed by coarse-graining it using the Boltzmann-Ginzburg-Landau (BGL) approach. 
The starting point of the BGL approach is the one-particle probability density function (PDF) $P(\rv, \phi)$ giving the number density of microtubules with center of mass position $\rv$ and orientation angle $\phi$. Integrating this PDF over all angles and positions yields the total number of microtubules in the system. The Fourier modes of this PDF in angular space read:
\begin{equation*}
    P_k(\rv) = \frac{1}{2\pi}\int_0^{2\pi}\dd{\phi}P(\rv, \phi)e^{-ik\phi}.
\end{equation*}
These modes have a clear physical interpretation. 
The zeroth mode can be identified with a coarse-grained filament density $\rho(\rv):=2\pi P_0(\rv)$. 
The first mode is proportional to the local mean orientation of the filaments, for which we define a polarization field $\polv(\rv):=2\pi\expval{\rho \, \nv}$; 
the second mode corresponds to the nematic order, and so on. Starting from $P(\rv,\phi)$, the BGL procedure allows us to obtain hydrodynamic equations for these coarse-grained fields.

The full calculation, including spatial dependence, is detailed in App.~\ref{app:BGL}. However, the emergence of global order can already be captured by studying a spatially homogeneous system. In this case, the PDF reduces to $P(\rv, \phi)=P(\phi)$. Its time evolution is described by a Boltzmann-like kinetic equation, which reads:
\begin{align}
\label{eq:kineticPhi}
\delt P(\phi)&= -D_r\partial^2_\phi P(\phi) + \int_{-\wcrit}^{\wcrit}\dd{\omega}P(\phi_-) P(\phi_+)g(\omega)\nonumber\\
&\phantom{{}={}}+ \int_{\wcrit}^{2\pi{-\wcrit}}\dd{\omega}P\left(\phi_- +\frac{\pi}{2}\right)P\left(\phi_+ +\frac{\pi}{2}\right)g(\omega)\nonumber\\
&\phantom{{}={}}-\int_{-\pi}^{\pi}\dd{\omega}P(\phi+\omega)P(\phi)g(\omega).
\end{align}
Here, the first term describes rotational diffusion with a rotational diffusion constant $D_r$. 
The three integrals (collision integrals) result from the motor-mediated interactions: the first two represent gain terms, respectively from the polar and antipolar interactions (with ${\phi_\pm=\phi \pm \omega/2}$), while the third integral is a loss term. 
The integrands are proportional to ${g(\omega) = G \, m \, \abs{\sin\omega}L^2}$, which is the rate of interaction of a given microtubule with filaments oriented at an angle $\omega$ with respect to it. It results from the integration over all possible relative positions of two interacting partners (see App.~\ref{app:BGL}). Its dependence on $\omega$ reflects the fact that collinear filaments need to be very close to intersect, whereas perpendicular filaments intersect over a large range of positions.

In the following, we measure space in units of the microtubule length $L$ and time in units of $D_r^{-1}$. Moreover, we introduce the parameter $\alpha=D_r/G$, which can be read as a passive-to-active ratio ($D_r$ being the rate of thermal diffusive rotation and $G$ the rate of motor-mediated interaction). This is a crucial dimensionless quantity in our model. By comparing the time scales of passive and active processes, it yields a measure of their relative importance in the dynamics of the motor-filament mixture. Since we absorbed the mean motor density $\bar{m}_\mathrm{ph}$ into the interaction rate $G$, higher motor concentrations result in a lower value of $\alpha$, reflecting the increased activity of the system. We use this parameter to rescale the probability density as $P\to \alpha^{-1}L^2 P$, making the resulting equations dimensionless.

Using the above rescalings, we decompose $P(\phi)$ into its Fourier components $P_k$ and project Eq.~\eqref{eq:kineticPhi} onto these components. This gives their time evolution, which reads:
\begin{align}
\label{eq:kineticPk}
\delt P_k&= -k^2P_k + m\sum_{q} f(k,q)P_qP_{k-q},
\end{align}
where the first term reflects rotational diffusion and the second term results from the collision integrals in Eq.~\eqref{eq:kineticPhi}. 
The full expression for the factor $f(k,q)$, depending on the critical angle $\wcrit$, is given in App.~\ref{app:BGL}.

To identify ordering transitions, we study the stability of the isotropic homogeneous state with $P_0=\bar\rho/(2\pi)$, where $\bar\rho$ is the mean value of the microtubule density field $\rho$, and $P_k=0$ for all $\abs{k}>0$~\footnote{Note that, since we rescaled the PDF $P$ by $\alpha^{-1}\propto\bar{m}_\mathrm{ph}$, the mean value of the field $\rho$ is proportional to the product of the physical average concentrations of microtubules and motors, $\bar\rho\propto\bar{\rho}_\mathrm{ph}\cdot\bar{m}_\mathrm{ph}$.}.
Linearizing Eq.~\eqref{eq:kineticPk} around this state, we obtain for $k\neq 0$:
\begin{align}
\label{eq:decayPk}
\delt P_k&= k^2[m\bar\rho/\rho^{(k)}_c-1]P_k.
\end{align}
Here, we have introduced a critical density for every mode, ${\rho^{(k)}_c := 2\pi k^2/[f(k,0)+f(k,k)]}$. 
For ${m\bar\rho>\rho^{(k)}_c}$, the corresponding mode experiences exponential growth, and the isotropic state is unstable towards the emergence of orientational order. For our system, $\rho^{(k)}_c$ can be positive only for $k=\pm 1$ (corresponding to polar order) and $k=\pm 2$ (nematic order). This limits the possible ordering transitions to these two types of order. The corresponding critical densities read:
\begin{subequations}
\label{eq:rhoc}
\begin{align}
    \rho^{(p)}_c:=\rho^{(\pm 1)}_c &= \frac{3\pi}{2-6\cos(\wcrit/2)-2\cos(3\wcrit/2)},\\
    \rho^{(n)}_c:=\rho^{(\pm 2)}_c &= -\frac{12\pi}{1+3\cos(2\wcrit)}.
\end{align}
\end{subequations}
Figure~\ref{fig:micro}b shows the dependence of the two critical densities on the critical angle $\wcrit$. Depending on this parameter, different ordering transitions can take place, with polar order emerging at large $\wcrit$ and nematic order emerging for a range of critical angles around $\wcrit=\pi/2$. In particular, the choice $\wcrit=\pi$ taken in Refs.~\cite{AransonTsimring2006,MaryshevPRE2018} leads to a motor-mediated polar ordering transition, while the parity symmetric case $\wcrit=\pi/2$ studied in Ref.~\cite{MaryshevSM2019} leads to a nematic ordering transition.
The existence of polar and nematic order transitions in interacting filament mixtures mirrors what has been found in Ref.~\cite{Ahmadi2006} using a different coarse-graining approach, where the role of $\wcrit$ is replaced by the ratio of polar and nematic alignment rates $g$.

Intriguingly, there is a range of values of $\wcrit$ for which the isotropic state is unstable towards the emergence of both polar and nematic order (Fig. \ref{fig:micro}b). This can lead to the coexistence of patterns involving both kinds of order, as has been investigated for self-propelled spherical particles with aligning collisions \cite{Denk2020}. In this work, we limit ourselves to the case where the only kind of orientational order exhibiting an ordering transition is polar order. This corresponds to critical angles larger than $\wcritqp\approx 107.7^\circ $, and to densities lower than the nematic critical density, $m\rho<\rhocn$. Thus, from now on, only the polar critical density will be relevant in our theory, and for simplicity we will write $\rhoc$ instead of $\rhocp$. Importantly, this assumes the parity symmetry of the interaction is broken.

To obtain a finite set of equations, the infinite sum over all Fourier modes appearing in Eq.~\eqref{eq:kineticPk} has to be truncated. In App.~\ref{app:BGL}, we do this using the Ginzburg-Landau closure, which assumes that the system is close to the polar critical density, so that $\bar\rho-\rhoc=:\epsilon^2$ is small. Then, it can be shown that the Fourier modes scale as $P_k\sim \epsilon^\abs{k}$, so that higher Fourier modes of the PDF can be neglected. Furthermore, reintroducing spatial dependence into the PDF, a gradient expansion is performed, with gradients of the PDF scaling as $\nabla\sim\epsilon$. Finally, in the regime described above, i.e., for $\wcrit>\wcritqp$ and $m\rho<\rhocn$, the nematic order can be adiabatically eliminated. Truncating the equations at the lowest order in $\epsilon$, this procedure yields deterministic equations for the MT density and the polarization field.

\subsection{\label{sec:motorfield}Motor field}
The interaction rate in our model is proportional to the local motor density $m$, which can vary in space and time. In solution, molecular motors are subject to Brownian motion, which makes them diffuse. Additionally, they can bind and unbind from microtubules. In their bound state, they walk along filaments, experiencing directed transport in regions where the microtubules show polar order. For weak spatial dependence of the concentration profiles and rapid attachment and detachment dynamics, the free and bound motor populations are related linearly by a local reactive balance \cite{AransonTsimring2006}. Thus, they are both proportional to the (total) local motor concentration $m$.

Under these conditions, the dynamics of $m$ is determined by diffusion and advection along the mean orientation of the microtubules. Introducing a diffusive constant $D_m$ and an effective advective velocity $v_m$, it reads \cite{Lee2001}:
\begin{eqnarray}
\label{eq:m}
    \partial_t m = D_m\laplacian m - v_m\grad(m\polv).
\end{eqnarray}
For our choice of units, $D_m$ is the ratio between $D_r^{-1}$ and the diffusive time scale of the motors along one filament length, and thus it is a large number. Therefore, the dynamics of the motor field will relax relatively fast to its stationary configuration. Assuming flux balance, we can estimate the amplitude of the gradients of the motor field via $\grad m=\gamma m\polv$, with the motor Péclet number $\gamma:=v_m/D_m$. In the following, we focus on the case of small Péclet number $\gamma$, implying that the gradients in $m$ are negligible compared to the other terms in the hydrodynamic equations, as we discuss below.

\subsection{Derived continuum equations}
\begin{figure}
\includegraphics[width=.92\linewidth]{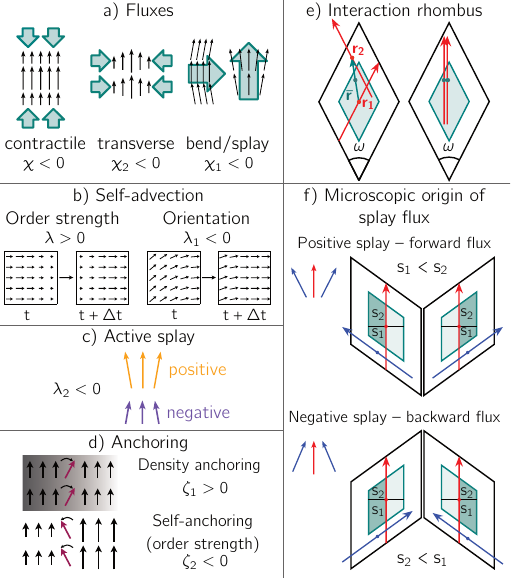}
\caption{\label{fig:terms}
Illustration of the behavior encoded in the terms of Eq.~\eqref{eq:der}. a) Polarization-induced fluxes in the $\rho$-equation. The black arrows indicate the polarization and the density flux is shown in teal. The contractile flux accumulates density in highly polar ordered regions along the order axis for $\chi=\chi_1+\chi_2<0$. Conversely, $\chi_2<0$ leads to a transverse flux into ordered regions perpendicular to the order axis. Splay deformations in the polarization vector cause a splay flux along the vector $(\div\nv)\nv$, and bend deformations induce a flux pointing into the inside of the bend, for $\chi_1<0$. b) The self-advection of the polarization amplitude and that of its direction are controlled by two different coefficients. The former is advected along $\polv$ for $\lambda=\lambda_1+\lambda_2-2\zeta_2>0$, while the latter is self-advected according to $\lambda_1$ alone, which is typically negative in our model. c) The active splay suppresses the order in regions of negative splay $\div\nv<0$, and enhances it for positive splay, for $\lambda_2<0$. d) The polarization tends to align along gradients in $\rho$ for $\zeta_1>0$ (lighter regions denote higher $\rho$), and against gradients of the order strength for $\zeta_2<0$. e) Definition of the interaction rhombus. For an intersection to occur, given a filament at $\rv_1$ and a second filament at an angle $\omega=\phi_2-\phi_1$ to the first one, the possible values of $\rv_2$ define a rhombus centered at $\rv_1$ of side length $L$ and aperture $\omega$ oriented along the bisector $\phi=\phi_2-\phi_1$. For $\abs{\omega}<\wcrit$, an interaction will align the filaments and slide them to their common center of mass at $\rv_1+\xiv/2$. The positions of the filaments after the interaction define a smaller rhombus of side length $L/2$, shown here in teal. f) The emergence of a splay flux from the microscopic interaction is explained using the interaction rhombus. For positive splay (top row), there are more filaments with $\omega>0$ to the left, and more filaments with $\omega<0$ to the right of a given filament. Due to this asymmetry, the rhombus of possible positions after the interaction acquires more weight on one side, shown in dark teal. In both cases, this results in an effective flux in the forward direction, since the area $s_2$ in the front of the original position is larger than the area $s_1$ in its back. For negative splay (bottom row) the argument is inverted, with $s_1>s_2$, so that the net flux is backward. Overall, this results in a splay flux along $(\div\nv)\nv$.}
\end{figure}

\label{sec:der_eq}
The full BGL procedure, elaborated in App.~\ref{app:BGL}, yields equations for the microtubule density field $\rho$ and the polarization field $\polv$:
\begin{subequations}
\label{eq:der}
\begin{eqnarray}
    \partial_t\rho &=& \laplacian(D_\rho\rho + \nu m\rho^2 + \alpha_2\rho^2+\alpha_3\rho^3)\nonumber\\
    &&{}+\partial_i\partial_j(\chi_1p_ip_j)+\laplacian(\chi_2p^2)\,,\label{eq:der_rho}\\
    \partial_t \polv &=& \left[(m\rho/\rhoc-1)-\beta m^2p^2\right]\polv\nonumber\\
    &&{}+\kappa_1\laplacian\polv+\kappa_2\grad(\div\polv)\nonumber\\
    &&{}-\lambda_1(\polv\vdot\grad)\polv-\lambda_2(\div\polv)\polv\nonumber\\ &&{}+\zeta_1\grad\rho+\zeta_2\grad p^2\,.\label{eq:der_p}
\end{eqnarray}
\end{subequations}
These two equations, together with Eq.~\eqref{eq:m}, constitute our continuum model for microtubule-motor mixtures. The coefficients appearing in these equations are all functions of the microscopic parameters $\wcrit$ and $\eta$, as well as the mean microtubule density $\bar\rho$ and the local motor density field $m$. The dependence of the coefficients on the microscopic parameters is detailed in App.~\ref{app:coefficients}. Note that, due to the rescaling of the PDF, both $\rho$ and $\polv$ are measured in units of $\alpha/L^2$.

Throughout the BGL procedure, we keep $m$ constant, promoting it to a field in the final equations. This assumes that the Péclet number $\gamma$ is small, so that the gradients of $m$ are of order $O(\gamma)$ and can be neglected in comparision with the rest of the equations. In the density equation \eqref{eq:der_rho}, we place $m$ inside the innermost gradient to ensure MT density conservation. A different choice would lead to a term of order $O(\gamma)$, and hence it doesn't affect the phenomenology of the equations for small $\gamma$. An alternative approach to the one proposed here, discussed in Ref.~\cite{AransonTsimring2006}, is to take into account the spatial dependence of $m$ throughout the BGL derivation and to evaluate the field at the center of mass of a filament pair in the collision integrals to ensure density conservation. This gives rise to terms involving $\grad m$, so the difference to our model is again of order $O(\gamma)$.

Equation \eqref{eq:der_rho} is a continuity equation for the conserved microtubule density $\rho$. The first line of that equation does not involve the polarization field. It consists of: i) a diffusion term with translational diffusion constant $D_\rho$; ii) a term emerging from the motor-mediated interaction proportional to $\nu$, which is typically (i.e., in most of parameter space, and in particular in the regions which will be relevant for this work) negative, thus resulting in an anti-diffusive isotropic contraction; iii) two terms resulting from excluded volume interactions, with $\alpha_2=\alpha/(32\pi)$ and $\alpha_3=\alpha^2/(192\pi)$, which are proportional to the passive-to-active ratio $\alpha=D_r/G$ defined earlier. To maintain the high-density effects that arise due to steric repulsion, we do not linearize these terms in $\rho$. Overall, the first line of Eq.~\eqref{eq:der_rho} gives rise to an effective isotropic diffusive flux, whose strength depends on the local values of $m$ and $\rho$.

The second line of Eq.~\eqref{eq:der_rho} consists of fluxes that arise due to inhomogeneities in the polarization field. We write the polarization vector $\polv$ as $\polv=p\nv$, with the polarization amplitude (or order strength) $p$ and the director $\nv$. Then, as shown in App.~\ref{app:terms_rho}, the polarization-dependent fluxes can be decomposed into four contributions (see Fig.~\ref{fig:terms}a). The first two contributions are due to gradients in the amplitude $p$: i) the \emph{contractile flux} depending on $\chi:=\chi_1+\chi_2$ (which is typically negative), resulting in the accumulation of microtubule density longitudinally to the polarization into regions of high polar order; ii) the \emph{transverse flux}, depending on $\chi_2$ (which is typically negative for sufficiently small critical angles $\wcrit$), accumulating density into ordered regions perpendicularly to the order axis. The other two contributions are due to gradients in the director field $\nv$: iii) the \emph{bend flux}, controlled by $\chi_1$ (which is typically negative), moving density into the inside of a bend (i.e., towards the center of its osculating circle), transversally to the order; iv) the \emph{splay flux}, also controlled by $\chi_1$, which advects density along the order for positive splay ($\div\nv>0$) and against it for negative splay. These fluxes are the polar equivalent of the active currents in an active nematic, where $\chi=\chi_1/2=-\chi_2$ holds \cite{Marchetti2013}. They arise due to the motor-mediated relative sliding of filament pairs.

The polarization equation, Eq.~\eqref{eq:der_p}, has a Toner-Tu-like form \cite{TonerTu1998}. Two important differences to the Toner-Tu equations should be noted: The first is the dependence of most coefficients on an additional field $m$, which introduces a local modulation of their strength (see App.~\ref{app:coefficients}). The second difference lies in the coupling to the density field. In the Toner-Tu model, the polarization vector has a dual role: it indicates the local mean orientation of the system's constituent particles, but also the velocity of their self-propulsion. This results in an advective term of the form $-\div\polv$ that appears in the density part of the Toner-Tu equations. Here, instead, the coupling of the density to the polarization is realized exclusively through the $\chi_{1,2}$ terms discussed above. The absence of an advective term in the density equation Eq.~\eqref{eq:der_rho} reflects that the only motion introduced by activity in our model is relative motion, as opposed to self-propulsion or net sliding.

The various terms in the polarization equation \eqref{eq:der_p} are discussed in depth in App.~\ref{app:terms_p}. Here, we note that the first two lines can be read as terms emerging from a Ginzburg-Landau free energy akin to model A dynamics \cite{Hohenberg1977}. The first term leads to the emergence of a non-zero order parameter in regions where $m\rho>\rhoc$ holds, whose saturation is controlled by the cubic $\beta$ term. For a homogeneous system, this results in an equilibrium polarization given by:
\begin{equation}
\label{eq:p0_GL}
p_0^2=\frac{m\rho-\rhoc}{\beta m^2\rho_c}.
\end{equation}
On the other hand, the terms in the second line of Eq.~\eqref{eq:der_p} penalize gradients in the polarization field. The stiffness $\kappa:= \kappa_1+\kappa_2$ controls the cost of splay deformations as well as longitudinal variations of the order strength, whereas bend deformations and transversal variations of the order strength are penalized by $\kappa_1$.

The rest of the terms in the polarization equation are all proportional to the antiparallel sliding coefficient $\eta$ (see discussion in App.~\ref{app:terms_micro}). Two self-advection effects must be distinguished (see Fig.~\ref{fig:terms}b): the self-advection of the order strength controlled by $\lambda:=\lambda_1+\lambda_2-2\zeta_2$, which is typically \emph{positive}, moves patterns in the polarization amplitude $p$ \emph{along} the direction of the polar order; on the other hand, the self-advection of the orientation, controlled by $\lambda_1$ only, which is typically \emph{negative}, advects patterns in the order direction $\nv$ \emph{against} the direction of the polarization. These self-advection effects break time-reversal symmetry, so that the polarization equation can only be derived from a free energy when both of them vanish, $\lambda=0=\lambda_1$ \cite{Marchetti2013}.

The term proportional to $\lambda_2$, which is equal to $\lambda_1$ in the derived model and thus typically negative, gives rise to an ``active splay'' effect (see Fig.~\ref{fig:terms}c). For $\lambda_2<0$, this enhances the polarization in regions of positive splay ($\div\nv>0$) and inhibits the polar order in regions of negative splay ($\div\nv<0$).

Finally, the last line of Eq.~\eqref{eq:der_p} leads to anchoring effects, i.e., the alignment of the polarization field with respect to gradients (see Fig.~\ref{fig:terms}d). The first term with coefficient $\zeta_1>0$ leads to the alignment of $\polv$ along gradients of $\rho$ (``density anchoring''), whereas the second term with coefficient $\zeta_2<0$ leads to the rotation of $\polv$ away from regions of strong polar order (with high $p^2$) and towards isotropic regions (where $p^2$ is small, ``self-anchoring''). 

While the quantitative dependence of the hydrodynamic coefficients in Eq.~\eqref{eq:der} on the parameters of the microscopic model can only be extracted by performing the full Boltzmann-Ginzburg-Landau derivation, it is instructive to motivate why they show the signs they do by heuristic microscopic arguments. Here, we explain the emergence of the splay flux with $\chi_1<0$ as an example and refer the reader to App.~\ref{app:terms_micro} for a discussion of the other terms. To understand the splay flux, it is useful to introduce the interaction rhombus (see Fig.~\ref{fig:terms}e). This rhombus is centered at the position $\rv_1$ of the center of mass of a given filament; it is defined by the possible positions $\rv_2$ a second filament can occupy such that the two filaments intersect, with a given intersection angle $\omega=\phi_2-\phi_1$. The resulting rhombus has side length $L$ and aperture $\omega$. Before an interaction, the two filament centers are separated by a vector $\xiv=\rv_2-\rv_1$, whereas after the interaction they will have moved relative to each other until they both lie at the common center of mass $(\rv_2+\rv_1)/2$, assuming $\abs{\omega}<\wcrit$ so that no antiparallel sliding is involved. The possible positions of the filaments after the interaction define a smaller rhombus of side length $L/2$ centered within the interaction rhombus (shown in teal in Fig.~\ref{fig:terms}e). Now, in a situation with positive splay, given a filament at $\rv_1$, on average there will be more filaments at a positive angle $\omega>0$ to the left of $\rv_1$ than to its right, and more filaments at a negative angle $\omega<0$ to the right than to the left. For geometric reasons, illustrated in Fig.~\ref{fig:terms}f, in both these situations it is more probable that the filament at $\rv_1$ slides forward than backward when an interaction occurs, because a larger portion of the side of the interaction rhombus that is favored by splay (left or right) lies in the front compared to the back. This results in an overall forward flux for positive splay. In the case of negative splay, an analogous argument leads to a backward flux, so that indeed we find that the microscopic interaction model implies the emergence of a flux along the splay vector $(\div\nv)\nv$, which entails $\chi_1<0$. The bend flux can be explained analogously, exchanging the roles of the left/right directions with those of the forward/backward directions.

\section{Phase Diagram of the Derived Model}
\label{sec:PD}
In this section, we focus on the derived model, Eqs.~\eqref{eq:m} and \eqref{eq:der}, and inspect its behavior by performing numerical simulations using finite element methods (see App.~\ref{app:sim} for details). To this goal, we initiate the system in the homogeneous isotropic state ($\polv=0$  with constant $\rho=\bar\rho$ and $m=1$) and perturb it with small-amplitude noise. Initiating the system above criticality, i.e., with $\bar\rho>\rho_c$, the isotropic state rapidly develops non-zero orientational order at early times. As time progresses, different patterns emerge depending on the choice of the parameters.
\subsection{Branching instability and active foams}
\begin{figure}
\includegraphics[width=\linewidth]{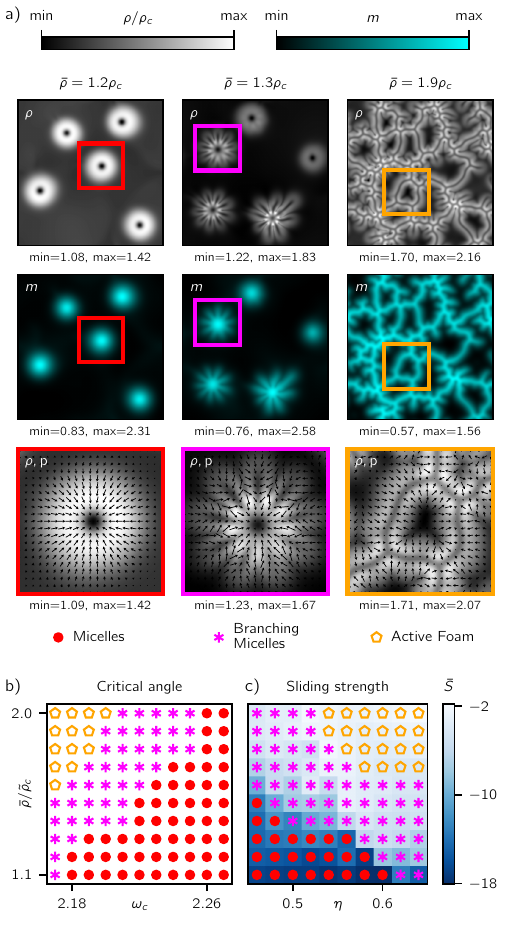}
\caption{\label{fig:PD}Transition between active supramolecular assemblies. a) Snapshots (top: $\rho$, center: $m$, bottom: zoomed detail of $\rho$ and $\polv$)  from numerical simulations run for $t=500$ at parameter values $\wcrit=2.18$, $\eta=0.58$, $\alpha=0.7$, $D_m=0.2$, $v_m=0.004$ in a system of size $60\times 60$ for different mean microtubule densities $\bar\rho = 1.2, 1.5, 2.0$ (in units of $\rhoc$). See Videos S1-3 for the time evolution \bibnotemark[SI].  b) Phase diagram as a function of the critical angle $\wcrit$ and the mean density $\bar\rho$ for fixed $\eta=0.58$ at the parameter values listed above. We measure the density in units of the critical density $\rhoc$ at $\wcrit=2.18$, denoted as $\tilde\rho_c$. The phases were determined by visual inspection, averaging over five initial conditions. The Branching Micelle phase has at least one micelle developing branches, and the Active Foam phase shows at least one closed loop. c) $\eta$ and $\rho$ phase diagram for fixed $\wcrit=2.18$ at the parameter values listed above. The color in the background gives the value of the parameter $\bar{S}$ defined in the text, which is a proxy for the emergence of structure on short length scales.}
\end{figure}

For sufficiently small critical angles $\wcrit$ and intermediate antiparallel sliding strengths $\eta$, we observe a transition between different inhomogeneous states as we increase the initial microtubule density $\bar\rho$. At low densities, radially symmetric aster-like structures take shape (see left column of Fig.~\ref{fig:PD}a and Video S1 \bibnote[SI]{See Supplemental Material at (link provided after publication)}). At the center of these structures, a defect is found, where the polarization vanishes and the microtubule density has a minimum. Around this defect, a ring of elevated microtubule density forms, in which the polarization field points inward. In contrast, the motor density shows a peak in the middle of the aster. In the following, we refer to these structures as active \emph{micelles}, in analogy to lipid monolayer rings \cite{Terentjev2015}. As will become clear below, this terminology accounts for the fact that the microtubule-depleted zone in the center can be quite large, in contrast to what is normally referred to as an aster (e.g., in Ref.~\cite{AransonTsimring2006}).

For higher initial mean densities $\bar\rho$, while micelles form at early times, they become unstable and lose their rotational symmetry. The high-density ring breaks apart, and several branches form that extend radially outward. Along the branches, the polarization field orients itself perpendicularly to the branch on each side, in a bilayer-like fashion. These branching micelles are well-separated at intermediate densities (see middle column of Fig.~\ref{fig:PD}a and Video S2 \bibnotemark[SI]).

At the highest densities, the branches emerging from different unstable micelles connect, forming bilayers that give rise to a foam-like network, extending throughout the simulated system (see right column of Fig.~\ref{fig:PD}a and Video S3 \bibnotemark[SI]). Each cell of the foam is thus encapsulated by bilayer edges which join together at the vertices of the foam. Each cell center is depleted of both microtubules and motors, with the polarization field pointing outward. As time progresses, the active foam undergoes constant reconfiguration, with new bilayers branching out of its edges and connecting to other parts of the network. This leads to the formation of new vertices and new cells.

In Figures \ref{fig:PD}b-c, we show the phase diagram for different initial densities $\bar\rho$ and varying sliding strengths $\eta$ as well as critical angles $\wcrit$. We distinguish the phases as follows: In the Micelle phase, a population of stable micelles is formed. If at least one of these micelles exhibits branching, we assign the parameter set to the Branching Micelles phase. Finally, we define Active Foams as networks of bilayers with at least one closed loop. To assemble the phase diagrams, we identified these phases via visual inspection and averaged our categorization over five simulations. As a quantitative measure of the transition between the various active supramolecular assemblies, complementing visual inspection, we use the structure factor $S(\vc{q})=\abs{\hat{\rho}_{\vc{q}}}^2$, with $\hat\rho_{\vc{q}}$ being the Fourier transform of the MT density. For branching micelles and foams, this structure factor will have heavy tails for large $\abs{\vc{q}}$ due to the short-wavelength detail introduced by the bilayers. The integral $\bar{S} = \int\dd[2]{q}\log{S(\vc{q})}$ is a measure of the strength of these heavy tails, which we plot in Fig. \ref{fig:PD}c to show the correspondence with the micelle instability transition found via visual inspection.

The phase diagrams in Fig.~\ref{fig:PD}b-c show that both the critical angle $\wcrit$ and the antiparallel sliding strength $\eta$ are important parameters for the instability that drives the transition from micelles to branching micelles, and further on to active foams. Indeed, the transition is only found for sufficiently small values of $\wcrit$: this highlights the relevance of \textit{antiparallel aligning} interactions for the instability. Antiparallel alignment alone is not sufficient, however: for sufficiently small $\eta$, micelles are always stable, hinting at the role \textit{antiparallel sliding} plays in the transition. These two ingredients, antiparallel alignment and sliding, are thus crucial for the branching instability, which leads to the formation of the bilayers that constitute the elemental building block of the active foam networks. On the microscopic level, this can be explained if we think of the bilayer as two opposing ordered monolayers that partially overlap. Parallel alignment stabilizes the polar order within one monolayer, ensuring that all filaments point in the same direction. Conversely, as pairs of opposing filaments belonging to different monolayers interact with each other, antiparallel alignment is essential to guarantee that they stay aligned along the axis perpendicular to the bilayer. Indeed, if they were to interact exclusively via parallel alignment ($\wcrit=\pi$), they would rotate away from their initial configurations towards the bisector, thus disrupting the antiparallel orientational order of the two opposing monolayers. Thus, a sufficiently small value of $\wcrit$ preserves the \emph{orientational} arrangement of the bilayer. Antiparallel sliding, on the other hand, guarantees that pairs of opposite filaments separate upon interaction, sliding back to their original positions on either side of the bilayer. This polarity sorting mechanism keeps the two monolayers well-defined, preserving the \emph{spatial} arrangement of the bilayer, i.e., its separation into opposing monolayers.

While this argument explains why $\wcrit$ should be sufficiently smaller than $\pi$ to ensure bilayer stability, the fact that the bilayer is an intrinsically polar object (with the polarization having opposite sign on either side) requires that it should also be sufficiently large for polar order to survive over the dominance of nematic order, as emerges from the discussion in sec.~\ref{sec:coarsegrain} (see also Fig.~\ref{fig:micro}b). Indeed, the parity symmetric case $\wcrit=\pi/2$ studied in Ref.~\cite{MaryshevSM2019} only showed the emergence of nematic patterns, precluding the formation of polar bilayers. Hence, the parity symmetry breaking in the interaction rules is essential: A sufficiently large range of intersection angles leading to parallel alignment ($\wcrit\gg\wcritqp$) is needed for polar order to be dominant, whereas a sufficiently large range of intersection angles with antiparallel alignment ($\wcrit\ll\pi$) ensures the preservation of the bilayer structure, thus allowing for the formation of active foams.

In the simulations presented in Fig.~\ref{fig:PD}, we kept $D_m=0.2$ and $v_m=0.004$ fixed. While changing the overall magnitude of these parameters does not significantly affect the phenomenology (determining the relative time scales of the microtubule and motor dynamics), the ratio between them, i.e., the motor Péclet number $\gamma=v_m/D_m$, does change the observed patterns. In particular, for $\gamma = 0$ (no motor advection), instead of well-separated micelles, an aster network emerges, similar to the ones observed in earlier studies \cite{Lee2001,AransonTsimring2006,MaryshevPRE2018,Besse2022}. In these networks, the filament density does not fall off outside of the aster. Instead, it plateaus to a constant value until the boundary to a neighboring aster is reached, forming an aster network. At zero motor advection, no bilayers form and no active foams are observed in the part of phase space probed here. As $\gamma$ is increased away from zero, the aster network splits up into separated micelles. The width of the high-density ring around the center of each micelle decreases as $\gamma$ is increased. Likewise, for the branching micelle and active foam phase, increasing $\gamma$ leads to a decrease in the total width of the bilayer. Concurrently, the intermediate region between the micelles, as well as between different bilayers, is depleted of both microtubules and motors, with almost no polar order. Thus, the inhomogeneity of the motor field introduces a spatial organization of the system, with regions of locally increased activity (high $m$) exhibiting the formation of ordered supramolecular assemblies like micelles and bilayers, and regions of decreased activity (low $m$) constituting the disordered background separating the ordered structures.

\subsection{Fingering instability}
Finally, we investigated the role of the passive-to-active ratio $\alpha$ in the micelle phase. Physically, decreasing $\alpha$ corresponds to increasing the ratio between the motor-mediated interaction rate $G$ and the rotational diffusion rate $D_r$. Recalling that $G\propto\bar{m}_\mathrm{ph}$ and $\rho\propto\bar{m}_\mathrm{ph}\bar\rho_\mathrm{ph}$, this implies that decreasing $\alpha$ while keeping $\rho$ constant amounts to increasing the mean motor concentration $\bar{m}_\mathrm{ph}$ in the system (thereby making the interaction rate $G$ for a given pair of intersecting filaments larger) while decreasing the microtubule density $\bar\rho_\mathrm{ph}$, such that the overall rate of interaction events in the system (which is proportional to $\bar{m}_\mathrm{ph}\bar\rho_\mathrm{ph}$) stays constant.

\begin{figure}
\includegraphics[width=\linewidth]{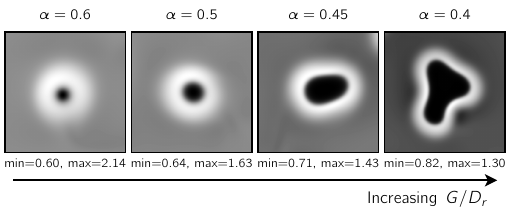}
\caption{Micellar shape instability. \label{fig:PDalpha}Snapshots ($12\times 12$) from simulations at $\wcrit=2.18$, $\eta=0.5$, $D_m=0.2$, $v_m=0.002$ with initial mean density $\bar\rho=1.1\rhoc$, run for $t=500$ in a $60\times 60$ geometry, for different values of $\alpha$. As $\alpha$ is decreased, the micelles seen in the simulations ``open up'', leading to a larger depleted region in the center ($\alpha=0.5$). For even lower values of alpha, the micelles lose their circular shape, elongating along one axis at first ($\alpha=0.45$) and developing more pronounced interface modulations at the lowest values probed ($\alpha=0.4$), see also Video S4 \bibnotemark[SI]. Since $\alpha=D_r/G$, decreasing its value corresponds to increasing the importance of active processes (motor-mediated interaction with rate $G$) with respect to passive ones (rotational diffusion with rate $D_r$), for example by increasing the mean motor concentration.}
\end{figure}

Figure~\ref{fig:PDalpha} shows snapshots from numerical simulations for different $\alpha$. Decreasing $\alpha$ from the value $\alpha=0.7$ used for Fig.~\ref{fig:PD}, we observe an enlargement of the central depleted region, which is why we use the more general term ``micelle'' rather than ``aster''. Interestingly, similar hollow micelles were observed for high motor densities (corresponding to small $\alpha$) in agent-based simulations \cite{Ansari2022}. For even lower values of $\alpha$, the micelle loses its symmetric shape, elongating along one axis. For the lowest values of $\alpha$, it shows a more pronounced modulation of curvature along its perimeter, with the formation of protruding lobes that extend out of the micelle. This shape instability, which we refer to as fingering instability, is distinct from the branching instability shown in Fig.~\ref{fig:PD}, since the high-density ring delimiting the micelle is not broken up into bilayers. Instead, the ring itself is deformed, exhibiting a higher number of lobes as $\alpha$ is decreased. The shapes keep evolving dynamically as time progresses (see Video S4 \bibnotemark[SI]).

In summary, the numerical simulations we have performed show two distinct micelle instabilities: the first one is the branching instability, which leads to the formation of bilayers around the perimeter of the micelle, which at higher MT density connect together to form active foam networks; the second one is the fingering instability, where the high-density ring exhibits a shape modulation instead of breaking up. While the branching instability crucially relies on both antiparallel alignment (i.e., sufficiently small values of $\wcrit$) and antiparallel sliding (high $\eta$), the fingering instability appears at sufficiently small values of the passive-to-active ratio $\alpha$, corresponding to high motor concentrations.

\section{The Phenomenological Model}
\label{sec:phen}
In the model presented in Eq.~\eqref{eq:der}, all the coefficients are functions of the microscopic parameters, i.e., the critical angle $\wcrit$ and the sliding strength $\eta$, as well as the mean microtubule density $\bar\rho$ and the local motor density $m$. As we have seen, each model parameter (such as $\chi$, $\lambda_2$, etc.) is associated to a certain emergent mechanism (contractile flux, active splay, etc.). In the following, we will see how the interplay of these mechanisms controls the behavior of the system, selecting length scales and driving instabilities. To better understand the role of each mechanism involved, it is useful to generalize the model we have derived, allowing for independent variation of all the continuum model parameters. This generalization allows us to validate the analytical calculations we will perform in the following sections, by varying the importance of the various terms separately from each other. Furthermore, it allows the exploration of a larger fraction of parameter space, beyond the one defined by the functional relationship to the microscopic parameters. 

Physically, this abstraction beyond the derived model is motivated by the fact that a different set of microscopic interaction rules compared to the one proposed in this work would lead to a set of equations that may have different functional relationships of the coefficients, but that share the same structure regarding the terms appearing in the equation. This is true even for interaction rules that are too complex to allow for a derivation of the corresponding continuum model coefficients by hand. The reason for this is that the model includes all terms up to a certain order in the Ginzburg-Landau expansion (i.e., in the gradients and fields) that are allowed by symmetry for a system governed by the fields $\rho$, $\polv$ and $m$. The only exception is the advection term $-\div\polv$ in the $\rho$-equation Eq.~\eqref{eq:der_rho}, which is absent in our theory, since it doesn't arise in a system involving only relative motion of filament pairs. Studying the equations with independent coefficients allows for a more complete exploration of the physical behavior they can give rise to, extending the analysis to a broader class of models and actual physical systems. This bottom-down approach is phenomenological in nature, as it acquires generality in exchange for the loss of a connection to an interaction picture, thus complementing bottom-up approaches as the one discussed so far.

In general, the parameters involved in such a model can be arbitrary functions of $m$. For simplicity, we reduce that dependence by rewriting all the parameters that have more complicated functional relationships in the derived model ($\chi_{1,2}$, $\kappa_{1,2}$, $\lambda_{1,2}$, $\zeta_{1,2}$) as linear functions of the form $\chi_1=m\hat{\chi}_1$, where we denote the proportionality constants with hats. Furthermore, by rescaling the fields we can set $\rhoc=\beta=1$. We refer to the resulting equations as the \emph{phenomenological model}, which is given in App.~\ref{app:ph}.

\begin{figure}
\includegraphics[width=\linewidth]{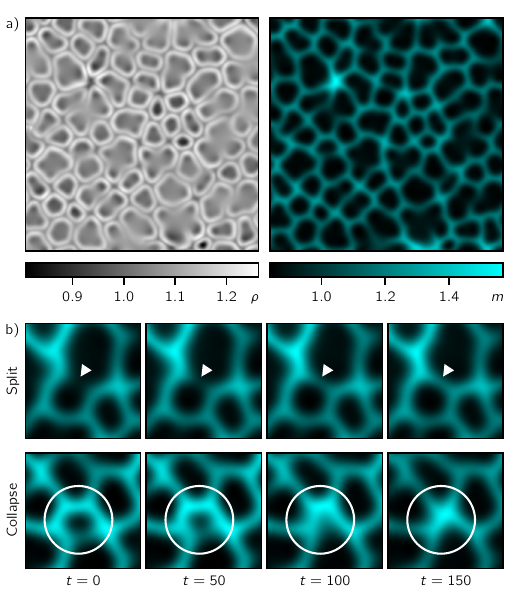}
\caption{\label{fig:phen_af}a) Snapshot from a simulation of the phenomenological model where foams appear (parameters listed in App.~\ref{app:sim}). The parameters of the model can be varied such that the foam is much smoother than those seen in the derived model, and longer-lived (see Video S5 \bibnotemark[SI]). b) Detailed snapshots of the motor field, showing the activity of the foam. This activity manifests itself in cell splitting events, where new bilayers form connecting different parts of the network, as well as cell collapse events, where bilayers collide with each other, leading to the annihilation of existing cells.}
\end{figure}
Since the phenomenological model is a generalization of the equations derived in sec.~\ref{sec:derivedmodel}, all phases described in sec.~\ref{sec:PD} can be reproduced in the model, including micelles and active foams. Furthermore, the freedom posed by the phenomenological model allows us to reproduce the experimental phases more closely. For example, in Fig.~\ref{fig:phen_af}a and Video S5 \bibnotemark[SI] we show numerical simulations of this model in the active foam phase: these foams are less rough and more active compared to the ones seen in the derived model. The active foam cells evolve over time, showing cell division and cell collapse events that drive the sustained reconfiguration of the active foam for very long times (see Fig.~\ref{fig:phen_af}b) and have a very close resemblance to those seen in experiment \cite{Dogic2021}. These active foams exist over large regions of parameter space.

In the remainder of this paper, we study the various phases we have presented in sec.~\ref{sec:PD} by means of analytical methods, which will enable us to explain their phenomenology in terms of the interplay of different physical mechanisms. We will do this in two steps: In section \ref{sec:profiles}, we take a static perspective of the elementary structures we have observed – bilayers and micelles – and study what determines their concentration and polarization profiles and the selection of length scales. In section \ref{sec:instab}, on the other hand, we will turn to a dynamic standpoint, inspecting the mechanisms that underlie the two micellar instabilities described above. The phenomenological model will be used as a tool to validate the analytical derivations presented in those sections, by performing numerical simulations of the model while tuning the strength of the various mechanisms independently from one another.

\section{Stationary Profiles}
\label{sec:profiles}
In this section, we inspect the stationary concentration and polarization profiles of the micelle and bilayer solutions of the equations. This will allow us to understand the role of activity in selecting both the shape of these profiles and their characteristic length scales. We proceed as follows: in section \ref{sec:inprofile}, we examine the profiles of the interior of the bilayer. We explain how the microtubule density in the interior is depleted due to the contractile flux and how the interplay of the passive and active terms in the equations determines the width of the depleted region. We also inspect the micelle profile close to its center and show that the density dip seen there emerges through a similar mechanism as for the bilayer, up to the effect of splay. Then, in section \ref{sec:extprofile}, we turn to the region outside of the assembled structures, and investigate the role of motor inhomogeneity. Throughout this section and the next, the results apply both to the derived and the phenomenological models.

\begin{figure*}
\includegraphics[width=\textwidth]{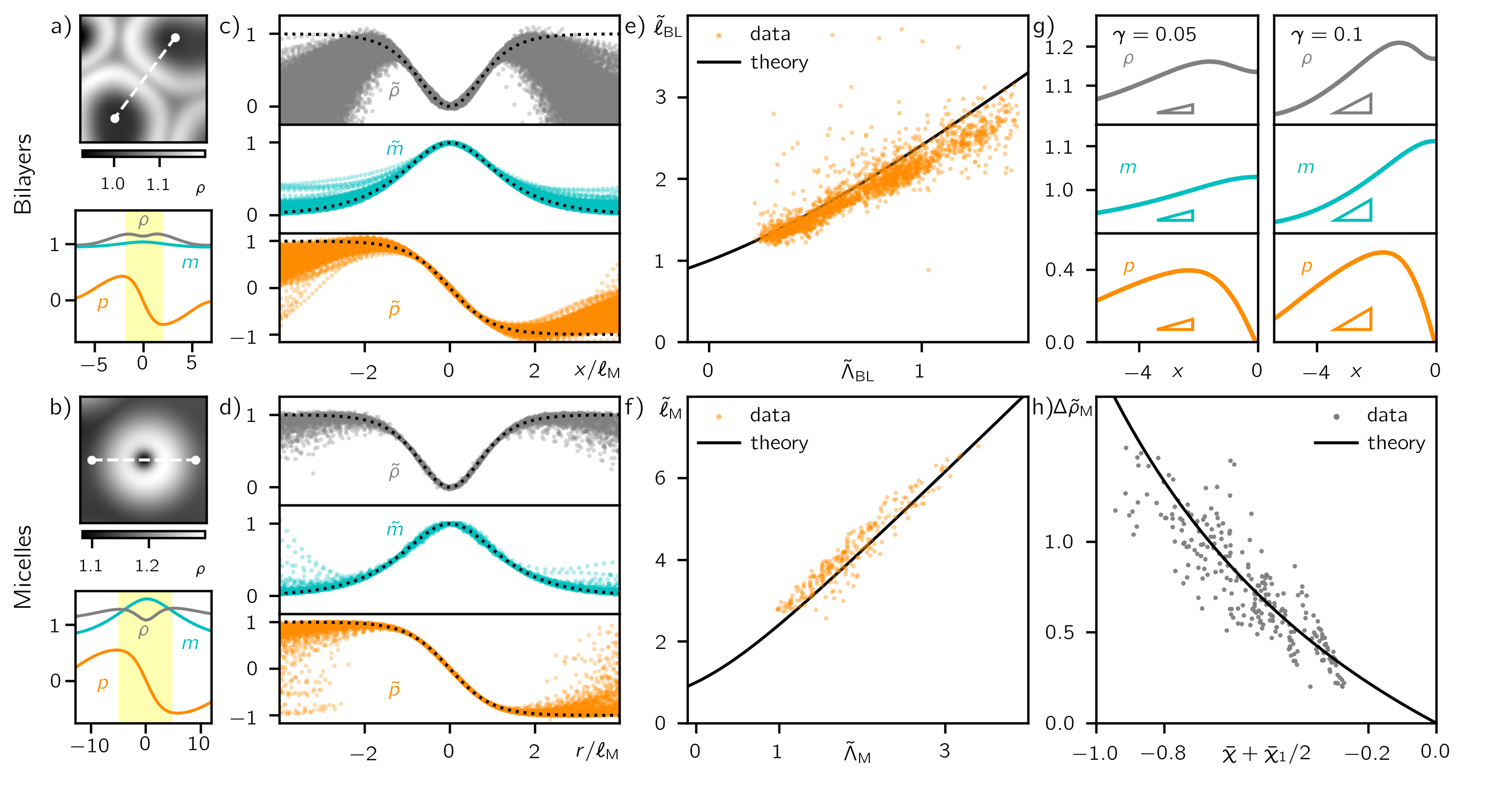}
\caption{\label{fig:profiles} Analysis of the bilayer and micelle cross-sections. a-b) Example for bilayer (a) and micelle (b) cross-section profiles. The shaded yellow regions mark the areas between the density and polarization peaks that delimit the interior of the bilayer and micelle. c-d) Simulating the phenomenological model for different values of anisotropic flux strengths $\hat\chi_{1,2}$, stiffness $\hat\kappa_1$, active splay coefficient $\hat\lambda_2$, anchoring coefficients $\hat\zeta_{1,2}$ and motor advection $v_m$ (see App.~\ref{app:sim} for the parameter values), cross-section profiles for bilayers (c) and micelles (d) can be extracted. Fitting these profiles with the functions given in Eqs.~\eqref{eq:rho_1Dsol}, \eqref{eq:p_1Dsol} and \eqref{eq:m_1Dsol}, one obtains the values for the fitting parameters $\rho_-$, $\Delta\rho$, $p_+$, $\ell$, $\gamma$ and $m_-$. Rescaling space with $\ell$ and the fields as $\tilde\rho=(\rho-\rho_-)/\Delta\rho$, $\tilde{m}=(m/m_-)^{1/\gamma p_+\ell}$, and $\tilde{p}=p/p_+$, the profiles (colored dots) all collapse on the same curves predicted by theory (black dotted lines). e-f) Using $p_+$, $\rho_-$ and $m_-$ from the fit in the Ginzburg-Landau term, as well as the mean values $\rho_0$ and $m_0$ in the interior of the bilayers (e) and micelles (f) for the dependencies in the other coefficients, the values of $\Lambda$ can be calculated from Eqs.~\eqref{eq:LBL} and \eqref{eq:Lmic}. Here, we plot the length scale $\ell$ extracted from the polarization fit against $\Lambda$. Rescaling the axes by $\tilde{\ell}=\sqrt{(m_-\rho_-/\rhoc-1)/(2\kappa)}\ell$ and $\tilde{\Lambda}=\sqrt{(m_-\rho_-/\rhoc-1)/(2\kappa)}\Lambda$, the data from different measurements collapse onto one curve, which theory predicts to be $\tilde\Lambda+\sqrt{\tilde\Lambda^2+1}$ for both bilayers (top) and micelles (bottom). g) Bilayer profiles for different values of $\gamma$. In the outer region (outside of the polarization and density maxima), the slopes of all three fields increase with $\gamma$. The triangles indicate the slopes predicted by theory using Eq.~\eqref{eq:ext_res} with the values of $\rho_0$, $p_0$, and $m_0$ at $x=-3.2$. h) Strength of density phase separation in the interior of the micelles versus $\bar\chi+\bar\chi_1/2$. Rescaling $\Delta\rho$ as $\Delta\tilde\rho=\Delta\rho/(\rho_0-\rho_c/m_0)$, where $\rho_0$ is the mean between the maximum and the minimum, the data collapses on a curve, as predicted by Eq.~\eqref{eq:rad_PS_Delta}.}
\end{figure*}

\subsection{Inner profiles}
\label{sec:inprofile}
\subsubsection{Bilayers}
The bilayer is a solution of Eqs.~\eqref{eq:der} in which the fields vary only in one direction (perpendicular to the bilayer) and the polarization is oriented along this direction, which we choose to coincide with the $x$-axis. The bilayer is delimited by maxima in the microtubule density $\rho$ on either side, with the polarization changing sign in the middle. We focus on the region between these maxima, which we refer to as the interior of the bilayer (see Fig.~\ref{fig:profiles}a). For small motor Péclet number $\gamma$, we can assume the value of the motor field to be constant in the area within the bilayer, so that $m=m_0$. Furthermore, assuming weak phase separation of the microtubule density, we can linearize it around a reference value $\rho_0$ in Eq.~\eqref{eq:der_rho}. With these simplifications, the stationary equations can be solved explicitly. Here we present the main results, directing the reader to App.~\ref{app:blprof} for the full calculation.

The density profile we obtain has the form:
\begin{equation}
\label{eq:rho_1Dsol}
    \rho(x)=\rho_-+\Delta\rho\tanh^2(x/\ell),
\end{equation}
where $\ell$ is a characteristic length scale which will be specified below [see Eq.~\eqref{eq:ell}]. The strength of phase separation $\Delta\rho$ determines the depth of the density dip, with the minimum value $\rho_-$ at the center of the bilayer and the maximum value $\rho_+$ at its boundaries given by $\rho_\pm=\rho_0\pm\Delta\rho/2$. For the bilayer, it reads:
\begin{equation}
\label{eq:1D_PS_Delta}
    \DrBL=\frac{-2\bar\chi}{2+\bar\chi}(\rho_0-\rhoc/m_0).
\end{equation}
Here, we have introduced the effective contractility ${\bar\chi=\chi/(\beta m_0\rho_c\Deff)}$, giving the ratio between the contractile flux coefficient $\chi$ and the effective isotropic diffusivity ${\Deff=D_\rho+2(\nu m_0+\alpha_2)\rho_0+3\alpha_3\rho_0^2}$ resulting from the various terms appearing in the first line of Eq.~\eqref{eq:der_rho}. When $\chi=0$, the phase separation $\Delta\rho$ in Eq.~\eqref{eq:1D_PS_Delta} vanishes, while it grows monotonously as $\chi$ increases in the negative direction. Thus, we conclude that the dip in the $\rho$ field at the center of the bilayer is due to the contractile flux, that accumulates microtubules into the ordered regions on either side of the bilayer. In Fig.~\ref{fig:profiles}c, we plot the bilayer profiles from numerical simulations. Shifting the $\rho$-profile by $\rho_-$ and rescaling it by $\Delta\rho$, and rescaling space by $\ell$, the microtubule density profiles all collapse on a $\tanh^2$-curve, in accordance with Eq.~\eqref{eq:rho_1Dsol}.

As the bilayer is crossed from left to right, the polarization changes sign from positive to negative. Due to the continuity of the field, this means it has to cross zero at the center of the bilayer. The positive and negative values are connected by a kink profile:
\begin{align}
\label{eq:p_1Dsol}
    p(x) &= -p_+\tanh(\frac{x}{\ell}).
\end{align}
This $\tanh$-profile is confirmed by numerical simulations; see Fig.~\ref{fig:profiles}b. In the equation above, $p_+$ is the Ginzburg-Landau equilibrium value imposed by Eq.~\eqref{eq:p0_GL} at the density maximum $\rho_+$, i.e., $p_+ = \sqrt{(m_0\rho_+-\rhoc)/(\beta m_0^2\rhoc)}$, and $\ell$ is the characteristic length scale of the kink, determining its width and being the same as in Eq.~\eqref{eq:rho_1Dsol}. It reads:
\begin{align}
\label{eq:ell}
    \ell &= \Lambda+\sqrt{\Lambda^2+\frac{2\kappa}{m_0\rho_-/\rhoc-1}}.
\end{align}
$\Lambda$ collects the contribution from the active terms. For the bilayer, it is given by:
\begin{align}
\label{eq:LBL}
   \Lambda_\mathrm{BL}=\frac{-\bar\lambda p_+}{2(m_0\rho_-/\rhoc-1)},
\end{align}
with $\bar\lambda = \lambda+2\zeta_1\chi/\Deff$ an effective self-advection coefficient, including the self-advection of the order strength discussed in sec.~\ref{sec:der_eq} as well as the effect of the gradient in the $\rho$ field emerging due to the contractility $\chi$, which the polarization field couples to via the density anchoring coefficient $\zeta_1$.

For $\Lambda=0$, Eq.~\eqref{eq:ell} reduces to the well-known length scale of the kink solution for domain walls in passive systems \cite{CahnHilliard}, which emerges via the competition of the Ginzburg-Landau term $m_0\rho_-/\rhoc-1$, that strives to impose a non-zero order parameter everywhere (and in particular at the center of the bilayer, thus preferring a short interfacial length scale) and the stiffness term $\kappa$, which evens out gradients in the polarization and favors a wider interface. In contrast, the purely active contributions giving rise to the effective self-advection make $\Lambda$ non-vanishing. They shift the polarization pattern in the forward or backward direction depending on its sign, thus closing the bilayer further for $\bar\lambda>0$ while opening it up for $\bar\lambda<0$, changing $\ell$ accordingly. In the derived model, $\bar\lambda$ is typically negative, so that the active terms lead to a widening of the bilayer. Rescaling the $p$-profiles by $p_+$ and space by $\ell$, we find that the polarization profiles from numerical simulations collapse onto a tanh-curve, confirming the validity Eq.~\eqref{eq:p_1Dsol}; see Fig.~\ref{fig:profiles}c. Furthermore, in Fig.~\ref{fig:profiles}e, we plot the value for $\ell$ extracted by fitting the $p$-profiles against $\LBL$. Rescaling both axes by $\sqrt{(m_-\rho_-/\rhoc-1)/(2\kappa)}$, with $m_-$ the value of the motor field at the center of the bilayer, all data points from the simulations collapse onto one curve, confirming the behavior predicted by Eq.~\eqref{eq:ell}.

Finally, in the small $\gamma$ approximation, we can investigate how the motor profile deviates from a constant by integrating Eq.~\eqref{eq:m_1Dsol} and inserting the polarization profile \eqref{eq:p_1Dsol} obtained for $\gamma=0$. This yields:
\begin{equation}
\label{eq:m_1Dsol}
    m(x) = m_-\cosh(x/\ell)^{-\gamma p_+\ell},
\end{equation}
where $m_-$ is the motor density at the center of the bilayer and $\ell$ is given in Eq.~\eqref{eq:ell}. In Fig.~\ref{fig:profiles}c, we confirm this result with numerical simulations. One should expect this profile to be valid only close to the bilayer center at $x=0$, as the decay of $m$ for $\abs{x}\to\infty$ breaks the initial assumption of constant motor density used in the polarization profile. We will investigate the effect of motor inhomogeneity in the region outside of the bilayer below in sec.~\ref{sec:extprofile}.

In summary, we have found that in the interior part of the bilayer, the polarization follows a $\tanh$-profile interpolating between opposing orientations, similar to the kink solution found in equilibrium systems. However, the characteristic length scale of this solution is modified by the contribution of the active terms, i.e., self-advection and anchoring. As a result of the polarization gradient, the microtubule density field $\rho$ develops peaks on either side of the bilayer. The strength of this phase separation is controlled by the contractile flux, which accumulates density in regions of stronger polar order. Finally, we have seen that the motor density profile shows a peak at the center of the bilayer, whose shape is given by Eq.~\eqref{eq:m_1Dsol}. This peak develops as a consequence of the advection of motors along the polarization field, from the outside into the inside of the bilayer.

\subsubsection{Micelles}
Micelles are radially symmetric stationary solutions of Eqs.~\eqref{eq:der} with the polarization pointing in the inward radial direction. As for the bilayer, the polarization vanishes and the density shows a density dip at the micelle center. In this section, we inspect this correspondence closer and identify the effects that the splay in the polarization field has on the profile. Again, we focus on the inner part of the micelle, i.e., the region inside the ring of maximum density (see Fig.~\ref{fig:profiles}b). The calculation is analogous to the bilayer case but requires a few additional approximations due to the non-vanishing splay. We refer the reader to App.~\ref{app:micprof} for the details, while discussing the main results here.

In the limit of weak phase separation, the profiles are given by the same functions as for the bilayer, given in Eqs.~\eqref{eq:rho_1Dsol}, \eqref{eq:p_1Dsol}, and \eqref{eq:m_1Dsol}, where the cross-section coordinate $x$ is substituted by the radial coordinate $r$. Figure \ref{fig:profiles}d shows micelle profiles extracted from numerical simulations, which indeed collapse onto the curves predicted by theory upon shifting and rescaling, like for the bilayer case.

However, the scaling parameters appearing in the previous section, such as the strength of the phase separation $\Delta\rho$ and the length scale $\ell$, are modified as a consequence of the splay of the micelle. The quantity $\Delta\rho$ appearing in Eq.~\eqref{eq:rho_1Dsol} obtains a new contribution due to the splay flux. For $\chi_1<0$, this flux is directed outward, enhancing phase separation by depleting the center of the micelle. Thus, Eq.~\eqref{eq:1D_PS_Delta} is replaced by:
\begin{equation}
\label{eq:rad_PS_Delta}
    \DrMIC=\frac{-2\bar\chi-\bar\chi_1}{2+\bar\chi+\bar\chi_1/2}(\rho_0-\rhoc/m_0),
\end{equation}
where $\bar\chi_1=\chi_1/(\beta m_0\rho_c\Deff)$ is the ratio between the splay flux strength and effective isotropic diffusivity. In Fig.~\ref{fig:profiles}h, the values for $\DrMIC$ obtained from fitting simulated micelle profiles are plotted against $\bar\chi_\mathrm{M}=\bar\chi+\bar\chi_1/2$. Rescaling the $y$-axis by $\rho_0-\rhoc/m_0$, they collapse onto a curve given by $-2\bar\chi_\mathrm{M}/(2+\bar\chi_\mathrm{M})$, as predicted by Eq.~\eqref{eq:rad_PS_Delta}.

On the other hand, the characteristic length scale $\ell$ is modified as well. The active contribution $\Lambda$ appearing in Eq.~\eqref{eq:LBL} changes to:
\begin{equation}
\label{eq:Lmic}
    \LMIC = \frac{- (\bar\lambda+\zeta_1\chi_1/\Deff+\lambda_2)p_+}{2(m_0\rho_-/\rhoc-1)}.
\end{equation}
This reflects the coupling to the additional splay flux $\chi_1$ via the density anchoring $\zeta_1$, as well as a contribution from the active splay term controlled by $\lambda_2$. For inward-pointing microtubules, the splay in the micelle is negative; for $\lambda_2<0$, this leads to a suppression of the order in the inside of the micelle, which results in a larger length scale. We check the prediction by plotting the values of $\ell$ extracted from the polarization profiles of simulated micelles in Fig.~\ref{fig:profiles}f, obtaining the same collapse as for the bilayer using the modified expression for the active contribution $\Lambda$.

In summary, we have found that the profile of the micelle solution is the radial counterpart of the bilayer. Indeed, both bilayers and the micelles have a defect at their center, where the polarization vanishes and the density is depleted due to the contractile flux. In both cases, the motor field shows a maximum at the center due to its advection along the polarization. However, in contrast to the bilayer, the polarization field in the micelle solution is splayed. This splay has two consequences: it enhances the phase separation, depleting the center of the micelle more strongly, due to the splay flux; it affects the characteristic length scale of the solution, which is larger than that of the bilayer due to the effect of the active splay.

\subsection{Outer profiles}
\label{sec:extprofile}
The $\rho$- and $\polv$-profiles of the interior of the bilayer discussed above plateau away from the center. This constant asymptotic behavior, however, relies on the assumption of a constant motor field, which is no longer fulfilled for $\gamma>0$. 
Indeed, for finite values of $\gamma$, the motor field is advected along the polarization, acquiring a non-vanishing slope in ordered regions. Choosing a reference point outside of the bilayer with microtubule and motor densities $\rho_0$ and $m_0$, to lowest order in $\gamma$ the motor field will have a slope $\delta m = \gamma m_0 p_0$, where $p_0$ is the equilibrium polarization given by Eq.~\eqref{eq:p0_GL}. In this section, we discuss how this motor inhomogeneity gives rise to sloped microtubule and polarization profiles in the outside region of the bilayer.

The $m$-dependence in Eq.~\eqref{eq:p0_GL} implies that the slope in $m$ gives rise to a slope in $p$: regions closer to the bilayer are more ordered due to the higher motor concentration there. The polarization gradient gives rise to contractile fluxes, thus resulting in a gradient in the MT density $\rho$, with higher concentrations close to the bilayer. This slope, in turn, feeds back into the polarization equation due to the $\rho$-dependence in Eq.~\eqref{eq:p0_GL}.

In App.~\ref{app:extprofile}, we calculate the expressions for the slopes in the three fields $\rho$, $p$ and $m$ to first order in $\gamma$, obtaining:
\begin{subequations}
\label{eq:ext_res}
\begin{align}
    \delta \rho &= -\frac{2\chi}{\Deff}p_0\delta p-\gamma p_0\tilde{D}+O(\gamma^2),\label{eq:ext_rho}\\
    \delta \pol &=\gamma \frac{(2-m_0 \rho_0/\rho_c) - m_0\tilde D/\rho_c}{2\beta m_0^2+2 m_0\chi/(\rho_c\Deff)}+O(\gamma^2),\label{eq:ext_p}\\
    \delta m &=\gamma m_0p_0+O(\gamma^2).\label{eq:ext_m}
\end{align}
\end{subequations}
In addition to the effect from the contractile flux, controlled by $\chi<0$, Eq.~\eqref{eq:ext_rho} includes a second term due to the gradient of $m$. The inhomogeneous motor density introduces a spatial variation of the activity, which affects the terms that arise from motor-mediated interactions in the $\rho$-equation \eqref{eq:der_rho}. We have defined $\tilde{D}=m_0(\nu\rho_0^2+\pdv{\chi}{m}p_0^2)/\Deff$, which is negative for most of parameter space in the derived model, resulting in a slope $\delta\rho$ that follows $\delta m$ due to contraction of the microtubules into areas of increased activity. The slope in the polarization amplitude given in Eq.~\eqref{eq:ext_p} encodes the effect of the coupling to both densities $m$ and $\rho$.
In Fig.~\ref{fig:profiles}g, we compare these quantitative theoretical predictions with bilayer profiles extracted from numerical simulations of the phenomenological model.

In summary, we have shown that the introduction of motor inhomogeneity due to a non-vanishing Péclet number $\gamma$ leads to the emergence of a slope in all three fields in the outside region of the bilayer due to the coupling between them. Overall, we find that introducing an inhomogeneous motor field leads to a depletion of the microtubule density and a lower polar order strength far from the bilayer, segregating the bilayer from its isotropic background. We expect a similar mechanism to control the outside profile of the micelles as well, up to contributions from the splay. This explains why a finite $\gamma$ is required to obtain well-separated micelles. We conclude that the motor advection leads to the emergence of high-activity regions in the system, allowing for the assembly of ordered supramolecular structures such as bilayers and micelles.

\section{Stability analysis}
\label{sec:instab}
The goal of this section is to understand the branching and fingering instabilities of the micelle solutions observed in sec.~\ref{sec:PD}, explaining both their location in the phase diagram and the mechanisms involved in their activation. To achieve this, we proceed by first studying the stability of the homogeneous ordered state, where we will see that all the relevant mechanisms are already at work. In the second half of this section, we extend the insights we have gained for those states to the more complicated micelle solutions.

\subsection{Instabilities of the homogeneous ordered state}
\begin{figure}
    \centering
    \includegraphics{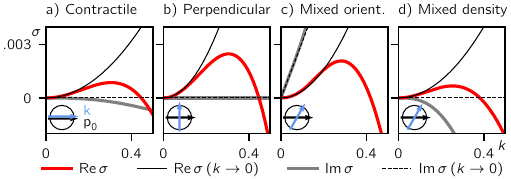}
    \caption{Dispersion relations for the instabilities of the homogeneous state. The thick lines show the real and imaginary parts of the largest eigenvalue of the full Jacobian (see App.~\ref{app:LSA}), while the thin lines show the analytical approximations up to second order in $k$. For all plots, $\rho_0=1.1$, $\rho_c=\beta=m_0=1$, $D_\rho=1$, $\mu=\alpha_1=\alpha_2=\kappa_2=0$, $\chi_2=-\chi_1/2$. a) Contractile instability, arising for a wave vector along the direction of polarization ($\varphi=0$) for $\bar\chi<-1$. Here, we chose $\chi_1=-.24$, $\kappa_1=0.2$, $\lambda_1=0.05$, $\lambda_2=\zeta_1=\zeta_2=0$. b) Perpendicular instability ($\varphi=\pi/2$), arising when one of Eqs.~\eqref{eq:perp_instab} is fulfilled. $\kappa_1=0.2$, $\lambda_2=-1.7$, $\zeta_2=-0.2$, $\chi_1=-.1$, $\lambda_1=-1$, $\zeta_1=0$. c) Mixed orientational instability, arising for a mixed wave vector ($\varphi=\pi/3$) according to Eq.~\eqref{eq:mixed_rho_instab}. $\kappa_1=0.2$, $\lambda_2=-2$, $\zeta_2=-0.2$, $\chi_1=-.04$, $\lambda_1=-0.2$, $\zeta_1=0$. d) Mixed density instability, arising for a mixed wave vector ($\varphi=\pi/3$) when Eq.~\eqref{eq:mixed_or_instab} is fulfilled. $\chi_1=-.1$, $\lambda_1=-0.2$, $\zeta_1=-1$, $\kappa_1=0.15$, $\lambda_2=0$, $\zeta_2=0$.}
    \label{fig:disp}
\end{figure}
\label{sec:hom_instab}
Equations \eqref{eq:m} and \eqref{eq:der} imply the existence of two stationary states with homogeneous microtubule and motor densities $\rho=\rho_0$ and $m=m_0$. One is the isotropic state with $\polv=0$; the other, emerging for $m_0\rho_0>\rhoc$, is the homogeneous ordered state with a polarization amplitude given by Eq.~\eqref{eq:p0_GL} and a polarization direction selected by spontaneous symmetry breaking.

The stability of the isotropic state has already been inspected analytically in previous works for similar models \cite{Liverpool2003,Liverpool2005,Ahmadi2006,MaryshevPRE2018}. In App.~\ref{app:LSA}, we extend the analysis to our model. For $m_0\rho_0>\rhoc$, the two polarization directions show a type-III instability \cite{CrossHohenberg1993}, whose dispersion relation has a maximum at zero wave vector $k=0$, which corresponds to the emergence of global order. Additionally to this ordering instability, for small passive-to-active ratio $\alpha$, the system exhibits a density (bundling) instability at high $\rho$, which requires the introduction of a bilaplacian term to the $\rho$-equation to be regularized \cite{ZiebertZimmermann2005,AransonTsimring2006,MaryshevPRE2018}. In this work, we limit ourselves to sufficiently large $\alpha$, so that the density instability is not relevant.

In the remainder of this subsection, we study the linear stability of the homogeneous ordered state. We choose the coordinate system such that the initial polarization lies along the $x$-axis, $\polv=p_0\vcu{e}_x$, with $p=\sqrt{(m_0\rho_0/\rhoc-1)/(\beta m_0^2)}$. Then, we apply a periodic perturbation to this state, which has the form $\rho=\rho_0+\delta\rho e^{i\kv\vdot\rv}$, $\polv=p_0\vcu{e}_x + (\delta p_\parallel, \delta p_\perp)e^{i\kv\vdot\rv}$. It can be shown that $\delta p_\parallel$ controls the perturbation of the order strength, whereas $\delta p_\perp$ leads to a variation of the order direction \bibnote[delp]{Writing $\polv=p\nv$, to lowest order, we find $\delta\polv=\delta p\vcu{e}_x + p_0\delta \nv$. Multiplying this equation with $\vcu{e}_{x,y}$, using $\delta \nv\vdot\vcu{e}_x=0$ (due to the unit vector condition), and comparing with the expression for $\polv$ in the main text, we obtain $\delta p = \delta p_\parallel e^{i\kv\vdot\rv}$ and $\vcu{e}_y\vdot\delta \nv = \delta p_\perp e^{i\kv\vdot\rv}/p_0$.}. We keep the motor field $m$ constant, as the role of its perturbations should be negligible in the small $\gamma$ limit.

In the long wavelength limit, the linear stability analysis for this state can be performed analytically. In this limit, the perturbations in the order strength can be expressed in terms of density and orientational perturbations, reducing the linearized dynamics to a two-dimensional Jacobian of the form:
\begin{equation}
\label{eq:J_instab}
    \partial_t\mqty(\delta\rho\\ \delta p_\perp) = \mqty(J_{11} & J_{12}\\ J_{21} & J_{22})\mqty(\delta\rho\\ \delta p_\perp),
\end{equation}
where the entries $J_{ij}$ of the Jacobian are given in App.~\ref{app:LSA}. Studying the eigenproblem of this Jacobian in dependence of the choice of the wave vector $\kv=(k_\parallel, k_\perp)=k(\cos\varphi,\sin\varphi)$ allows us to characterize the instabilities of the homogeneous state, which are all of type-II \cite{CrossHohenberg1993}. Different choices of $\kv$ affect both the eigenvalues of the Jacobian, determining whether the base state is stable, and the eigenvectors, which specify the type of perturbation involved in the instability. Indeed, the eigenvectors of this matrix mix density perturbations ($\delta\rho$) and orientational perturbations ($\delta p_\perp$) to different degrees, resulting in distinct instabilities. We refer the reader to App.~\ref{app:LSA} for the details of the analysis, while we present the main results here.

\begin{figure}
\includegraphics[width=\linewidth]{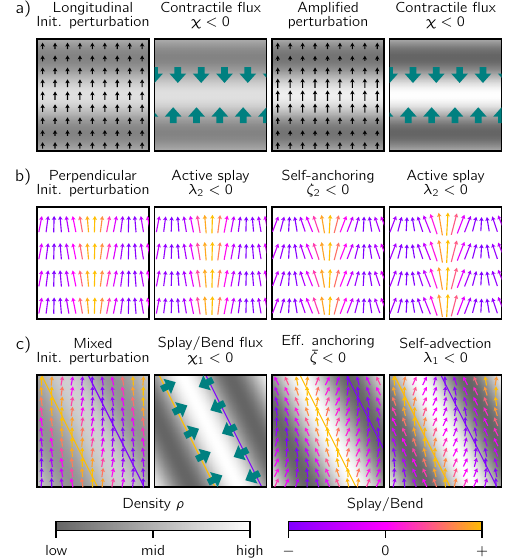}
\caption{\label{fig:LSAcartoons}Feedback mechanisms underlying the instabilities of the homogeneous ordered state. The small arrows indicate the polarization field, whereas density fluxes are shown in teal. a) Contractile instability. Perturbing the density longitudinally, regions of higher and lower density emerge, resulting in a modulation of the polar order strength. The contractile flux controlled by $\chi$ accumulates density into the region of stronger polar order. The feedback is positive if this flux overcomes the effective diffusivity, i.e., for $\bar\chi<-1$. b) Orientational instability. Perturbing the initial state with an orientational perturbation in the perpendicular direction, regions with splay of different signs (given by $\div\nv$) will arise. For $\lambda_2<0$, the active splay increases the polarization in regions of positive splay and decreases it in regions with negative splay. The self-anchoring couples to this modulation, rotating the polarization away from regions of stronger order for $\zeta_2<0$, amplifying the splay even more. Now, the active splay can act anew, resulting in positive feedback for $\lambda_2\zeta_2$ positive and sufficiently large. This feedback drives the perpendicular and mixed orientational instabilities. c) Mixed density instability. A mixed wave vector perturbation where the density and orientational components are in (anti-)phase gives rise to a modulation in splay and bend, whose extrema (yellow and purple lines) lie between the extrema of the density wave. The combined action of splay and bend flux ($\chi_1<0$) accumulates density between the two lines. As a consequence, the effective anchoring ($\bar\zeta<0$) rotates the orientational field away from the density maximum, thus moving the splay/bend maximum forward. Counteracting this, the self-advection ($\lambda_1>0$) shifts the orientational pattern backward, so that we obtain the initial configuration, but with an amplified perturbation in both the density and the polarization. The instability is activated if $-\bar\zeta\chi_1/\lambda_1$ is large enough to overcome the effective transversal flux in Eq.~\eqref{eq:mixed_rho_instab}.
}
\end{figure}

For a fully longitudinal wave vector ($k_\perp=0$), the density and orientational perturbations decouple. Then, we find that an instability arises for sufficiently strong effective contractility, $\bar\chi<-1$ (see Fig.~\ref{fig:disp}a). To understand this instability, we recall that the contractile flux decreases diffusion in the direction of the polar order (see Fig.~\ref{fig:terms}a). When this flux becomes strong enough to overcome the effective isotropic diffusivity $\Deff$, the total flux in the direction of the order becomes anti-diffusive, resulting in accumulation of density along $\polv$ (see Fig.~\ref{fig:LSAcartoons}a). This results in the formation of ordered bands extensing transversally to the polarization. The corresponding eigenvector in Eq.~\eqref{eq:J_instab} lies entirely in the direction of the density perturbation $\delta\rho$, since the direction of the order is not modulated ($\delta p_\perp=0$). This \emph{contractile instability} was previously described for a similar model in Ref.~\cite{Ahmadi2006}, where it was referred to as the bundling instability.

The other limiting choice of wave vector $\kv$ is exactly perpendicular to the order of the initial state ($k_\parallel=0$); see Fig.~\ref{fig:disp}b. Then, an instability arises if any of the following inequalities are fulfilled (see App.~\ref{app:LSA}):
\begin{subequations}
\label{eq:perp_instab}
\begin{align}
\label{eq:perp_instab1}
    \beta m_0^2[\kappa+\Deff(1+\bar\chi_2)]&<\lambda_2\zeta_2,\\
\label{eq:perp_instab2}
    \beta m_0^2\kappa+\frac{\bar\zeta\chi_2}{\Deff(1+\bar\chi_2)}&<\lambda_2\zeta_2.
\end{align}
\end{subequations}
These conditions correspond to a positive trace and a negative determinant of the Jacobian in Eq.~\eqref{eq:J_instab}, respectively. Here, $\bar\chi_2=\chi_2/(\Deff\beta m_0\rhoc)$ indicates the strength of the transversal flux compared to the isotropic effective diffusivity and we assumed $1+\bar\chi_2>0$. Furthermore, $\bar\zeta=\zeta_1+\zeta_2/(\beta m_0\rhoc)$ is an effective anchoring to density interfaces, reflecting that higher density correlates with more polar order, which the self-anchoring ($\zeta_2$) couples to. In the derived model, $\bar\zeta$ is negative in most of parameter space.

The product $\lambda_2\zeta_2$ on the right-hand side of these inequalities can be interpreted as a feedback mechanism between active splay and self-anchoring (see Fig.~\ref{fig:LSAcartoons}b). As the order orientation is perturbed transversely, regions of positive and negative splay form. For $\lambda_2<0$, the active splay enhances the polarization in regions of positive splay and reduces it in regions of negative splay. As a consequence, the self-anchoring rotates the polarization away from the former regions towards the latter for $\zeta_2<0$, resulting in an even stronger splay. For sufficiently strong $\lambda_2$ and $\zeta_2$, this mechanism overcomes the left-hand sides in Eqs.~\eqref{eq:perp_instab}, giving rise to a positive feedback loop, making the homogeneous ordered state unstable. Crucially, this \emph{perpendicular instability} requires that the two coefficients have the same sign, which is the case for most of the parameter space of our derived model. The corresponding eigenvector has $\abs{\delta p_\perp}\gg\abs{\delta\rho}$ for small $k$, reflecting the orientational character of this instability.

For general wave vectors mixing both longitudinal and perpendicular components, two new instabilities emerge in addition to the ones discussed above. These are strongest for wave vectors that are almost perpendicular to the order, with a small non-vanishing component in the longitudinal direction, s.t. $k_\parallel^2\gg k_\perp^2> 0$. The first such instability arises for (see App.~\ref{app:LSA}):
\begin{align}
\label{eq:mixed_or_instab}
    \kappa-2\frac{\bar\zeta\chi_1}{\lambda_1}&< \frac{\lambda_2\zeta_2}{\beta m_0^2}.
\end{align}
As for the perpendicular instability, the corresponding eigenvector is predominantly orientational for $k\to 0$. For this reason, we refer to it as the \emph{mixed orientational instability}. The right-hand side shows the same feedback between active splay ($\lambda_2$) and self-anchoring ($\zeta_2$) discussed above, which is counteracted by the stiffness $\kappa$ on the left-hand side. Thus, this instability is intimately related to the perpendicular one. However, the skewed wave vector has two consequences that make the mixed orientational instability distinct from the latter. The first consequence is a non-vanishing imaginary part of the eigenvalue, which is linear in $k$ for $k\to 0$ and proportional to $\lambda_1$, so that the instability is associated with a propagation controlled by the self-advection (it is a type-II-o instability in the nomenclature of Ref.~\cite{CrossHohenberg1993}, see Fig.~\ref{fig:disp}c). The second consequence is the second term on the left-hand side, which is stabilizing for the typical signs in the derived model (but it can lead to positive feedback for the opposite signs, see App.~\ref{app:LSA}).

The second instability with mixed wave vector arises for (see App.~\ref{app:LSA}):
\begin{align}
\label{eq:mixed_rho_instab}
    &\Deff(1+\bar\chi_2)<-2\frac{\bar\zeta\chi_1}{\lambda_1}.
\end{align}
The eigenvector associated with this instability involves both density and orientational components that are of the same order in $k$. The two components are in phase or anti-phase, depending on the choice of the wave vector. For strong $\abs{\lambda_1}$, the density component dominates, so we refer to the instability as the \emph{mixed density instability}. In contrast to the mixed orientational instability, here the imaginary part of the eigenvalue vanishes faster than $k^2$ as $k\to 0$ (see Fig.~\ref{fig:disp}d).

Similarly as before, we can interpret the factor $\bar\zeta\chi_1$ appearing on the right-hand side of the inequality \eqref{eq:mixed_rho_instab} as a feedback mechanism between the effective anchoring to density interfaces, controlled by $\bar\zeta$, and the splay/bend flux, controlled by $\chi_1$. Figure \ref{fig:LSAcartoons}c illustrates this feedback. For an initial perturbation involving both the density and the order direction in (anti-)phase, a maximum and a minimum in the splay and bend arise on either side of each density maximum. As a consequence, for $\chi_1<0$, the splay flux and the bend flux advect the $\rho$ field towards the density maximum (along the $x$-direction and $y$-direction, respectively). This leads to an increase of the density perturbation, which the order orientation couples to via the effective anchoring $\bar\zeta<0$. This further increases the splay and bend, while shifting the orientational pattern to the front; the self-advection of the orientation, controlled by $\lambda_1<0$, brings it back in place. The increased orientational perturbation leads to an even stronger splay and bend flux, so that repeating the loop results in positive feedback, where the density and orientational perturbations grow together, for negative $\bar\zeta\chi_1/\lambda_1$. On the left-hand side of Eq.~\eqref{eq:mixed_rho_instab}, the total transversal flux appears (as the sum of the isotropic and transverse anisotropic contributions), which tends to even out any gradients in density, thereby counteracting the feedback mechanism that drives the mixed density instability.

In summary, we have identified four different instabilities that arise for the homogeneous ordered state, each relying on a feedback mechanism rooted in the interplay of different terms in the dynamical equations \eqref{eq:der}, as is illustrated in Fig.~\ref{fig:LSAcartoons}. The instabilities are: i) The contractile instability, which arises for sufficiently strong contractility $\chi$ and makes the initial state unstable by accumulating density longitudinally to the initial order. ii) The perpendicular instability, which relies on the feedback between the active splay, controlled by $\lambda_2$, and the self-anchoring, controlled by $\zeta_2$. When the two coefficients have the same sign and are sufficiently large, their interplay results in the growth of orientational perturbations (i.e., splay) in the direction perpendicular to the initial order. iii) The mixed orientational instability, which relies on the same mechanism as the perpendicular one, but involves a wave vector that mixes longitudinal and perpendicular components. iv) Finally, the mixed density instability emerges from a feedback controlled by the effective anchoring $\bar\zeta$, the splay/bend flux $\chi_1$, and the orientational self-advection $\lambda_1$, and gives rise to a simultaneous growth of density and orientational perturbations for mixed wave vectors.

In the following subsection, we discuss the role these instabilities play for micelle solutions, connecting them with the micellar instabilities observed in sec.~\ref{sec:PD}.

\subsection{Micelle instabilities}
\label{sec:mic_instab}

The micelle solution is an inhomogeneous base state with radial symmetry and radial inward-pointing polarization ($\polv=p_0\vcu{e}_r$ with $p_0<0$). To study its stability, we introduce a perturbation that is periodic in the angular coordinate, with node number $\kn$, and in the radial coordinate, with wave vector $\kr$. In App.~\ref{app:LSAmic}, we show that in the limit of large $r$ (corresponding to weak curvature), the mechanisms behind the instabilities of the homogeneous ordered state are extended to the micelle solutions. The role of the longitudinal direction is taken by the radial coordinate, while the perpendicular direction is substituted by the angular coordinate. However, due to the splay of the initial base state, the angular direction is perpendicular to the order only locally, so that the perpendicular instability of the homogeneous state is absent for the micelles, while the mixed instabilities survive even with non-vanishing initial splay.
\begin{figure}
\includegraphics[width=.95\linewidth]{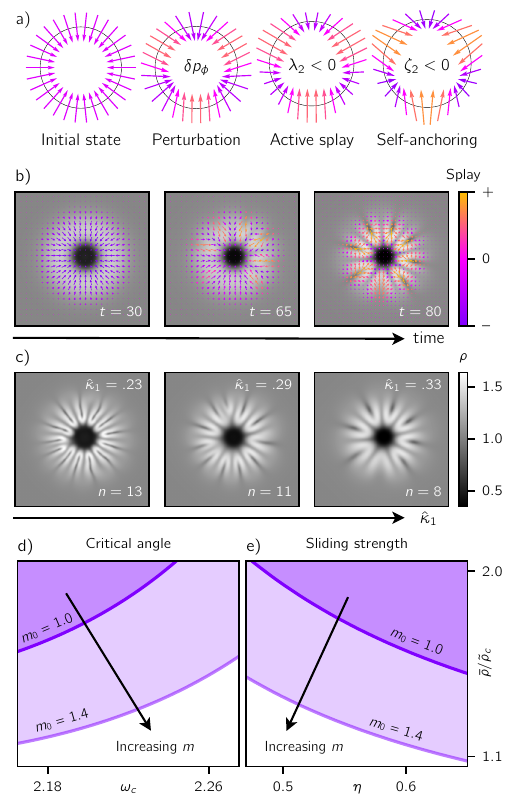}
\caption{\label{fig:BrInst} Micellar branching instability. a) The feedback mechanism between active splay and self-anchoring from Fig.~\ref{fig:LSAcartoons}b extends to an initially splayed configuration. Adding an angular perturbation results in the modulation of splay around the ring, which leads to a modulation of the polar order strength due to the active splay. The self-anchoring then rotates the polarization away from regions with stronger order, thereby leading to a positive feedback loop. b) Simulations of the phenomenological model (parameters in App.~\ref{app:sim}; see Video S6 \bibnotemark[SI]) in a region where $\lambda_2\zeta_2>0$ show that an initially prepared micelle becomes unstable via branching. At early times, the splay and polar order strength are modulated around the ring, showing an initial dynamics governed by the orientational feedback mechanism. At later times, the variation of the polarization amplitude leads to contractile density fluxes, that accumulate density into ordered regions, thus forming bilayer-like branches. c) By increasing the strength of the stiffness coefficient $\hat\kappa_1$, the number of branches could be decreased. d, e) Plotting the regions of validity of the inequality \eqref{eq:mixed_or_instab} in terms of the derived model parameters $\bar\rho$, $\omega_c$ and $\eta$, we see that the mixed orientational instability arises in the same regions where micelle branching was observed in Fig.~\ref{fig:PD}b-c. The dark purple region shows the validity regime of the inequality for $m_0=1$. Higher values of $m_0$ (light purple regime, $m_0=1.4$) lead to an increase of the area where the instability arises, showing closer correspondence to Fig.~\ref{fig:PD}b-c. This reflects the fact that the motor density is locally increased inside micelles due to motor advection.
}
\end{figure}

Heuristically, the fact that the instabilities of the homogeneous state extend to the micelles can be made plausible in light of the underlying feedback mechanisms discussed in the previous subsection. Indeed, these feedback mechanisms do not rely on the initial order being homogeneous and can be extended to situations with non-vanishing initial splay, like the micelle solutions. An example is the feedback between the active splay and the self-anchoring, which drives the mixed orientational instability (recall Fig.~\ref{fig:LSAcartoons}b). In Fig.~\ref{fig:BrInst}a, we show how it generalizes to the micelle. While the initial splay is now negative, this does not affect the mechanism: an angular perturbation leads to a modulation in the splay around the micelle. This gives rise to a modulation of polarization amplitude due to the active splay. The self-anchoring couples to this modulation, amplifying the splay, so that we obtain the same feedback as before. Similarly, the mechanism underlying the mixed density instability also generalizes to splayed configurations. 

To inspect the connection between the linear instabilities and the phenomenology discussed in sec.~\ref{sec:PD}, we resort to numerical simulations of the phenomenological model. There, we first prepare a stable micelle in a regime where the instabilities do not arise. Then, we take this micelle as an initial condition but change the phenomenological model parameters, so as to activate either of the two mixed instabilities and probe their time evolution in the nonlinear regime.

\subsubsection{Branching instability}
First, we change the phenomenological model parameters to activate the mixed orientational instability, by choosing the product $\lambda_2\zeta_2$ large enough (see App.~\ref{app:sim}). Figure \ref{fig:BrInst}b shows a time series of snapshots from such a simulation (see also Video S6 \bibnotemark[SI]). After some time, the instability sets in, resulting first in a modulation of the orientation (and thus the splay) around the micelle, and later in a redistribution of density. Indeed, as a consequence of the contractile flux (see Fig.~\ref{fig:terms}a), density is accumulated into the regions of positive splay, corresponding to stronger polar order, and depleted from the regions in between. The result is the formation of bilayer-like branches, with the order pointing in opposite directions on either side.

In Fig.~\ref{fig:BrInst}c, we vary the stiffness $\hat\kappa_1$, which results in a modulation of the number of branches that form. Increasing $\hat\kappa_1$ corresponds to a shift in the maximum of the dispersion relation (Fig.~\ref{fig:disp}c) to the left, as the left-hand side of the inequality \eqref{eq:mixed_or_instab} becomes larger and the term of the eigenvalue that is quadratic in the wave vector $k$ is reduced. The shift of the maximum towards smaller wave vectors corresponds to smaller node numbers $n$, so that the number of branches is reduced.

The bilayer-like branches emerging in Fig.~\ref{fig:BrInst} are reminiscent of the micellar branching instability we have discussed for the derived model in sec.~\ref{sec:PD}. This correspondence is confirmed by inspecting the location of the mixed orientational instability in the phase space of the derived model. In Fig.~\ref{fig:BrInst}d-e, we plot the regions where inequality \eqref{eq:mixed_or_instab} holds in terms of the critical angle $\wcrit$, the antiparallel sliding $\eta$ and the mean microtubule density $\bar\rho$. Indeed, the instability is activated for large $\bar\rho$ and $\eta$ and sufficiently small $\wcrit$, corresponding to the observations made in Fig.~\ref{fig:PD}b-c.

It is important to note that the analysis in the previous section was performed for the homogeneous ordered state, where $m_0=1$ by definition. When plotting the regime of validity of Eq.~\eqref{eq:mixed_or_instab} for this value of $m_0$, the resulting region (shown in dark purple in Fig.~\ref{fig:BrInst}d-e) is smaller compared to the region of the phase diagram where branching occurred in Fig.~\ref{fig:PD}b-c. For non-vanishing motor Péclet numbers $\gamma>0$, however, the local value of the motor field might deviate from its mean: inside the micelles, it is increased due to motor advection. Since the instability is activated locally, this local value should be taken as the initial motor density instead. Indeed, for larger values of $m_0$, we observe an enlargement of the instability region (light purple area in Fig.~\ref{fig:BrInst}d-e), in closer agreement with the observations of Fig.~\ref{fig:PD}b-c. Thus, we can identify the mixed orientational instability as the mechanism driving micelle branching.

\subsubsection{Fingering instability}
\begin{figure}
\includegraphics[width=\linewidth]{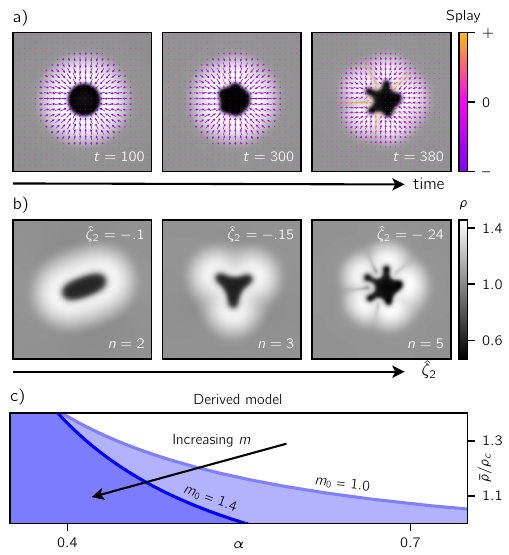}
\caption{\label{fig:ShInst}Micellar fingering instability. a) Simulations of the phenomenological model in a region where $\bar\zeta\chi_1/\lambda_1<0$ show that an initially prepared micelle becomes unstable via the mixed density instability (parameters in App.~\ref{app:sim}; see Video S7 \bibnotemark[SI]). At early times, both the orientational order and the density are modulated around the ring, reflecting the simultaneous growth of the perturbations predicted for the mixed density instability. At later times, the variation of splay around the ring leads to splay fluxes, which lead to the protrusion of regions with positive splay into the center of the micelle, whereby fingers are formed. b) By increasing the strength of the self-anchoring coefficient $\hat\zeta_2$, the number of fingers can be increased.
c) We plot the regions of validity of the inequality \eqref{eq:mixed_rho_instab} in terms of the microtubule density $\bar\rho$ and the passive-to-active ratio $\alpha$ from the derived model. The mixed orientational instability arises for small $\alpha$, mirroring the observation made in Fig.~\ref{fig:PDalpha}. The blue regions show the validity regime of the inequality for $m_0=1.0$ (lighter) and $m_0=1.4$ (darker).}
\end{figure}

Starting from the same stable micelle as the initial condition, but choosing the parameters to activate the mixed density instability (see App.~\ref{app:sim}), we find a different time evolution, shown in Fig.~\ref{fig:ShInst}a. After some time, the density and polar order perturbations grow simultaneously, in accordance to the eigenvector of the mixed density instability. As time progresses, the interface to the depleted center of the micelle is modulated in its shape, resulting in the formation of fingers. This can be understood in terms of the splay flux (see Fig.~\ref{fig:terms}a), which advects regions of more positive splay along the polar order, invading the center of the micelle more than those with more negative splay. This difference in the splay flux in the angular direction results in the interface modulation.

In Fig.~\ref{fig:BrInst}b, we vary the self-anchoring strength $\hat\zeta_2$, whereby the number of fingers is changed. Again, this can be explained in terms of the dispersion relation (Fig.~\ref{fig:disp}d). As $\hat\zeta_2$ is increased, the right-hand side of Eq.~\eqref{eq:mixed_rho_instab} becomes larger, enlarging the $k^2$-dependent term of the real part of the eigenvalue and shifting its maximum to the right. This corresponds to higher modes $n$ being activated by the instability.

Just as the mixed orientational instability results in micellar branching, we can now connect the mixed density instability to the micellar fingering seen in Fig.~\ref{fig:PDalpha}. To this end, we investigate the domain of validity of Eq.~\eqref{eq:mixed_rho_instab} in terms of the parameters of the derived model. In Fig.~\ref{fig:ShInst}c, we plot this region in dependence of the passive-to-active ratio $\alpha$ and the mean microtubule density $\bar\rho$. As before, we choose a higher value for $m_0$ to mimic the local increase of the motor field in the micelle. Indeed, we find that the mixed density instability emerges for small values of $\alpha$, as in Fig.~\ref{fig:PDalpha}. Hence, we identify the mixed density instability as the mechanism driving micellar fingering in the derived model.

This insight allows us to interpret the instability as a competition between passive and active mechanisms. Indeed, the left-hand side of Eq.~\eqref{eq:mixed_rho_instab} is proportional to the effective diffusivity $\Deff$, which incorporates the effect of diffusion and steric interactions. On the right-hand side, instead, we find the factor $\bar\zeta\chi_1$, which encodes a feedback between the splay flux and anchoring, two mechanisms that arise due to the active motor-mediated interactions. When decreasing the passive-to-active ratio $\alpha$, $\Deff$ becomes smaller, until eventually the active right-hand side prevails and the instability sets in. Crucially, the dominating passive mechanism here is the steric interaction (whose contribution in the equation is the only $\alpha$-dependent term). As a consequence, in contrast to the branching instability, the fingering instability is suppressed at high microtubule densities.

In conclusion, we have shown that the two micellar instabilities we have seen in sec.~\ref{sec:PD} in numerical simulations of the derived model can be explained in terms of positive feedback between different mechanisms appearing in the continuum equations. The branching instability was shown to arise as a consequence of the mixed orientational instability, which relies on the feedback between active splay and self-anchoring to introduce splay along the micelle ring, leading to the formation of bilayer-like branches. On the other hand, the fingering instability is intimately related to the mixed density instability of the homogeneous state, which arises due to the interplay of anchoring, splay and bend fluxes as well as self-advection, which lead to the simultaneous growth of density and orientational perturbations along the ring, resulting in the formation of fingers.

\section{Discussion}
\label{sec:discussion}

In this work, we have introduced a field theory for active filament systems driven by relative filament motion, resulting in the emergence of active micelles and active foams.
Our theory applies to microtubule-motor mixtures, and more generally, to systems comprised of stiff cytoskeletal filaments and molecular motors that are transported along these filaments, inducing alignment and sliding interactions. 
By constructing the field theory from the bottom up, starting from the microscopic interactions between filaments, we have established an explicit link between processes at the microscopic level and the emergent collective behavior at macroscopic scales.

At the microscopic level, our theory accounts for several generic features for the interaction between microtubules and molecular motors. 
Firstly, the motors walk along individual filaments in a directed manner. 
On the coarse-grained level, this results in collective advection of the motors in regions of strong polar order, giving rise to an inhomogeneous motor concentration whose gradients are proportional to the motor's Péclet number ${\gamma = v_m/D_m}$. 
Secondly, the motors' movement induces relative alignment and sliding of pairs of microtubules. 
The alignment can be parallel or antiparallel based on the initial crossing angle of the filaments. 
In the former case, the filaments slide together, whereas in the latter case they slide apart. 
In our theory, we break the parity symmetry between these two cases by considering general critical angles $\wcrit$, determining the bias towards parallel alignment. 
Furthermore, we take into account stalling effects that impede antiparallel sliding, controlling its strength via the parameter $\eta$. 
Our analysis shows that the motor-induced alignment interaction induces orientational order at high microtubule and motor densities, while the relative sliding gives rise to non-equilibrium polarization-dependent density fluxes as well as self-advection, active splay and anchoring effects.

Based on our field theory, we predict the formation of various active supramolecular assemblies: Firstly, radially symmetric \emph{active micelles}, which can undergo two distinct instabilities, a \emph{fingering} and a \emph{branching} instability. The latter leads to the formation of \emph{bilayers}, where the polar order switches signs from one side to the other.
Additionally, we have identified a new active phase characterized by the formation of large-scale interconnected microtubule bilayer networks, which we refer to as \emph{active foams}. 
Our bottom-up approach allows us to determine the phase diagram of the derived continuum model in terms of the microscopic parameters. 
In particular, we find a transition between active micelles and active foams controlled by the microtubule density and relying on the branching instability. 
We determine the relevant features of the interactions for this transition to occur.

Furthermore, through a comprehensive analysis of our field theory, we identify the mechanisms that determine the inhomogeneous stationary solutions and the instabilities of the active filament system. We demonstrate our analytical results using the phenomenological model, which keeps the same structure of the dynamical equations as in the derived model, but allows for free variation of the coefficients of the individual terms in the equations.
This approach expands the applicability of our results to a wide range of field theories characterized by similar field equations. These encompass theories pertaining to different underlying microscopic processes or to theories derived through different coarse-graining methods.

\subsection*{Active micelles and morphological transitions}

We explored the phase diagram of our coarse-grained field theory using numerical simulations. 
Over a wide region of parameter space, radially symmetric structures form, which we termed active micelles, in analogy to the micelles formed by lipid molecules. 
The microtubules in active micelles arrange themselves into a high density ring around the center, just as the lipids in their passive counterpart. 
In this monolayer ring, the microtubule plus-ends correspond to the hydrophobic tails of lipids, pointing toward the micelle center, while the minus-ends extend outward, like the hydrophilic heads in lipid micelles. 
As the motors move along the microtubules towards the plus-end, the motor concentration accumulates inside the active micelles.

The key parameter determining the shape of the active micelles in our model is the passive-to-active ratio ${\alpha = D_r/G}$. 
This is the ratio between $D_r$, determining the time scale of rotational filament diffusion, and the rate of active filament-filament interaction $G$, which is proportional to the mean motor concentration. 
Besides the motors' Péclet number $\gamma$, this is a second important dimensionless quantity characterizing the dynamics of the system.
Akin to a Péclet number it characterizes the relative importance of thermal diffusion and transport processes.
For large values of $\alpha$, the active micelles have a structure similar to the asters found in previous studies \cite{AransonTsimring2006, Gopinath2012, Husain2017}, which show a weak depletion of microtubule density at the center. 
However, as $\alpha$ is lowered, corresponding to a prevalence of active processes, the depleted region at the micelle center expands.

Intriguingly, this result closely resembles the observations made in agent-based simulations of filaments and motors~\cite{Ansari2022}, which showed that spatially confined asters become hollow at high motor concentrations, where the active interaction dominates over passive processes. 
Similar hollow structures formed by microtubules were also seen in \textit{Xenopus} egg extracts \cite{Mitchison2013}. 
The connection between our results and these experiments deserves the attention of future research.

For very low passive-to-active ratios $\alpha$, we observe that the micelles undergo a fingering instability. 
In the course of  this instability, the micelles start deviating from their circular shapes, as the microtubule monolayer ring starts to bend, forming lobes (Fig.~\ref{fig:PDalpha}). 
The instability is characterized by a simultaneous increase in both density and orientational fluctuations, departing from the originally radially symmetric base state.
This is due to the interplay between the splay flux, which advects the microtubule density in the presence of splay, and the effective anchoring of the polarization to density gradients. 
The coupling between these two effects leads to the growth of the splay of the microtubule polarization as well as the redistribution of density around the monolayer ring. 
Once this modulation is strong enough, the splay flux leads to the invasion of regions of positive splay towards the inside of the micelle, while regions of negative splay recede towards the outside of the micelle, leading to the bending of the monolayer ring and the formation of fingers (Fig.~\ref{fig:ShInst}). 
The fact that both the splay flux and the effective anchoring are active mechanisms in our theory, resulting from the motor-mediated interaction, explains why this instability arises at small passive-to-active ratios $\alpha$.

Interestingly, a phenomenologically similar fingering instability has been observed in active nematic droplets~\cite{Alert2022}.
However, the underlying feedback mechanism responsible for this instability is distinctly different from the one proposed in the present work for active micelles. The nature of our micellar fingering instability is fundamentally polar, involving mechanisms such as anchoring along density gradients and self-advection, that can't be captured by a nematic order parameter.
Moreover, the instability observed for active nematic droplets is primarily interfacial in nature, driven by contractile stresses in tandem with perpendicular anchoring at the droplet's boundary~\cite{Alert2022}.
In contrast, the fingering instability in our active micellar structures originates within the monolayer's bulk, giving rise to a splay and density modulation that subsequently leads to the emergence of fingers from the interface.
A common aspect of both fingering instabilities is the fact that the growth of a finger is driven by contractility, which manifests itself as a splay flux in our active micelles and as active stress in active nematic droplets~\cite{Alert2022}.
The invasion of active interfaces due to active stresses in nematic models was first described in Ref.~\cite{Kempf2019}.

Phenomenologically similar morphological transitions involving the growth of fingers at interfaces have been observed in the spreading of bacterial colonies~\cite{XuNejad2022} and epithelial tissue~\cite{Poujade2007,Alert2019}. 
In the bacterial system, the fingering instability is driven by the formation of $-1/2$ nematic defects. 
Once these defects are formed, the growth of the protrusions in the boundary is again driven by the active stresses arising from the splay deformation. 
As the interfaces of the finger become unstable, the interface morphology evolves into a branch-like structure. 
In contrast, the instability of polar interfaces described for active nematic droplets~\cite{Alert2019} relies on self-propulsion. 
Leading regions of the advancing fronts move faster than trailing regions, resulting in a modulation of the interface. 
These examples demonstrate the ubiquity of fingering instabilities at deformable boundaries in active matter systems, resulting from different underlying mechanisms. 
While our work has active filament mixtures in mind, we expect the new mechanism for fingering transitions proposed in our work to generalize to other systems where the polar order doesn't manifest itself in self-propulsion, but instead drives the redistribution of density via splay-dependent fluxes.

\subsection*{From micelles to bilayers: the branching instability}

Active micelles also exhibit a branching instability, markedly different from the fingering instability, both in terms of phenomenology and the underlying mechanism. 
It is manifested by a breakup of the microtubule monolayer ring surrounding the center of the micelle into layered branches that extend radially outward.
Each of these branches consists of a bilayer, in which the microtubule polarization points in opposite directions on either side, like lipid molecules in lipid bilayers. 
At the center of each bilayer, the microtubule density has a dip, which arises due to contractile fluxes.

Differing from the fingering instability, this branching instability primarily results from the amplification of orientational perturbations.
This arises from two factors affecting how the orientational order changes over time: active splay, which modulates the strength of the polar order according to the sign of its splay, and self-anchoring, which reorients alignment according to the gradients of its strength.
The feedback between these two effects results in the amplification of splay along the micelle ring, ultimately causing it to split into bilayers (Fig.~\ref{fig:BrInst}).

Since this instability involves only the feedback between terms in the polarization equation, we anticipate its applicability across a broad spectrum of active matter systems involving polar order.  
Unlike the splay instability described for collections of self-propelled agents \cite{Mishra2010, Gopinath2012}, the instability presented here does not rely on self-propulsion.
The mechanism underlying our instability can thus also be expected to play a role in systems where the polar alignment of the agents does not result in collective propagation, for which microtubule-motor mixtures are but one example~\footnote{The branching instability does not require the system to be active, since for $\lambda_2=2\zeta_2$ and $\lambda_1=0$ the polarization equation obeys gradient dynamics, and for sufficiently large $\lambda_2$ Eq.~\eqref{eq:mixed_or_instab} holds true.}.
 
\subsection*{From bilayers to active foam networks}

In the derived coarse-grained model, the branching instability occurs for large antiparallel sliding strength $\eta$ and values of the critical angle $\wcrit$ that are small compared to the fully polar case $\wcrit=\pi$, but large enough to remain in the range where polar order dominates.  
Within this parameter range, stable micelles form at small microtubule densities. 
As the microtubule density $\rho$ is increased, a branching instability eventually occurs, causing bilayers to grow from the micelles.

At even higher microtubule densities, we find that the bilayers from different branching micelles interconnect, resulting in the formation of an active foam-like network.
The bilayers reconfigure over time, leading to the splitting and merging of active foam cells (see Video S3 \bibnotemark[SI]).
Similar active foams have been experimentally observed in mixtures of microtubules and kinesin-4 motors \cite{Dogic2021}.
As in our theory, asters formed at the lowest microtubule densities in those experiments, while foams emerged at high microtubule concentrations. 
In addition to replicating this experimentally observed transition controlled by the microtubule density $\rho$, our study underscores those features of the microscopic interaction that are essential for the formation of bilayers and active foams.
Firstly, the regime where the branching instability and the formation of large-scale foam-like bilayer networks occur at large $\eta$ and small $\wcrit$ demonstrates the importance of the antiparallel contribution to the alignment interaction in these processes. Heuristically, antiparallel alignment stabilizes the orientational order within the bilayer, as it aligns the microtubules on either side along the bilayer normal, while antiparallel sliding ensures the two opposing microtubule monolayers that constitute the bilayer have only partial overlap so that the polar order is non-vanishing. Secondly, the parity symmetry of the alignment interaction has to be broken, i.e., the critical angle $\wcrit$ has to be larger than $\pi/2$ to promote the emergence of polar order over nematic order. 
Finally, the inhomogeneity of the motor concentration due to the procession of motors along the filaments has to be taken into account, as foams cannot form for vanishing motor Péclet numbers $\gamma$.
Our work comprehensively addresses all of these properties and their interplay, a consideration that has been disregarded in studies using the BGL approach \cite{AransonTsimring2006, Ziebert2007, MaryshevPRE2018, MaryshevSM2019}.

It should be noted that the experiments on microtubule\--motor mixtures~\cite{Dogic2021} were performed in a three-dimensional setting.
The emergence of the micelle-foam transition in our two-dimensional model suggests that the mechanisms responsible for active foam formation may not be inherently reliant on the three-dimensional nature of the system.
Nevertheless, exploring extensions of our model into three dimensions could offer valuable insights and remains an intriguing direction for future investigations, e.g., by using generalizations of the BGL approach to three dimensions~\cite{Mahault2018}.
Furthermore, our model neglects hydrodynamic interactions, which could play an additional role in the experimental system. 
To study the effect of these interactions, coarse-graining methods that take them into account explicitly, such as the one proposed in Ref.~\cite{Reinken2018}, could be applied in future studies.

\subsection*{Dynamic control parameter fields}
Our work highlights a fundamental principle shared by many non-equilibrium systems far from equilibrium, including models of active matter and pattern-forming systems: self-organization is governed by dynamic fields of order (or bifurcation) parameters. This principle is essential in mass-conserving reaction-diffusion systems, where the dynamic bifurcation parameters of total protein densities regulate local reactive equilibria~\cite{Halatek2018, Brauns2020, Wurthner2022}.
It also applies to active systems where conserved densities are coupled to orientational order~\cite{Denk2020}.
In these systems, the conserved densities (here, the microtubule density $\rho$ and the motor density $m$) have a dual role as control parameters and dynamic variables. 
On the one hand, they drive pattern formation on a local scale by governing local reactive equilibria or the emergence of orientational order. 
On the other hand, the emerging pattern gives rise to fluxes that redistribute density throughout the system.

This dual role of the control parameters can organize the system into well-separated spatial domains where phase transitions occur locally. 
In our system, for example, the motor density $m$ accumulates inside bilayers and micelles. 
These active supramolecular assemblies constitute islands above criticality ($m\rho>\rho_c$) where polar order develops, lying in an isotropic background below criticality.
A general strategy to study these systems is to apply insights from homogeneous states to the local scale, relating the mean densities of the former to the local densities of the latter. 
Here, this idea motivates applying the linear stability analysis of homogeneous states to understand micelle instabilities. 
Similarly, for mass-conserving reaction-diffusion systems~\cite{Frey2018}, it justifies the local equilibrium theory~\cite{Halatek2018, Brauns2020} using regional dispersion relations, which was recently applied to forecast complex patterns in spatially varying geometries~\cite{ Wurthner2022}.

\subsection*{Towards a general theory of active foams}
We believe that the present study takes an important step towards formulating a general theory of active foams, encompassing distinct types of order.
While previous theoretical studies have explored foams governed by scalar fields \cite{Fausti2021} and nematic fields \cite{MaryshevSM2020}, our work expands the horizon by providing a comprehensive theory for active foams with polar order, which connects for the first time the continuum theory with an underlying physical system.

All these active foams exhibit sustained dynamics, characterized by the continuous collapse of cells and the concurrent formation of new edges within the foam network.
This leads to the existence of non-equilibrium steady states with characteristic length and time scales.
These features are fundamentally distinct from passive foams, whose dynamics are governed by a free energy, and rely solely on interfacial forces, such as surface tension \cite{Isenberg1992}.
Importantly, passive foams are not maintained for indefinite times, but exhibit coarsening behavior, meaning smaller cells tend to shrink over time while larger ones grow at their expense \cite{Glazier1992}.

Within the different classes of active foams, notable qualitative distinctions emerge, particularly concerning their sustaining mechanisms.
In the scalar active foams of Active Model B+, the active terms in the dynamical equation of the density field make the capillary tension negative, leading to an interfacial instability \cite{Fausti2021}. In contrast, the dynamics of the nematic foams of Active Model C ~\cite{MaryshevSM2020} and the polar foams discussed here crucially depends on the orientational order, which drives phase separation and governs the instabilities in the system. In addition, the various order parameters involved lead to differences in the structure of the foam edges, which are nematic bands in active model C~\cite{MaryshevSM2020} but polar bilayers in the present work.
Exploring and systematically understanding these essential differences, along with identifying common underlying principles, necessitates the development of a broader and unified framework -- a promising direction for future research.

Furthermore, the existence of active foams with these three types of order opens up exciting possibilities for investigating foams with higher ($p$-atic) symmetry~\cite{Giomi2022} or even coexisting multiple types of order~\cite{Denk2020}.
Such a line of research promises to unravel general principles in active matter systems, highlighting the emergence of complex non-equilibrium steady states from the dynamics of relatively few interacting fields that break the requirements of detailed balance.

\section*{Acknowledgements}
The authors acknowledge Rémi Boros and Zvonimir Dogic for insightful discussions. 
This work was funded by the Deutsche Forschungsgemeinschaft (DFG, German Research Foundation) through the Collaborative Research Center (SFB) 1032 – Project-ID 201269156 – and the Excellence Cluster ORIGINS under Germany’s Excellence Strategy – EXC-2094 – 390783311.
\appendix
\section{\label{app:BGL}Derivation of the hydrodynamic equations}
\subsection{Kinetic equation}
The starting point of the Boltzmann-Ginzburg-Landau procedure is the one-particle probability density function $P(\rv,\phi)$. $P(\rv,\phi)\dd[2]\rv\dd\phi$ indicates the probability of finding a microtubule inside the infinitesimal surface element at position $\rv$ and at an orientation along the unit vector $\nv(\phi)=(\cos\phi ,\sin\phi)$ with an angle in the interval $[\phi,\phi+\dd\phi)$. The time evolution of the probability density function follows a Boltzmann-like kinetic equation, including diffusive terms ($I_{\text{diff}}$), a term arising from excluded volume interactions ($I_{\text{excl}}$), as well as gain and loss terms from motor-mediated alignment and sliding interactions (``collisions'', $I_{\text{coll}}$)):
\begin{eqnarray}
\label{eq:boltzmanneq}
	\partial_t P(\rv,\phi) = I_{\text{diff}} + I_{\text{excl}} + I_{\text{coll}}.
\end{eqnarray}

Normalizing the one-particle PDF such that integrating it over all angles and positions yields the total number of microtubules in the system, we can reinterpret $P(\rv,\phi)$ as a local density of microtubules with a given orientation. In this setting, the Fourier transform of $P(\rv,\phi)$ in angular space yields modes with a clear physical interpretation. Using the following convention:
\begin{eqnarray}
\label{eq:fourier}
P(\rv;\phi)=\sum\limits_{k=-\infty}^\infty P_k(\rv)\exp(ik\phi),
\end{eqnarray}
we can identify the three lowest modes with the coarse-grained density, mean polarization vector, and nematic tensor fields (the definition of the latter two absorbs the density):
\begin{subequations}
\label{eq:fields}
\begin{align}
\rho(\rv) &=\int_{0}^{2 \pi}\dd{\phi}P(\rv, \phi)=2 \pi P_{0}(\rv),\\
\polv(\rv) &= \int_{0}^{2 \pi}\dd{\phi}\vcu{n}(\phi) P(\rv, \phi) 
=2\pi\mqty(\Re P_{-1}\\\Im P_{-1}),\\
 Q_{ij}(\rv) &= \int_{0}^{2 \pi}\dd{\phi}(2n_i n_j-\delta_{ij}) P(\rv, \phi)\nonumber\\
 &=2\pi\mqty(\Re P_{-2}&\Im P_{-2}\\\Im P_{-2}&-\Re P_{-2}).
\end{align}
\end{subequations}
Note that we chose to include a factor of $2\pi$ into the definition of the polarization and nematic tensor, to make the final equations more readable.

The diffusive term in Eq.~\eqref{eq:boltzmanneq} includes both rotational and translational diffusion. With the Einstein summation convention, they read:
\begin{eqnarray}
\label{eq:Idiff}
I_{\text{diff}}=D_r\partial_\phi^2P(\rv,\phi)+\partial_i[D_{ij}\partial_jP(\rv,\phi)],
\end{eqnarray}
where $D_r$ is the rotational diffusion constant and the anisotropic diffusion tensor is $D_{ij}=D_\parallel n_in_j+D_\perp(\delta_{ij}-n_in_j)$. We follow Ref.~\cite{AransonTsimring2006} in using the values for rigid rods so that $D_\parallel = 2D_\perp = D_r\cdot L^2/24$.

On the other hand, we write the excluded volume term as a Smoluchowski-like advection term:
\begin{eqnarray}
I_{\text{excl}}=\partial_i[P(\rv,\phi)D_{ij}\partial_jU(\rv)],
\end{eqnarray}
where the potential $U(\rv)$ arising from steric exclusion was derived for rigid rods in Ref.~\cite{MaryshevPRE2018}, and reads (taking into account two-filament and three-filament interactions):
\begin{eqnarray}
U(\rv) = \frac{2}{\pi}L^2\rho(\rv)+\frac{1}{4\pi}L^4\rho(\rv)^2.
\end{eqnarray}
Thus, the excluded volume term yields a $\rho$-dependent contribution to the isotropic diffusivity. In Ref.~\cite{Ahmadi2006}, the effect of the steric interactions on the orientational order was considered, which we neglect here for simplicity.

\subsection{Interaction kernel}
The collision term $I_{\text{coll}}$ in Eq.~\eqref{eq:boltzmanneq} depends on the motor-mediated interaction rules we have specified in sec.~\ref{sec:micro}. These are encoded into the interaction kernel, which determines how the interaction rate between two microtubules depends on their positions and orientational configurations. We model the microtubules as ideal rods of length $L$ which interact only if they intersect. For two microtubules with center of mass positions $\rv_{1,2}$ and angles $\phi_{1,2}$, with distance vector $\xiv=\rv_2-\rv_1$, intersection angle $\omega=\phi_2-\phi_1$ and bisector angle $\phi=(\phi_1+\phi_2)/2$, this intersection condition defines a rhombus centered at $\rv_1$, oriented along the bisector unit vector $\vcu{n}(\phi)$, of side length $L$ and aperture $\omega$, as shown in Fig.~\ref{fig:terms}e. The center of the second microtubule $\rv_2$ has to reside within the rhombus in order for an interaction to be possible.

To parametrize this rhombus, two sets of base vectors are useful. The first one is the orthonormal basis $\{\npv,\nv\}$ defined by the bisector unit vector and its clockwise perpendicular unit vector $\npv(\phi)=\vcu{n}(\phi-\pi/2)$; the second one is the non-orthogonal basis $\{\vcu{u}, \vcu{v}\}$ pointing along the sides of the rhombus, defined by $\vcu{u}=\sin(\omega/2)\npv+\cos(\omega/2)\nv$ and $\vcu{v}=-\sin(\omega/2)\npv+\cos(\omega/2)\nv$. Using the latter basis, we decompose the distance vector as $\xiv=\xicu\vcu{u}+\xicv\vcu{v}$, and define the interaction kernel $K=K(\rv_1,\rv_2;\phi_1,\phi_2)$ as:
\begin{align}
\label{eq:kernel}
    K:= Gm\cdot\Theta\left(L-2\abs{\xicu}\right)\Theta\left(L-2\abs{\xicv}\right).
\end{align}
The Heaviside theta functions introduced here delimit the interaction rhombus. They depend on $\phi_{1,2}$ via the base vectors $\{\vcu{u}, \vcu{v}\}$. We will use the following shorthand notation for an arbitrary function $F(\xiv)$ (which can  also depend on variables other than $\xiv$):
\begin{equation}
\int\limits_{K(\phi,\omega)}\!\dd[2]\xiv F(\xiv) := \int\!\dd[2]\xiv K(\rv_1,\rv_2;\phi_1,\phi_2)F(\xiv),
\end{equation}
with the positions and orientations given by $\rv_{1,2}=\rv\mp\frac{\xiv}{2}$, $\phi_{1,2}=\phi\mp\frac{\omega}{2}$.

With the interaction kernel defined above, we can turn to the collision term from Eq.~\eqref{eq:boltzmanneq}. It can be decomposed into gain and loss parts, where the gain takes different forms for polar ($\upuparrowsnarrow$) and antipolar ($\updownarrowsnarrow$) interactions:
\begin{align}
I_\text{coll}=J_\text{gain}^{\upuparrows}
+J_\text{gain}^{\updownarrows}
-J_\text{loss}.
\end{align}
The three terms read:
\begin{subequations}
\label{eq:coll_ints}
\begin{align}
	&J_\text{gain}^{\upuparrows} = \int\limits_{-{\wcrit}}^{\wcrit}\dd{\omega}\int\limits_{K(\phi,\omega)}\dd[2]\xiv P(\rv_1;\phi_1)P(\rv_2;\phi_2),\\
	&J_\text{gain}^{\updownarrows} =\! \int\limits_{{\wcrit}}^{2\pi-{\wcrit}}\dd{\omega}\hspace{-1em}\int\limits_{K(\phi+\frac{\pi}{2},\omega)}\hspace{-1em}\dd[2]\xiv P\big(\rv_1^\prime;\phi^\prime_1\big)P\big(\rv_2^\prime;\phi_2^\prime\big),\\
	&J_{\text{loss}} =\! \int\limits_{-\pi}^{\pi}\dd{\omega}\hspace{-1em}\int\limits_{K(\phi-\frac{\omega}{2},\omega)}\hspace{-1em}\dd[2]\xiv P(\rv;\phi)P(\rv-\xiv;\phi-\omega).
\end{align}
\end{subequations}
Note that the interaction rhombus is oriented along different angles in the three cases: the bisector of the microtubule pair before the interaction corresponds to the final microtubule angle $\phi$ for the polar gain, while it lies perpendicular to it for the antipolar gain; in the loss term, the bisector is at an angle $\omega/2$ with the filament before the interaction. We have used the shorthands $\phi_{1,2}^{\prime}=\phi_{1,2}+\frac{\pi}{2}$ and $\rv_{1,2}^\prime=\rv_{1,2}+\frac{\eta L}{2}\vcu{n}(\phi)$, where the latter accounts for antiparallel sliding.

\subsection{Homogeneous system and polar regime}
As discussed in the main text, the emergence of global order can be studied by inspecting the homogeneous state, where $P(\rv;\phi)=P(\phi)$. Under this assumption, the PDFs can be taken out of the spatial part of the collision integrals in $I_\text{coll}$, which then simply give the area of the interaction rhombus:
\begin{equation}
\label{eq:rhombus_area}
    \int\limits_{K(\phi,\omega)}\!\dd[2]\xiv=L^2Gm\abs{\sin\omega}=:g(\omega).
\end{equation}
Using this relation in the kinetic equation \eqref{eq:boltzmanneq} yields Eq.~\eqref{eq:kineticPhi}. Rescaling time, space and the PDF as indicated in the main text, and Fourier transforming the PDF, we obtain Eq.~\eqref{eq:kineticPk}, with the coefficients $f(k,q)$ given by the expression:
\begin{align}
f(k,q)&:=\int_{-\omega_c}^{\omega_c}\dd\omega \abs{\sin\omega}e^{i(k/2-q)\omega}-\int_{-\pi}^{\pi}\dd\omega \abs{\sin\omega}e^{-iq\omega}\nonumber\\
&\phantom{{}:={}}+\int_{\omega_c}^{2\pi-\omega_c}\dd\omega \abs{\sin\omega}e^{ik(\omega+\pi)/2-iq\omega}.
\end{align}
These integrals are analytically solvable. Using the properties of $f(k,q)$, one can show that the only modes that can show exponential growth beyond a certain critical density are those corresponding to the polar and nematic order, given in Eqs.~\eqref{eq:rhoc}. The $P_k$ with $\abs{k}\geq 3$, on the other hand, are always stable.

\subsection{Spatial integrals}
To calculate the collision integrals in the general case, including the spatial dependence of the PDF, we perform a gradient expansion. This assumes that the probability density function doesn't vary much on the scale of the microtubule length $L$, and will be justified in the Ginzburg-Landau expansion. We expand the PDFs appearing in Eqs.~\eqref{eq:coll_ints} around $\rv$ to second order in the gradients \bibnote{While a term $\laplacian(\div\polv)$ in the $\rho$-equation would be of the right order in the Ginzburg-Landau parameter $\epsilon$, its coefficient vanishes in our model; however, it is non-zero for an anisotropic bound motor profile, see Refs.~\cite{AransonTsimring2006,MaryshevPRE2018}.}, yielding:
\begin{equation}
P(\rv+\vc{a};\phi)\approx P(\rv;\phi)+(\vc{\vc{a}}\vdot\grad) P+\frac{1}{2}(\vc{\vc{a}}\vdot\grad)^2 P,
\end{equation}
where the shift vector $\vc{a}$ always fulfills $\abs{\vc{a}}\leq 1$ (i.e., $\leq L$) due to the Heaviside kernel \eqref{eq:kernel} appearing in the integrals. With this expansion, the spatial integrals reduce to integrals over the interaction rhombus involving polynomials of the displacement vector $\xiv$. Indeed, using the basis $\{\npv,\nv\}$, the gradients reduce to $\omega$-independent directional derivatives, and we are left with spatial integrals of the form:
\begin{equation}
    \int\limits_{K(\phi,\omega)}\!\dd[2]\xiv\;\xic_\perp^j\xic_\parallel^k,
\end{equation}
where $\xic_{\perp,\parallel}$ are the components of $\xiv$ in the basis $\{\npv,\nv\}$. Changing coordinates to $\{\vcu{u}, \vcu{v}\}$, this integral turns into:
\begin{equation}
    c_{jk}(\omega)Gm\int_{-L/2}^{L/2}\dd\xicu\int_{-L/2}^{L/2}\dd\xicv\;(\xicu-\xicv)^j(\xicu+\xicv)^k,
\end{equation}
where the prefactor resulting from the coordinate transformation reads $c_{jk}(\omega)=\abs{\sin\omega}\sin^{j}(\omega/2)\cos^{k}(\omega/2)$. The integral is analytically solvable for all $j$ and $k$. Note that the case $j=k=0$ reduces to Eq.~\eqref{eq:rhombus_area}.

After substituting the Fourier transform \eqref{eq:fourier} and projecting the kinetic equation on the individual Fourier modes, the only integrals left to calculate are trigonometric integrals over $\omega$, which are also analytically solvable. This procedure allows us to circumvent the numerical approach taken in Ref.~\cite{MaryshevPRE2018} by making use of the basis $\{\vcu{u}, \vcu{v}\}$.

\subsection{Ginzburg-Landau closure}
The equations we are left with contain an infinite number of Fourier modes (cf. the infinite sum in Eq.~\eqref{eq:kineticPk}). A closure must be chosen to truncate the equations. In the polar regime, where $\rhocp>0$ and all other Fourier modes decay (i.e., $\rhoc^{(k)}<0$ in Eq.~\eqref{eq:decayPk} of the main text), the Ginzburg-Landau closure can be used for mean densities $\bar\rho$ close to the critical density $\rhoc=\rhocp$. Under these conditions, the dynamics of Fourier modes $P_k$ with $\abs{k}\geq 2$ is much faster than the dynamics of $\rho$, which is a conserved field, and of $\polv$, whose growth rate $\epsilon^2:=\bar\rho-\rhoc$ is small. Then, we can adiabatically eliminate the higher Fourier modes in favor of $\rho$ and $\polv$ by requiring stationarity in the respective equations.

Neglecting gradient terms for the moment, the adiabatic elimination of the nematic mode $P_2$ in Eq.~\eqref{eq:kineticPk} yields a term proportional to $P_1^2$. Inserting this back into the equation for $P_1$, which contains a term proportional to $P_2P_{-1}$, this gives rise to a cubic term $P_{-1}P_1^2\sim p^2P_1$, which leads to the saturation of the polarization at small values. Indeed, as emerges from Eq.~\eqref{eq:p0_GL}, the equilibrium value of the polarization $p^2$ scales like $\epsilon^2$, which in turn implies $p\sim \epsilon$. From Eq.~\eqref{eq:kineticPk}, this allows us to obtain the scaling behavior of all other Fourier modes as $P_{k}\sim\epsilon^{k}$.

Now we can reintroduce the gradient terms. Balancing the terms arising via diffusion (i.e., the $\kappa_{1,2}$ terms in Eq.~\eqref{eq:der}) with the Ginzburg-Landau terms yields a scaling $\nabla\sim\epsilon$. In other words, the typical length scales of patterns in the system are  expected to scale like $\epsilon^{-1}$. Finally, balancing the terms arising in the diffusion equation, one obtains that deviations of the density field $\rho$ from its mean value scale like $\rho-\bar\rho\sim \epsilon^2$. Having obtained the scaling of all fields, we can truncate the density and polarization equations by keeping terms up to $O(\epsilon^3)$.

Strictly, the BGL procedure is only rigorously motivated for small values of $\epsilon$. However, the equations obtained can be extrapolated to finite $\epsilon$ values. While the predictions made in this extrapolated regime should not be expected to be quantitatively accurate, this leap of faith gives the possibility of making qualitative statements formulated in terms of the parameters of the microscopic interaction model. This would be impossible to do in a fully top-down approach, where the continuum equations are posited using symmetry arguments, while the functional dependence of the coefficients on microscopic parameters is left undetermined.

In terms of the fields introduced in Eq.~\eqref{eq:fields}, the adiabatic elimination of $P_{\pm 2}$ yields the following expression for the nematic tensor:
\begin{equation}
\label{eq:adiabaticQ}
    Q_{ij}=-12B\cos\wcrit m(p_ip_j)\TS,
\end{equation}
with the traceless symmetric tensor $(p_ip_j)\TS=p_ip_j-\delta_{ij}p^2/2$, and
\begin{equation}
    B^{-1}=12\pi+\bar\rho(1+3\cos{2\wcrit}).
\end{equation}

\subsection{Final equations}
\label{app:coefficients}
As we have discussed above, the full calculation of the collision term $I_{coll}$ can be performed analytically. Since it is quite lengthy, we perform it using Wolfram Mathematica, providing the final results here. Using the adiabatic elimination of the nematic order $Q_{ij}$ given in Eq.~\eqref{eq:adiabaticQ}, and introducing the shortcuts:
\begin{align}
    \chi_Q &=\frac{1}{96}+\left(\frac{\sin^2\wcrit}{16\pi}\eta^2-\frac{1}{72\pi}\right)m\bar\rho,\nonumber\\
    \lambda_Q &=\frac{\sin^2\wcrit}{4\pi}\eta\bar\rho,
\end{align}
the coefficients in Eq.~\eqref{eq:der} read:
\begin{eqnarray*}
D_\rho&=&\frac{1}{32},\;\alpha_2=\frac{\alpha}{32\pi},\;\alpha_3=\frac{\alpha^2}{192\pi},\\
\nu&=&\frac{\cos^2\!\frac{\wcrit}{2}}{8\pi}\eta^2-\frac{1}{48\pi},\\
\chi_1&=&-\frac{\cos^2\!\frac{\wcrit}{2}}{4\pi}\eta^2m-12B\cos\wcrit\chi_Qm,\\
\chi_2&=&\frac{\cos\wcrit\cos^2\!\frac{\wcrit}{2}}{8\pi}\eta^2m+6B\cos\wcrit\chi_Qm,\\
\beta&=&\frac{4\cos\wcrit\left(7-15\cos\frac{\wcrit}{2}+3\cos\frac{5\wcrit}{2}\right)}{5\pi}B,
\end{eqnarray*}
\begin{eqnarray}
\kappa_1&=&\frac{5}{192}+\frac{24-20\cos\frac{\wcrit}{2}-5\cos\frac{3\wcrit}{2}+\cos\frac{5\wcrit}{2}}{960\pi}\bar\rho m,\nonumber\\
\kappa_2&=&\frac{1}{96}+\frac{8-5\cos\frac{3\wcrit}{2}-3\cos\frac{5\wcrit}{2}}{1440\pi}\bar\rho m,\nonumber\\
\lambda_1&=&\lambda_2=\frac{\cos^2\!\frac{\wcrit}{2}}{\pi}\eta m+12B\cos\wcrit\lambda_Qm^2,\nonumber\\
\zeta_2&=&\frac{\cos\wcrit\cos^2\!\frac{\wcrit}{2}}{2\pi}\eta m+6B\cos\wcrit\lambda_Qm^2,\nonumber\\
\zeta_1&=&\frac{\cos^2\!\frac{\wcrit}{2}}{\pi}\eta m\bar\rho.
\end{eqnarray}
\section{Interpreting the equations}
\subsection{Terms in the density equation}
\label{app:terms_rho}
The density equation \eqref{eq:der_rho} is a continuity equation of the form $\partial_t\rho+\div\vc{J}=0$, with the conserved flux:
\begin{align}
    \vc{J}&=-(D_\rho+2(m\nu+\alpha_2)\rho+3\alpha_3\rho^2)\grad\rho\nonumber\\
    &\phantom{{}={}}-(\polv\vdot\grad)(\chi_1\polv)-[\div(\chi_1\polv)]\polv-\grad(\chi_2p^2).
\end{align}
The first line of this flux represents an effective isotropic diffusive flux resulting from diffusion, motor-mediated interactions and steric repulsion. The second line, on the other hand, is polarization-dependent. To understand the different fluxes it involves, we write $\polv=p \nv$ with unit vector $\nv$, and define a unit vector $\npv$ which is clockwise perpendicular to the latter. For simplicity, we assume a homogeneous motor field here, s.t. $\chi_{1,2}$ are constant. Then, the polarization-dependent part of $\vc{J}$ reads
\begin{align}
    &-\chi(\dpar p^2)\nv-\chi_2(\dperp p^2)\npv\nonumber\\
    &-\chi_1p^2\dpar\nv-\chi_1p^2(\div\nv)\nv,
\end{align}
where we have defined $\dpar=\nv\vdot\grad$ and $\dperp=\npv\vdot\grad$, as well as $\chi=\chi_1+\chi_2$. These four terms correspond to the four fluxes shown in Fig.~\ref{fig:terms}a: i) the contractile flux, advecting density along $\nv$ into regions of high polar order for $\chi<0$; ii) the transverse flux, advecting density along $\npv$ into regions of high polar order for $\chi_2<0$; iii) the bend flux, which is perpendicular to the polarization and advects density to the inside of a bend for $\chi_1<0$; iv) the splay flux, which is parallel to the polarization and advects density along the splay vector $(\div\nv)\nv$ for $\chi_1<0$.

\subsection{Terms in the polarization equation}
\label{app:terms_p}
The polarization equation \eqref{eq:der_p} can be divided into an equation for its amplitude and one for its orientation. To this goal, we write $\polv=p \nv$ as before. By taking the scalar product of Eq.~\eqref{eq:der_p} with $\nv$ and $\npv$ respectively, and using that $\nv\vdot\partial_t\nv=0$ to preserve the unit vector length, we find:
\begin{subequations}
\begin{align}
    \partial_t p &= \left[(m\rho/\rhoc-1)-\lambda_2p\div\nv-\beta m^2p^2\right]p\nonumber\\
    &-\lambda p\dpar p+\zeta_1 \dpar \rho\nonumber\\
    &+\kappa_1 (\nv\vdot\laplacian \polv)+\kappa_2 \dpar(\div\polv),\label{eq:ph_amp}\\
    p\npv\vdot\partial_t\nv&=-\lambda_1p^2\npv\vdot(\dpar\nv)\nonumber\\
    &+\zeta_1\nabla_{\perp}\rho+\zeta_2\nabla_{\perp}p^2\nonumber\\
    &+\kappa_1 (\npv\vdot\laplacian \polv)+\kappa_2 \nabla_{\perp}(\div\polv).\label{eq:ph_or}
\end{align}
\end{subequations}
We can now read off the effect of the various terms. The first line of Eq.~\eqref{eq:ph_amp} describes an ordering transition taking place in regions with $m\rho>\rhoc$, whose exponential growth is saturated by the $\beta$ term. The term proportional to $\lambda_2$ amplifies (reduces) the order in regions with positive (negative) splay for $\lambda_2<0$, and vice versa for $\lambda_2>0$, an effect which we term \emph{active splay}. $\lambda_2$ has no effect on the orientational dynamics.

Both equations include a term advecting the pattern along the polarization field. For the polarization strength, this self-advection is controlled by $\lambda=\lambda_1+\lambda_2-2\zeta_2$, while for the orientational field, it is controlled by $\lambda_1$ alone.

The other terms appearing in the equations represent anchoring and stiffness contributions. These are best understood as deriving from a free energy. Indeed, the polarization equation \eqref{eq:der_p} can be written as a model A with additional active contributions that cannot be written in terms of a free energy \cite{Marchetti2013}: 
\begin{equation}
\label{eq:modelA}
    \delt\polv=-\fdv{\mathcal{F}}{\polv}-\lambda_1(\polv\vdot\grad)\polv-(\lambda-\lambda_1)(\div\polv)\polv.
\end{equation}
The two active terms arise due to self-advection, stemming from the fact that the latter breaks time-reversal symmetry. For vanishing self-advection, $\lambda=0=\lambda_1$, these terms disappear and the polarization equation reduces to pure gradient dynamics. Since the density equation and the motor equation cannot be written in terms of the same free energy, however, the system is still active even in this case.

The free energy $\mathcal{F}$ in Eq.~\eqref{eq:modelA} reads:
\begin{align}
\label{eq:free_energy}
    \mathcal{F}=\int\dd[2]{x}&\left[-\frac{1}{2}\left(\frac{m\rho}{\rhoc}-1\right)p^2+\frac{\beta m^2}{4}p^4\right.\nonumber\\
    &\phantom{\biggl[}-\zeta_1(\polv\vdot\grad)\rho-\zeta_2(\polv\vdot\grad)p^2\nonumber\\
    &\phantom{\biggl[}\left.+\frac{\kappa_1}{2}(\partial_ip_j)(\partial_ip_j)+\frac{\kappa_2}{2}(\div\polv)^2\vphantom{\frac{1}{2}}\right]\!.
\end{align}
The first line is responsible for the Ginzburg-Landau transition, while the second line favors a certain alignment of the order with respect to density and order strength gradients. For $\zeta_{1,2}>0$, parallel alignment to these gradients is favoured, while $\zeta_{1,2}<0$ favours antiparallel alignment. For this reason, we refer to $\zeta_1$ as \emph{density anchoring} and $\zeta_2$ as \emph{self-anchoring}.

The last line penalizes spatial variations of the polarization field. For constant $m$, it can be written as follows, up to boundary terms:
\begin{align}
    &\frac{\kappa}{2}[p^2(\div\nv)^2+(\dpar p)^2+(\div\nv)\dpar p^2]\nonumber\\
    +&\frac{\kappa_1}{2}[p^2(\dpar\nv)^2+(\dperp p)^2-(\dpar\nv)\vdot(\grad p^2)],
\end{align}
where $\kappa=\kappa_1+\kappa_2$. Thus, $\kappa$ penalizes splay deformations of the order as well as longitudinal variations of its strength; in contrast, $\kappa_1$ penalizes bend deformations as well as transversal variations of the order strength. Finally, the third term in each bracket denotes an additional anchoring effect that couples gradients in the polarization to splay and bend deformations.

\subsection{Microscopic origin of the terms in the $p$-equation}
\label{app:terms_micro}
\begin{figure}
    \centering
    \includegraphics{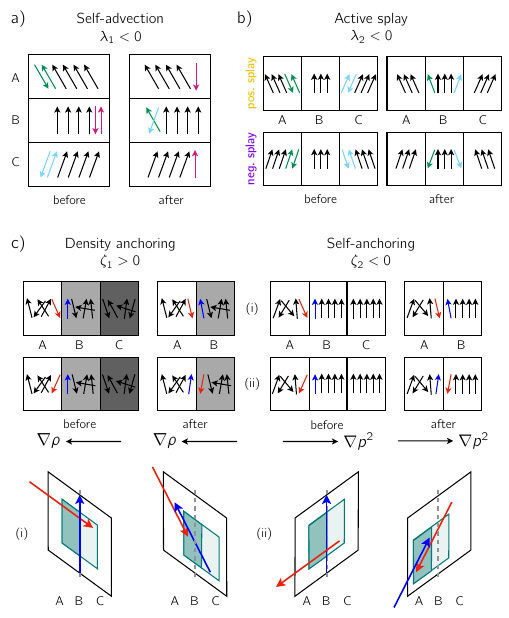}
    \caption{
    \label{fig:terms_micro} Microscopic origin of one-gradient terms in the polarization equation. a) Discretizing space in the longitudinal direction with respect to the polarization in region B, a bend causes the mean orientation to rotate to the left as one goes from C to B to A. In each compartment, there will be pairs of antiparallel filaments (shown in different colors). Upon interaction, antiparallel sliding separates the pairs, moving one filament up and one down. Thus, for example, the red filaments from B are separated into A and C; the blue filaments from C into B and a compartment below C which is not shown here. This separation leads to the rotation of the average polarization in each compartment to the left, explaining $\lambda_1<0$. b) Splay causes variation of the average polar orientation in the perpendicular direction. Upon interaction, pairs of antiparallel filaments are slid to the left and right, e.g., the green filaments from A separate into the compartment left of A (not shown) and into B. Depending on the sign of the splay, the order in B is amplified or reduced, explaining $\lambda_2<0$. c) A density gradient from the right to the left and a polar order gradient from the left to the right both cause the rate of antipolar interactions to be higher to the left than to the right. When a blue filament from the interface region (B) interacts with an almost antiparallel red filament from A, two things can happen depending on the angle between the filaments. In a situation like (i), the interaction rotates the blue filament in B to the left, where sliding causes little displacement compared to the initial position. In (ii), on the other hand, the interaction rotates the blue filament from B to the right, but slides it into A. Thus, overall, filaments at the interface rotate to the left, explaining $\zeta_1>0$ and $\zeta_2<0$.}
\end{figure}
In the main text, we have used a microscopic argument to explain the sign of the splay flux (see Fig.~\ref{fig:terms}f). Here, we use similar arguments to explain the signs of the coefficients $\lambda_{1,2}$ and $\zeta_{1,2}$ in the polarization equation. Note that all these coefficients are proportional to $\eta$ (cf. App.~\ref{app:coefficients}). Indeed, we will see that antiparallel sliding is responsible for the breaking of symmetries underlying all the signs of these coefficients.

We start by explaining the negative sign of the orientational self-advection coefficient $\lambda_1$. Suppose we start with a leftward bend, shown in Fig.~\ref{fig:terms_micro}a (the rightward bend is analogous). It is useful to discretize space into blocks, shown in the Figure as A, B and C. Each block corresponds to one coarse-grained volume element. Taking block B as our reference, A and C constitute its neighboring regions above and below, respectively. Now, due to the bend, the polarization in block A will point more to the left than in block B, and in block B more to the left than in block C. In each block, orientational fluctuations lead to the existence of a small number of filaments pointing into directions other than the average, and in particular in the opposite direction. Thus, there will be pairs of antiparallel filaments (colored in the figure) that can interact via antiparallel interactions. As they are already antiparallel, the only effect of that interaction will be to slide the filaments apart, shifting them to the blocks above and below, respectively. As shown in the figure, this results in the rotation of the average polarization to the left in all blocks. Analogously, for a rightward bend, the average polarization would be rotated to the right. This is equivalent to a backward propagation of the orientational order direction ($\lambda_1<0$). The crucial element is given by antiparallel sliding, which slides filaments to the \emph{back} with respect to the common center of mass, thus breaking the symmetry between the front and the back of a polarized region and allowing for a non-vanishing $\lambda_1$. Importantly, this self-advection mechanism does not arise as a consequence of a mass flux (in contrast to the self-advection term in the Navier-Stokes equations, for example), since microtubules are slid in two opposite directions and the total flux is zero. Instead, it should be seen as a purely orientational emergent phenomenon.

The negative sign of the active splay coefficient $\lambda_2$ can be explained similarly (see Fig.~\ref{fig:terms_micro}b). Here, we divide space into blocks perpendicular to the order direction, where B is our reference block, A lies to its left, and C to its right. In a situation with positive splay, filaments in A will be turned to the left and filaments in C to the right with respect to those in B. As before, due to orientational impurities, there will be pairs of antiparallel filaments in each block. Having such pairs interact in block B doesn't affect the order, since they will just slide along the order direction and stay in block B. Due to the splay, however, pairs of antiparallel filaments in blocks A and C will be slid into their neighboring blocks upon interaction by means of the antiparallel sliding. In a situation with positive splay, this will lead to the accumulation of filaments pointing almost in the right direction in block B, thereby increasing the average polarization in that block; on the other hand, in a situation with negative splay, these interactions will move filaments pointing in the opposite direction into block B, thereby disrupting the order there. This explains why $\lambda_2<0$. Again, antiparallel sliding is crucial.

Finally, we explain the anchoring coefficients $\zeta_{1,2}$ (see Fig.~\ref{fig:terms_micro}c). Again, we discretize space in three blocks. For the density anchoring, we impose a density gradient from right to left, so that A has many more filaments than C; for the self-anchoring, we impose a gradient in order from left to right, so that A is isotropic and C is ordered, while B is an intermediate region. From the perspective of B, in both these situations, it will be more probable to interact by means of an antiparallel interaction with filaments in A than with filaments in C. Indeed, for a density gradient, there are more filaments in A than in C regardless of the orientation, so that interaction is most likely with the former; for an order gradient, filaments in C are mostly parallel with those in B, so that there are relatively more antiparallel filaments in A. This imbalance results in a rotation of the filaments in B towards A. Indeed, an interaction with antiparallel filaments in A pointing to the right with respect to the filament in B (case (i) in Fig.~\ref{fig:terms_micro}c) will result in rotation of the latter to the left and little sliding, so that the filament stays in B and is rotated towards A; on the other hand, an interaction with antiparallel filaments in A pointing to the left with respect to the filament in B (case (ii) in Fig.~\ref{fig:terms_micro}c) results in strong sliding, so that the latter is moved into compartment A. Thus, overall, filaments at a density interface are rotated towards high-density regions ($\zeta_1>0$), whereas filaments at a disorder-order interface are rotated towards the isotropic domain ($\zeta_2<0$). Both mechanisms rely on antiparallel sliding.

\section{Phenomenological model}
\label{app:ph}
As illustrated in the main text, in the generalization of our derived model to a phenomenological model we make the choice of keeping a linear dependence of the active coefficients on $m$, and rescale the fields such that $\beta=1=\rhoc$. Thus, the phenomenological equations read:
\begin{subequations}
\label{eq:ph}
\begin{eqnarray}
    \partial_t\rho &=& \laplacian(D_\rho\rho+\nu m\rho^2+\alpha_2\rho^2+\alpha_3\rho^3)\nonumber\\
    &&{}+\hat\chi_1\partial_i\partial_j(mp_ip_j)+\hat\chi_2\laplacian (mp^2),\label{eq:ph_rho}\\
    \partial_t \polv &=& \left[(m\rho-1)-m^2p^2\right]\polv\nonumber\\
    &&{}+\hat\kappa_1 m\laplacian \polv+\hat\kappa_2 m\grad(\div\polv)\nonumber\\
    &&{}-\hat\lambda_1m(\polv\vdot\grad)\polv-\hat\lambda_2m(\div\polv)\polv\nonumber\\
    &&{}+\hat\zeta_1 m\grad\rho+\hat\zeta_2 m\grad p^2\label{eq:ph_p},\\
    \partial_t m &=& D_m\laplacian m - v_m\div(m\polv).
\end{eqnarray}
\end{subequations}
The quantities with the hat, as well as the other parameters $D_\rho$, $\nu$, and $\alpha_{2,3}$, are now no longer functions of microscopic parameters, but free parameters of the model that can be varied independently.

\section{Stationary profiles}
\subsection{One-dimensional (bilayer) profile}
\label{app:blprof}
Assuming that the fields vary only in one direction and that the polarization is oriented along this direction, Eqs.~\eqref{eq:m} and \eqref{eq:der} are reduced to the following one-dimensional equations:
\begin{subequations}
\label{eq:1D}
\begin{align}
    &\partial_t\rho = \partial_x^2(D_\rho\rho+m\nu\rho^2+\alpha_2\rho^2+\alpha_3\rho^3+\chi p^2),\label{eq:rho1D}\\
    &\partial_t p = (m\rho/\rhoc-1)p-\beta m^2p^3+\kappa \partial_x^2 p\nonumber\\
    &\phantom{\partial_t p ={}}+\zeta_1 \partial_x\rho-\lambda p\partial_x p,\label{eq:p1D}\\
    &\partial_t m = D_m\partial_x^2 m - v_m\partial_x(mp).\label{eq:m1D}
\end{align}
\end{subequations}

\subsubsection{Inner profile}
In this section, we derive the stationary profile of the interior of the bilayer. For small $\gamma$, in a first approximation, we can take the motor field to be constant, $m=m_0$. Then, we set the left-hand sides of the equations above to zero. We integrate Eq.~\eqref{eq:rho1D} twice, choosing the center of the bilayer (where $p=0$ and $\rho$ and $m$ are extremal) as the integration boundary. Assuming weak phase separation (i.e., small values of $\chi$, as we will see below), we can linearize the density around a reference value $\rho_0$. Then, we find:
\begin{equation}
\label{eq:rhovsp1D}
    \rho(x)= \rho_--\frac{\chi}{\Deff} p^2(x),
\end{equation}
where $\rho_-$ is the density at the center of the bilayer and
\begin{equation}
\label{eq:Deff}
    \Deff=D_\rho+2(\nu m_0+\alpha_2)\rho_0+3\alpha_3\rho_0^2
\end{equation}
is the effective isotropic diffusivity resulting from diffusion ($D_\rho$), motor-mediated interactions ($\nu$) and steric repulsion ($\alpha_{2,3}$). Equation \eqref{eq:rhovsp1D} encodes the depletion of the center of the bilayer as a consequence of contractile fluxes. Indeed, the polar order is non-vanishing at either side of the bilayer, while it must be zero at its center due to the sign change it undergoes as one crosses the bilayer. As a consequence, $\chi<0$ will accumulate density into the two ordered fronts of the bilayer, thereby depleting the center.

Assuming that at the density maximum of the bilayer the polarization profile is sufficiently flat, we can neglect the derivative terms in Eq.~\eqref{eq:p1D} at that point, such that the polarization at the maximum reads:
\begin{equation}
\label{eq:BL_p+}
    p_+^2=\frac{m_0\rho_+-\rhoc}{\beta m_0^2\rhoc},
\end{equation}
where $\rho_+$ is the density at the maximum. Inserting this into Eq.~\eqref{eq:rhovsp1D}, we find:
\begin{equation}
\label{eq:1D_PS}
    \frac{m_0\rho_--\rhoc}{m_0\rho_+-\rhoc}=1+\bar\chi,
\end{equation}
where we have defined the effective contractility as the ratio between the contractile flux in the ordered phase and the isotropic diffusive fluxes:
\begin{equation}
\label{eq:barchi}
    \bar\chi=\frac{\chi}{\beta m_0\rhoc\Deff}.
\end{equation}
The ratio in Eq.~\eqref{eq:1D_PS} is smaller than one (phase separation takes place) when the system shows contractile behavior, i.e., $\bar\chi<0$. When $\bar\chi$ reaches the critical value of $-1$, the center of the bilayer drops below criticality. In the following, we will assume $\bar\chi>-1$. By writing $\rho_\pm=\rho_0\pm\Delta\rho/2$, we can express the phase separation in terms of $\bar\chi$:
\begin{equation}
\label{eq:1D_PS_Delta_app}
    \Delta\rho=\frac{-2\bar\chi}{2+\bar\chi}(\rho_0-\rhoc/m_0).
\end{equation}

Inserting Eq.~\eqref{eq:rhovsp1D} into Eq.~\eqref{eq:p1D} at stationarity, we find:
\begin{align}
\label{eq:p1D_rhoel}
    0 &= ap-bp^3+\kappa \partial_x^2 p-\bar\lambda p\partial_x p,
\end{align}
where $a=m_0\rho_-/\rhoc-1$, $b=\beta m_0^2(1+\bar\chi)$, and we have defined the effective self-advection $\bar\lambda = \lambda+2\zeta_1\chi/\Deff$.

Inserting the ansatz \eqref{eq:p_1Dsol} into the stationarity condition \eqref{eq:p1D_rhoel} yields the conditions:
\begin{subequations}
\label{eq:statcond}
\begin{align}
    p_+^2 &= \frac{a}{b} = \frac{m_0\rho_+-\rhoc}{\beta m_0^2\rhoc},\\
    0&=a\ell^2+p_+\bar\lambda\ell-2\kappa.
\end{align}    
\end{subequations}
Here, the first equation is a consistency condition giving back Eq.~\eqref{eq:BL_p+}, while the second equation defines the length scale $\ell$.

The effect of the active terms is summarized in the coefficient $\Lambda$, which for the bilayer we define as follows:
\begin{align}
\label{eq:LBL_app}
    \LBL=\frac{-\bar\lambda p_+}{2a}.
\end{align}
This gives Eq.~\eqref{eq:LBL} in the main text. With this definition of $\LBL$, the second line of Eq.~\eqref{eq:statcond} gives the solutions:
\begin{align}
    \ellBL &= \LBL\pm\sqrt{\LBL^2+\frac{2\kappa}{a}},
\end{align}
where to obtain $\ell>0$ we select the $+$ sign, giving the relation \eqref{eq:ell} indicated in the main text.

Finally, the motor profile close to the center of the bilayer results by requiring stationarity in Eq.~\eqref{eq:m1D} and integrating the equation. This yields:
\begin{equation}
    \partial_x \log m = \gamma p.
\end{equation}
Integrating this condition and using the polarization profile from Eq.~\eqref{eq:p_1Dsol} gives an approximation for the $m$ profile which is valid for small $\gamma$, since we kept the motor field constant while deriving $p(x)$:
\begin{align}
    m(x) &= m_-\exp(\gamma\int_0^x p(x))\nonumber\\
    &= m_-\exp(-\gamma p_+\int_0^x\tanh\frac{x}{\ell}).
\end{align}
Solving the integral yields Eq.~\eqref{eq:m_1Dsol}.
 
\subsubsection{Outer profiles}
\label{app:extprofile}
In this section, we study the impact of motor inhomogeneities on the bilayer profile far from its center. Thus, we no longer assume constant motors. Then, requiring stationarity in the $\rho$-equation \eqref{eq:rho1D} gives:
\begin{align}
    \Deff\partial_x\rho + \nu\rho^2\partial_x m + \pdv{\chi}{m} \pol^2\partial_x m = -\chi\partial_x\pol^2.
\end{align}
On the other hand, Eq.~\eqref{eq:m1D} yields $\partial_x m = \gamma m p$. Inserting this into the equation above, we find:
\begin{align}
\label{eq:ext_rho_app1}
    \partial_x\rho = -\gamma\tilde{D}p-\chi\Deff^{-1}\partial_x\pol^2,
\end{align}
with $\tilde{D} = m(\nu\rho^2+\pdv{\chi}{m}\pol^2)/\Deff$. Substituting this result into the $p$-equation \eqref{eq:p1D}, we find:
\begin{align}
\label{eq:pgamma2}
    0 &= (m\rho/\rho_c-1)\pol-\beta m^2\pol^3\nonumber\\
    &\phantom{{}={}}-\bar\lambda p\partial_x p+\kappa\partial^2_x\pol-\gamma\zeta_1\tilde{D},
\end{align}

Now, we make a linear ansatz for the densities $\rho$ and $m$ as well as the polarization $p$:
\begin{subequations}
\label{eq:ext_ansatz}
\begin{align}
    \rho &= \rho_0 + \delta\rho (x-x_0),\\
    \pol &= \pol_0 + \delta\pol (x-x_0),\\
    m &= m_0 + \delta m (x-x_0),
\end{align}
\end{subequations}
where $\rho_0$, $\pol_0$ and $m_0$ are the values of the fields at some reference point $x_0$. We expect $\delta m$, $\delta\rho$ and $\delta\pol$ to be of order $O(\gamma)$, since they vanish for $\gamma=0$. Immediately, we obtain:
\begin{equation}
\label{eq:ext_m_app}
    \delta m = \gamma m_0 p_0+O(\gamma^2),
\end{equation}
and with this, from Eq.~\eqref{eq:ext_rho_app1}:
\begin{equation}
\label{eq:ext_rho_app}
    \delta \rho = -\frac{2\chi}{\Deff}p_0\delta p-\gamma p_0\tilde{D}+O(\gamma^2).
\end{equation}

Inserting the ansatz \eqref{eq:ext_ansatz} into Eq.~\eqref{eq:pgamma2} and setting $x=x_0$, we obtain:
\begin{align}
\label{eq:ext_p0}
    (m_0\rho_0/\rho_c-1)\pol_0-\beta m_0^2\pol_0^3-\bar\lambda p_0\delta p - \gamma\zeta_1\tilde{D} = 0.
\end{align}
Up to terms of order $O(\gamma)$, we recover the equilibrium polarization given by Eq.~\eqref{eq:p0_GL}. Inserting Eq.~\eqref{eq:ext_p0} back into Eq.~\eqref{eq:pgamma2}, we derive that equation with respect to $x$. Setting $x=x_0$ gives:
\begin{align}
    0&= \pol_0 m_0\delta \rho/\rho_c-2\beta m_0^2\pol_0^2\delta p\nonumber\\
    &\phantom{{}={}}+(\pol_0\rho_0/\rho_c-2\beta m_0\pol_0^3)\delta m+O(\gamma^2).
\end{align}
Using Eqs.~\eqref{eq:ext_m_app} and \eqref{eq:ext_rho_app}, we obtain:
\begin{align}
    \delta \pol =\gamma \frac{m_0 \pol_0 \rho_0/\rho_c - m_0\tilde D/\rho_c-2\beta m_0^2\pol_0^3}{2\pol_0 m_0\chi\Deff^{-1}/\rho_c+2\beta m_0^2\pol_0}+O(\gamma^2),
\end{align}
which, using Eq.~\eqref{eq:ext_p0}, yields the expression given in Eq.~\eqref{eq:ext_p} of the main text.

\subsection{Radial (micelle) profile}
\label{app:micprof}
In this section, we turn to the micelle profile, which we inspect close to its center. To this goal, we express Eqs.~\eqref{eq:m} and \eqref{eq:der} in polar coordinates, remove angular dependences and assume a radially oriented polarization $\polv=p\vcu{e}_r$. We obtain:
\begin{subequations}
\label{eq:Rad}
\begin{align}
    &\partial_t\rho=r^{-1}\partial_r[r\partial_r(D_\rho\rho + m\nu\rho^2 + \alpha_2\rho^2+\alpha_3\rho^3+\chi p^2)]\nonumber\\
    &\phantom{\partial_t\rho ={}}+r^{-1}\partial_r[r\chi_1p^2/r]\label{eq:rhoRad},\\
    &\partial_t p=\left[(m\rho/\rhoc-1)-\lambda_2p/r-\beta m^2p^2\right]p+\zeta_1\partial_r\rho\nonumber\\
    &\phantom{\partial_t p ={}}+\kappa [\partial_r^2 p+\partial_r p/r -p/r^2]-\lambda p\partial_rp,\label{eq:pRad}\\
    &\partial_t m=r^{-1}\partial_r[r(D_m \partial_rm-v_m m p)]\label{eq:mRad}.
\end{align}
\end{subequations}

We proceed similarly to the analysis of the inner bilayer profile. Thus, setting $m=m_0$ everywhere, linearizing the microtubule density around $\rho_0$ and integrating equation \eqref{eq:rhoRad} twice, using $r=0$ as an integration boundary, we find:
\begin{equation}
\label{eq:rhovsp1D_rad}
    \rho(r)= \rho_--\frac{\chi}{\Deff} p^2(r)-\frac{\chi_1}{\Deff}\int_0^r\dd{r^\prime}\frac{p^2(r^\prime)}{r^\prime}.
\end{equation}
The additional term proportional to $\chi_1$ enhances phase separation due to the additional splay flux emerging as a consequence of the radial symmetry. To deal with this non-local term, we expand the integral in Eq.~\eqref{eq:rhovsp1D_rad} for small $r$, using the Taylor expansion of $p^2$ to second order:
\begin{equation}
\label{eq:rad_papprox}
    p^2(r)\approx\frac{\partial_r^2p^2(r=0)}{2}r^2=:cr^2.
\end{equation}
Using this relation, the integral can be rewritten as follows, to second order in $r$:
\begin{align}
\label{eq:rad_intapprox}
    \int_0^r\dd{r^\prime}\frac{p^2(r^\prime)}{r^\prime}&\approx \left.\frac{p^2(r)}{r}\right|_{r=0}\hspace{-.5em}r +\left[\frac{\partial_rp^2(r)}{2r} - \frac{p^2(r)}{2r^2}\right]_{r=0}\hspace{-.5em}r^2\nonumber\\&= \frac{c}{2} r^2 = \frac{p^2(r)}{2}.
\end{align}
Using this approximation, the effect of the new term in Eq.~\eqref{eq:rhovsp1D_rad} is to shift $\chi\to \chi+\chi_1/2$. Thus, the phase separation strength $\Delta\rho$ is modified to the expression given in Eq.~\eqref{eq:rad_PS_Delta}.

Inserting Eqs.~\eqref{eq:rhovsp1D_rad} and \eqref{eq:rad_intapprox} into Eq.~\eqref{eq:pRad} at stationarity, we find:
\begin{align}
\label{eq:pRad_rhoel}
    &0 = (m_0\rho_-/\rhoc-1)p-\beta m_0^2(1+\bar\chi+\bar\chi_1/2)p^3+\kappa \partial_r^2 p\nonumber\\
    &\phantom{0 ={}}+\kappa[\partial_r p/r -p/r^2]\nonumber\\
    &\phantom{0 ={}}-\lambda_2 p^2/r-(\lambda+2\zeta_1(\chi+\chi_1/2)/\Deff)p\partial_r p.
\end{align}
To get this equation into a shape like Eq.~\eqref{eq:p1D_rhoel}, we again use a small $r$ approximation, using the Taylor expansion Eq.~\eqref{eq:rad_papprox}, which gives $p(r)\approx-\sqrt{c}r$. Then, to the lowest order in $r$, the second line of Eq.~\eqref{eq:pRad_rhoel} vanishes, and we can approximate the active splay term as follows:
\begin{equation}
    p^2/r\approx cr \approx p\partial_r p.
\end{equation}
Using this, the term $\bar\lambda$ arising for the bilayer in Eq.~\eqref{eq:LBL_app} is self-advection is shifted as $\bar\lambda \to \bar\lambda +\zeta_1\chi_1/\Deff+\lambda_2$, so that the active contribution to the length scale $\Lambda$ is modified to the expression given in Eq.~\eqref{eq:Lmic}.

Finally, requiring stationarity in Eq.~\eqref{eq:mRad}, the same steps as for the bilayer can be applied to obtain the motor profile given in Eq.~\eqref{eq:m_1Dsol}.

\section{Linear Stability Analysis}
\label{app:LSA}
\subsection{LSA of the isotropic homogeneous state}
To analyze the instabilities of the isotropic homogeneous state with $\rho=\rho_0$, $m=m_0$ and $\polv=0$, we perform a linear stability analysis of Eqs.~\eqref{eq:der} by introducing a periodic perturbation of the form:
\begin{subequations}
\label{eq:LSA_isotropic_pert}
\begin{align}
\rho(x,y,t)&=\rho_0+\delta\rho(t)e^{ikx},\\
p_i(x,y,t)&=\delta p_i(t)e^{ikx},
\end{align}
\end{subequations}
where we chose the coordinate system such that the $x$-axis is aligned with the periodic modulation, without loss of generality. For small $\gamma$, the motor field won't contribute significantly to the onset of the instability and we can keep it constant, $m=m_0$.

Inserting the ansatz \eqref{eq:LSA_isotropic_pert} into Eqs.~\eqref{eq:der}, and keeping terms up to linear order in the perturbations, we obtain the equations:
\begin{subequations}
\label{eq:LSA_isotropic_eqns}
\begin{align}
\partial_t\delta\rho &= -k^2\Deff\delta\rho,\\
\partial_t\delta p_x &= (m_0\rho_0/\rhoc-1)\delta p_x -\kappa k^2\delta p_x+i\zeta_1k\delta\rho,\\
\partial_t\delta p_y &= (m_0\rho_0/\rhoc-1)\delta p_y -\kappa_1k^2\delta p_y,
\end{align}
\end{subequations}
where $\Deff$ is given by Eq.~\eqref{eq:Deff}. The resulting eigenvalues are:
\begin{subequations}
\begin{align}
    \sigma_1&=-k^2\Deff,\\
    \sigma_2&=(m_0\rho_0/\rhoc-1)-\kappa k^2,\\
    \sigma_3&=(m_0\rho_0/\rhoc-1)-\kappa_1 k^2.
\end{align}
\end{subequations}
For $m_0\rho_0>\rhoc$, the two polarization directions show an instability even at $k=0$, corresponding to the emergence of global order. The instability takes place on long length scales, with $\sigma_2$ and $\sigma_3$ negative for large $k$, since variations of the order strength with short wavelengths are suppressed by the longitudinal stiffness $\kappa$ and the perpendicular stiffness $\kappa_1$, respectively.

In addition to this ordering instability, the system exhibits an instability with respect to density variations for $\Deff<0$, as discussed in Ref.~\cite{AransonTsimring2006,MaryshevPRE2018}. This density or ``bundling'' instability requires the introduction of a bilaplacian term of the form $-\nabla^4\rho$ in the density equation to be regularized at short wavelengths. In this work, we circumvent this by requiring $\Deff>0$, which can be achieved by choosing a sufficiently high value of $\alpha$. We postpone the analysis of the role of the density instability in our model to future work.

\subsection{LSA of the ordered state}
For $m_0\rho_0>\rhoc$, the homogeneous state with non-zero polarization $p_0=\sqrt{(m_0\rho_0-\rhoc)/(\beta m_0^2\rhoc)}$ in a certain direction is a stationary solution of equations \eqref{eq:der}. Starting from this base state, the instabilities that may arise are more convoluted due to the coupling between density and order perturbations. Again, we apply a periodic perturbation to this state:
\begin{subequations}
\label{eq:LSA_ordered_pert}
\begin{align}
\rho(x,y,t)&=\rho_0+\delta\rho(t)e^{i(k_\parallel x+k_\perp y)},\\
p_i(x,y,t)&=p_0\delta_{i,x}+\delta p_i(t)e^{i(k_\parallel x+k_\perp y)},
\end{align}
\end{subequations}
where we have chosen the $x$-axis to lie along the direction of the global polarization and kept the motor field constant. Thus, $k_\parallel$ and $k_\perp$ set the length scales of the perturbation longitudinally and perpendicularly to the order, respectively. Note that to the lowest order in the perturbation, the polarization strength is changed only by $\delta p_x=\delta p_\parallel$, while $\delta p_y = \delta p_\perp$ controls variations in the orientational direction \bibnotemark[delp].

By inserting the ansatz \eqref{eq:LSA_ordered_pert} into Eqs.~\eqref{eq:der} and keeping terms to linear order in the perturbations, we obtain:
\begin{subequations}
\label{eq:LSApolar}
\begin{align}
        \delt \delta\rho &= -\Deff k^2\delta\rho-\! 2p_0(\chi_1k_\parallel k_j\delta p_j+\chi_2k^2\delta p_\parallel),\\
        \delt \delta p_\parallel &=m_0p_0/\rhoc\delta\rho-2\beta m_0^2p_0^2\delta p_\parallel\nonumber\\
        &\phantom{{}={}}-\kappa_1 k^2\delta p_\parallel-\kappa_2k_\parallel k_j\delta p_j\nonumber\\
        &\phantom{{}={}}+i\zeta_1k_\parallel\delta\rho-i\lambda k_\parallel p_0\delta p_\parallel-i\lambda_2k_\perp p_0\delta p_\perp,\\
        \delt \delta p_\perp &=-\kappa_1 k^2\delta p_\perp-\kappa_2k_\perp k_j\delta p_j+i\zeta_1k_\perp\delta\rho\nonumber\\
        &\phantom{{}={}}+2i\zeta_2k_\perp p_0\delta p_\parallel-i\lambda_1k_\parallel p_0\delta p_\perp,
\end{align}
\end{subequations}
where summation over the index $j$ is implicit.

In the $\vc{k}\to 0$ (long wavelength) limit, only $\delta p_\parallel$ decays with a finite rate, as global variations in the amplitude of the order are exponentially suppressed, while the density field as a conserved quantity and the order direction as a Goldstone mode have slow dynamics. Therefore, we can adiabatically eliminate the perturbation in the order strength by setting $\delt \delta p_\parallel=0$ to obtain, to the lowest order in $k_{\parallel,\perp}$ we will require:
\begin{align}
    2\beta p_0\delta p_\parallel &= \frac{\delta\rho}{ m_0\rhoc}-i\frac{\lambda_2}{m_0^2} k_\perp\delta p_\perp.
\end{align}
The terms on the right-hand side reflect the dependence of the order strength on density variations due to the Ginzburg-Landau term, as well as on the splay induced by $\delta p_\perp$, due to the active splay term controlled by $\lambda_2$. Substituting this expression into Eqs.~\eqref{eq:LSApolar} yields the following two-dimensional Jacobian (see Eq.~\eqref{eq:J_instab}), to the lowest order in $k_{\parallel,\perp}$ in each component which we will need for the calculation of the eigenvalues:
\begin{subequations}
\label{eq:LSApolar2DJ}
\begin{align}
    &\hspace{-.3em}J_{11}=-\Deff(1+\bar\chi_2)k^2-\Deff\bar\chi_1k_\parallel^2,\\
    &\hspace{-.3em}J_{12}=-2\chi_1p_0k_\parallel k_\perp+i\lambda_2\chi_2/(\beta m_0^2)k_\perp^3,\\
    &\hspace{-.3em}J_{21}=i\bar\zeta k_\perp,\\
    &\hspace{-.3em}J_{22}=-i\lambda_1k_\parallel p_0+\lambda_2\zeta_2/(\beta m_0^2) k_\perp^2-\kappa_1k^2-\kappa_2k_\perp^2,
\end{align}
\end{subequations}
where we have introduced $\bar\zeta=\zeta_1+\zeta_2/(\beta m_0\rhoc)$, the effective anchoring to density interfaces, as well as $\bar\chi_2=\chi_2/(\Deff\beta m_0\rhoc)$. The eigenvalues of this Jacobian are given by:
\begin{equation}
\label{eq:ev_Tr_Det}
    2\sigma_{1,2} = \Tr \vc{J} \pm \sqrt{(\Tr \vc{J})^2-4\det \vc{J}}.
\end{equation}
In the following, we shall write $\vc{k}=(k\cos\varphi, k\sin\varphi)$.

\subsubsection{Longitudinal and mixed instabilities}
First, we study the case $k_\parallel\neq 0$, corresponding to wave vectors that are either fully longitudinal or that mix longitudinal and perpendicular components. Assuming $\lambda_1\neq 0$, the expression in the square root appearing in Eq.~\eqref{eq:ev_Tr_Det} reads as follows, up to third order in $k$:
\begin{equation}
\label{eq:LSA_ordered_discriminant}
-\lambda_1^2k_\parallel^2 p_0^2\left(1+\frac{2i}{\lambda_1k_\parallel p_0}(Xk_\parallel^2+Yk_\perp^2)\right).
\end{equation}
with 
\begin{subequations}
\begin{align}
X&= \Deff(1+\bar\chi)-\kappa_1,\\
Y&= \Deff(1+\bar\chi_2)+4\frac{\chi_1\bar\zeta}{\lambda_1}+\frac{\lambda_2\zeta_2}{\beta m_0^2}-\kappa.
\end{align}
\end{subequations}
In the long wavelength limit, the term $k_\perp^2/k_\parallel=k\sin\varphi\tan\varphi$ is much smaller than 1 as long as $\abs{\tan\varphi} \ll k^{-1}$, i.e., for sufficiently strong longitudinal admixture in the choice of the wave vector. If this condition is met, we can use the expansion $\sqrt{1+x}=1+\frac{x}{2}+O(x^2)$ to take the square root of Eq.~\eqref{eq:LSA_ordered_discriminant}, obtaining:
\begin{equation}
i\abs{\lambda_1k_\parallel}p_0-\sign(\lambda_1k_\parallel)\left[Xk_\parallel^2+Yk_\perp^2\right]+O(k^3).
\end{equation}
Inserting this result back into Eq.~\eqref{eq:ev_Tr_Det}, and absorbing $\sign(\lambda_1k_\parallel)$ into $\pm$, we find the real parts of the two eigenvalues (up to second order in $k$):
\begin{subequations}
\label{eq:LSA_ordered_mixed}
\begin{align}
    \Re\sigma_1 &= -\Deff(1+\bar\chi)k_\parallel^2\nonumber\\
    &\phantom{{}={}}-\left[\Deff(1+\bar\chi_2)+2\chi_1\bar\zeta/\lambda_1\right]k_\perp^2,\label{eq:LSA_ordered_rhomixed}\\
    \Re\sigma_2 &= -\kappa_1k_\parallel^2-\left[\kappa-\frac{\lambda_2\zeta_2}{\beta m_0^2}-2\chi_1\bar\zeta/\lambda_1\right]k_\perp^2.\label{eq:LSA_ordered_pmixed}
\end{align}
\end{subequations}
When at least one of these expressions is positive, the homogeneous state is unstable. The imaginary parts of the eigenvalues read:
\begin{subequations}
\begin{align}
    \Im \sigma_1 &= 0 + O(k^3),\\
    \Im \sigma_2 &= -\lambda_1 p_0k_\parallel + O(k^3).\label{eq:LSA_ordered_pmixed_im}
\end{align}
\end{subequations}
Thus, the instability corresponding to $\Re\sigma_1>0$ is stationary in the long wavelength limit, whereas the one corresponding to $\Re\sigma_2>0$ is oscillatory, with a wave velocity proportional to $\lambda_1$.

For $k_\perp = 0$, the perturbation is purely longitudinal. The Jacobian \eqref{eq:LSApolar2DJ} becomes diagonal to the lowest order in $k$ and the expressions \eqref{eq:LSA_ordered_mixed} are simply its diagonal entries. In this case, $\Re \sigma_1$ is positive for $1+\bar\chi<0$, i.e., when the contractile flux overcomes effective diffusion, giving rise to the \emph{contractile instability} described in the main text (see Fig.~\ref{fig:LSAcartoons}a). Since the Jacobian is diagonal, this instability is purely density-like, not involving orientational perturbations at all. On the other hand, a longitudinal orientational instability would require $\kappa_1<0$ and the introduction of a regularizing term, a case which we exclude here.

Now we turn to the case of general wave vectors with $k_\parallel\neq 0$. Taking the limit $\varphi\to\pi/2$ after the limit $k\to 0$, such that $\abs{\tan\varphi} \ll k^{-1}$, the $k_\perp^2$ terms in Eqs.~\eqref{eq:LSA_ordered_mixed} dominate. This corresponds to a perpendicular perturbation with a weak longitudinal component. In this limit, instabilities arise when the square brackets in Eqs.~\eqref{eq:LSA_ordered_mixed} become negative. This condition yields the inequalities \eqref{eq:mixed_or_instab} and \eqref{eq:mixed_rho_instab} discussed in the main text.

The eigenvector corresponding to the eigenvalue $\sigma_1$ reads:
\begin{align}
    \vc{v}_1 &= \mqty(J_{12}\\\sigma_1-J_{11}) = \mqty(-2\chi_1p_0k_\parallel k_\perp+O(k^3)\\-2\chi_1\bar\zeta/\lambda_1k_\perp^2).
\end{align}
Here, the components of $\vc{v}_1$ are of the same order in $k$, their ratio being $\bar\zeta p_0\cot\varphi/
\lambda_1$. Thus, the perturbation corresponding to this eigenvalue goes from predominantly density-like to predominantly orientational as $\varphi$ is tuned from $0$ to $\pm\pi/2$. For sufficiently large $\lambda_1$, the density component prevails for most wave vectors, so we refer to the instability corresponding to $\vc{v}_1$ as the \emph{mixed density instability}. Furthermore, to the lowest order in $k$, the two components are real, which implies that the phase shift between the orientational and density perturbations can be only $0$ or $\pi$. This means that the extrema in the splay and bend are in between the density extrema (see Fig.~\ref{fig:LSAcartoons}c).

On the other hand, the eigenvector associated to $\sigma_2$ reads:
\begin{align}
    \vc{v}_2 &= \mqty(\sigma_2-J_{22}\\J_{21}) = \mqty(O(k_\perp^2)\\i\bar\zeta k_\perp).
\end{align}
This eigenvector is predominantly orientational in the long wavelength limit, which is why we name the instability \emph{mixed orientational instability}. The density component is real to the lowest order, while the orientational component is purely imaginary, so the phase shift between the two perturbations is $\pm\pi/2$. This implies that the extrema in the density perturbation coincide with the extrema in the splay and bend of the orientational field.

\begin{figure}
    \centering
    \includegraphics{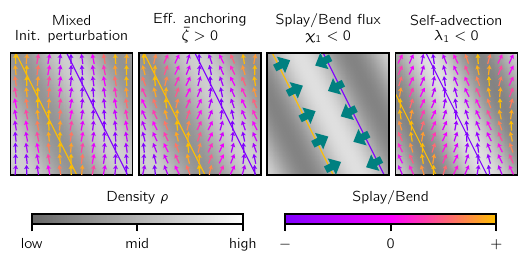}
    \caption{$\chi_1\bar\zeta/\lambda_1$ term in the mixed orientational instability. A mixed wave vector perturbation where the density and orientational components have a phase shift of $\pm\pi/2$ gives rise to splay and bend with the extrema (yellow and purple lines) coinciding with the extrema of the density wave. The effective anchoring rotates the orientational field along the density gradient, amplifying the orientational perturbation further. The splay/bend flux then accumulates density between the two lines, shifting the density pattern to the back. On the other hand, the self-advection $\lambda_1<0$ shifts the orientational pattern backward, returning an amplified initial perturbation, up to a backward propagation.}
    \label{fig:LSA_app}
\end{figure}

Note that the $\chi\bar\zeta/\lambda_1$ term in the eigenvalue \eqref{eq:LSA_ordered_pmixed} (or in the condition \eqref{eq:mixed_or_instab}) has a different sign compared to the mixed density instability, making it stabilizing for the signs dictated by the derived model. However, in general, its sign can switch, resulting in a destabilizing contribution by this term. In Fig.~\ref{fig:LSA_app}, we illustrate the underlying feedback mechanism. Starting from a perturbation in density and orientation, with the phase shift discussed above, the effective anchoring controlled by $\bar\zeta$ will amplify the perturbation of the orientational field. The amplified splay and bend advect the density into the region between their extrema, shifting the density pattern into a direction that depends on the sign of $\chi_1\bar\zeta$, like for the mixed density instability. In contrast to the latter, however, here $\lambda_1$ reinstates the initial phase shift by advecting the pattern into the same direction, instead of bringing it back in place (see Fig.~\ref{fig:LSAcartoons}c). This explains the oscillatory character of the mixed orientational instability, whose imaginary part is linear in $k$ (see Eq.~\eqref{eq:LSA_ordered_pmixed_im} and Fig.~\ref{fig:disp}c). Thus, the mixed orientational instability can emerge even in the absence of the feedback between $\lambda_2$ and $\zeta_2$, if $\chi\bar\zeta/\lambda_1$ has the proper sign.

To summarize, in the long wavelength limit the mixed density instability is stationary and has density and orientational components of the same order in $k$ in its eigenvector, which are in phase or antiphase with respect to each other. In contrast, the mixed orientational instability is oscillatory and has a dominating orientational component with a phase shift of $\pm\pi/2$ with respect to the density component. 

A special limit case is $\lambda_1=0$, since the second order term in $k$ in Eq.~\eqref{eq:LSA_ordered_discriminant} vanishes. Then, the leading order term in Eq.~\eqref{eq:ev_Tr_Det} is given by:
\begin{equation}
   2\sigma_{1,2}=\pm\sqrt{-8i\bar\zeta\chi_1 p_0 k_\parallel k_\perp^2},
\end{equation}
whose real and imaginary parts read:
\begin{subequations}
\begin{align}
   \Re\sigma_{1,2}&=\pm\sqrt{\abs{\chi_1\bar\zeta k_\parallel}p_0k_\perp^2},\\
   \Im\sigma_{1,2}&=\mp i\sign(\chi_1\bar\zeta k_\parallel)\sqrt{\abs{\chi_1\bar\zeta k_\parallel}p_0k_\perp^2}.
\end{align}
\end{subequations}
Hence, there is always an eigenvalue with a positive real part, making the homogeneous state always unstable with respect to mixed perturbations for $\lambda_1=0$. The mechanism responsible for this instability is the interplay between splay/bend flux and effective anchoring shown in Fig.~\ref{fig:LSAcartoons}c and Fig.~\ref{fig:LSA_app}.

\subsubsection{Perpendicular instability}
The calculation above breaks down for $k_\parallel = 0$, so we have to treat this case separately. For a purely perpendicular wave vector, the trace and the determinant of the Jacobian \eqref{eq:LSApolar2DJ} become real to the lowest order in $k$. Thus, Eq.~\eqref{eq:ev_Tr_Det} gives at least one positive eigenvalue if $\Tr \vc{J}>0$ or $\det \vc{J}<0$, resulting in the inequalities \eqref{eq:perp_instab}, where we are assuming that $\Deff(1+\bar\chi_2)>0$.

When any of those inequalities are fulfilled, at least one of the eigenvalues of the Jacobian \eqref{eq:LSApolar2DJ} will have positive real part, resulting in an instability which we refer to as the \emph{perpendicular instability}. As emerges from the feedback mechanism described in the main text (see Fig.~\ref{fig:LSAcartoons}b), the perpendicular instability hinges on orientational perturbations giving rise to splay, whereas density perturbations only have a secondary role. This is also seen in the eigenvectors that correspond to the eigenvalues given in Eq.~\eqref{eq:ev_Tr_Det}, which read:
\begin{equation}
\label{eq:ev_perp}
    \vc{v}_{1,2}=\mqty(J_{12}\\\sigma_{1,2}-J_{11}).
\end{equation}
Since both the eigenvalues and $J_{11}$ are of order $O(k_\perp^2)$, whereas $J_{12}$ is of order $O(k_\perp^3)$, in the long wavelength limit the perturbation will be chiefly orientational.

When $\Deff(1+\bar\chi_2)$ becomes negative, the sign in the inequality \eqref{eq:perp_instab2} is reversed. For the case of no effective anchoring $\bar\zeta=0$, this always results in an instability, which is density-like in nature (since the orientational component in Eq.~\eqref{eq:ev_perp} vanishes in that case). This is the perpendicular analog of the contractile instability discussed above: indeed, for $\Deff>0$, $\Deff(1+\bar\chi_2)<0$ for sufficiently strongly negative transversal flux coefficient $\chi_2$, which overcomes the isotropic effective diffusion in the direction perpendicular to the order. This results in an \emph{extensile instability} that breaks up the original homogeneous state in bands that extend along the order direction. For non-zero $\bar\zeta$, the instability becomes mainly orientational, as the self-anchoring will make the order deviate from its original direction. In our derived model, the extensile instability does not play any role.

\subsection{LSA of a radially symmetric solution}
\label{app:LSAmic}
In this section, we investigate how the instabilities derived above change when the initial state has non-vanishing splay, i.e., for micelle solutions. We start from a homogeneous, radially ordered solution of the equations in polar coordinates, and focus on a range of radii with $r\gg 1$. We perturb this solution as follows, again keeping $m=m_0$:
\begin{subequations}
\label{eq:LSA_rad_pert}
\begin{align}
\rho(r,\phi,t)&=\rho_0(r)+\delta\rho(t)e^{i\kn\phi}e^{i\kr r},\\
p_r(r,\phi,t)&=p_0(r)+\delta p_\phi(t)e^{i\kn\phi}e^{i\kr r}\\
p_\phi(r,\phi,t)&=\delta p_\phi(t)e^{i\kn\phi}e^{i\kr r}.
\end{align}
\end{subequations}
Here, $\kn$ is the node number of the angular perturbation, while $\kr$ controls the radial perturbation. We insert these equations into Eq.~\eqref{eq:der}, and use radial coordinates. Neglecting radial derivatives and keeping terms up to order $O(r^{-1})$ leads to the following condition on the solution $\rho_0(r)$, $p_0(r)$:
\begin{align}
\label{eq:LSA_rad_StatCond}
0&=m_0\rho_0/\rhoc-1 -m_0^2\beta p_0^2-\lambda_2p_0/r.
\end{align}
Differentiating this condition with respect to $r$ shows that $\partial_r \sim r^{-2}$, thus making it consistent to neglect the derivatives to this order.

Using Eq.~\eqref{eq:LSA_rad_StatCond} and keeping terms to lowest order in $r^{-1}$ for each power of $\kn$, the time evolution of the perturbations reads:
\begin{subequations}
\begin{align}
\partial_t\delta \rho&= -\frac{\kn^2}{r^2}\Deff\delta\rho-2\frac{\kn^2\chi_2-i\kn\chi_1}{r^2} p_0\delta p_r\nonumber\\
&\phantom{{}={}}+\left[\frac{i\kr}{r}-\kr^2\right]\Deff\delta\rho+2\chi\left[\frac{2i\kr}{r}-\kr^2\right]\pol_0\delta\pol_r\nonumber\\
&\phantom{{}={}}-2\chi_1\frac{p_0\kn\kr}{r}\delta\pol_\phi,\\
\partial_t\delta p_r&= \frac{m_0p_0}{\rhoc}\delta\rho-2\beta m_0^2 p_0^2\delta p_r-i\kn\frac{\lambda_2p_0}{r}\delta p_\phi,\\
\partial_t\delta p_\phi&= -\left[\frac{\kn^2\kappa}{r^2}+\frac{\lambda_1p_0}{r}\right]\delta p_\phi+i\kn{\zeta_1}{r}\delta\rho\nonumber\\
&\phantom{{}={}}+i\kn\frac{2\zeta_2p_0}{r}\delta p_r-i\kr\lambda_1\pol_0\delta\pol_\phi.
\end{align}
\end{subequations}
For $\kn=0$, the orientational perturbation $\delta\pol_\phi$ decouples. Its behavior is controlled by the sign of $\lambda_1 p_0$: for $\lambda_1\pol_0<0$, the perturbation is amplified over time, whereas it is suppressed for $\lambda_1\pol_0>0$. This is a consequence of splay, as emerges from the following argumentation. The coefficient $\lambda_1$ controls the self-advection of the polar order orientation. In a perfectly radial configuration, the order orientation does not change along the polarization, and thus the self-advection has no effect. In contrast, when the orientation is perturbed with $\kn=0$, the polarization acquires an angular component everywhere, and the non-vanishing splay becomes of relevance for the self-advection. For $p_0<0$, a negative $\lambda_1$ leads to a backward propagation of the perturbation along the ring, bringing the system back to its original state. In contrast, a positive $\lambda_1$ leads to a forward propagation of the perturbation, which consequently self-amplifies. This means that for $\lambda_1>0$, no stable micelles with inward-pointing polarization ($p_0<0$) can exist, since they are unstable to orientational perturbations with $\kn=0$.

Adiabatically eliminating $\delta\pol_r$ as we did in the linear stability analysis of the homogeneous ordered state, we can reduce the dynamics of the perturbations to a two-dimensional Jacobian, which, to the lowest order we will need in $r^{-1}$ and $\kr$, has the following entries:
\begin{subequations}
\label{eq:LSApolar2DJrad}
\begin{align}
    J_{11}&=\Deff \bigg[\frac{-\kn^2(1+\bar\chi_2)+i\kn\bar\chi_1}{r^2}+\frac{i\kr(1+2\bar\chi)}{r} \nonumber \\& \qquad
    -\kr^2(1+\bar\chi)\bigg],\\
    J_{12}&=-\frac{2p_0\chi_1\kn\kr}{r},\qquad J_{21}=i\bar\zeta\frac{\kn}{r},\\
    J_{22}&=-\lambda_1\left(i\kr+r^{-1}\right)p_0-\kappa\frac{\kn^2}{r^2}+\frac{\lambda_2\zeta_2}{\beta m_0^2}\frac{\kn^2}{r^2}.
\end{align}
\end{subequations}
To identify the unstable eigenvalues of this matrix, we use Eq.~\eqref{eq:ev_Tr_Det} again. The lowest order term in $r^{-1}$ inside the square root of that equation reads:
\begin{equation}
\lambda_1^2p_0^2(i\kr+r^{-1})^2.
\end{equation}
Note that, in contrast to the corresponding expression \eqref{eq:LSA_ordered_discriminant} in the linear stability analysis of the homogeneous state, this does not vanish for $\kr=0$. Thus, unlike for the perpendicular instability, no separate treatment is needed here. This is due to the finite splay of the base state, and will result in the exclusive relevance of the mixed instabilities for the micelle solutions. The next order term is given by:
\begin{align}
-2&\lambda_1p_0(i\kr+r^{-1}){}\cdot{}\nonumber\\
&{}\cdot{}\left[\left(\frac{\lambda_2\zeta_2}{\beta m_0^2}-\kappa+\frac{4i\kr\chi_1\bar\zeta}{\lambda_1(i\kr+r^{-1})}\right)\frac{\kn^2}{r^2}-J_{11}\right].
\end{align}
Factoring out $\lambda_1^2p_0^2(i\kr+r^{-1})^2$, we can use $\sqrt{1+x}\approx1+x/2$. Putting everything together, we find the real parts of the eigenvalues:
\begin{subequations}
\begin{align}
\Re\sigma_1&=-\Deff(1+\bar\chi)\kr^2\nonumber\\&\phantom{{}={}}-\left[2\frac{\chi_1\bar\zeta}{\lambda_1}s+\Deff(1+\bar\chi_2)\right]\frac{\kn^2}{r^2},\label{eq:sigma_shinstab}\\
\Re\sigma_2&=-\frac{\lambda_1p_0}{r}+\left[2\frac{\chi_1\bar\zeta}{\lambda_1}s+\frac{\lambda_2\zeta_2}{\beta m_0^2}-\kappa\right]\frac{\kn^2}{r^2}\label{eq:sigma_brinstab}.
\end{align}
\end{subequations}
Here, we have introduced the factor ${s=\left(1+(r \kr)^{-2}\right)^{-1}\in[0,1]}$, which depends on the ratio of the characteristic length scale of the micelle radius $r$ and the radial perturbation $\kr^{-1}$. For $r\gg \kr^{-1}$, $s\to 1$ and the expressions in the square brackets correspond to those appearing in Eq.~\eqref{eq:LSA_ordered_mixed}. Hence, we conclude that the mechanisms underlying the mixed instabilities discussed for the homogeneous ordered state extend to splayed solutions, up to the additional $\lambda_1/r$ term in $\Re\sigma_2$ (which is stabilizing for $\lambda_1p_0>0$).\\

\section{Numerical simulations}
\label{app:sim}
All simulations were performed using the finite element solver COMSOL Multiphysics, using square geometries with periodic boundary conditions. As initial condition, we started from the isotropic homogeneous state $\rho=\bar\rho$, $\polv=0$, $m=1$ and applied a random perturbation at every gridpoint, which was taken from a uniform distribution with widths $\delta\rho$, $\delta\polv$ and $\delta m$. In the simulations of the derived model presented in Figs.~\ref{fig:PD} and \ref{fig:PDalpha}, these were chosen as $\delta\rho=0.1\rhoc$, $\delta\polv=0.001$ and $\delta m=0.1$.

The simulation of the active foam shown in Fig.~\ref{fig:phen_af} was performed in a $120\times 120$ geometry, with $\bar\rho=1.1$, $D_\rho=0.1$, $\alpha_2=\alpha_3=0.05$, $\hat\chi_1=-0.2$, $\hat\chi_2=0.1$, $\hat\kappa_1=0.05$, $\hat\lambda_1=0.4$, $\hat\lambda_2=-0.7$, $\hat\zeta_1=0.1$, $\hat\kappa_2=\hat\zeta_2=\nu=0$, $D_m=0.2$, $v_m=0.04$. The snapshots in Fig.~\ref{fig:phen_af}b are of size $30\times 30$.

For the measurement of the bilayer profiles (top row of Fig.~\ref{fig:profiles}), we ran simulations in a geometry of size $80\times80$, for run times $t=2000$. The initial perturbation amplitudes were chosen as $\delta\rho=0.11$, $\delta p_{x,y} = 0.1$ and $\delta m = 0$. We varied the parameters
$\hat\chi_1\in\{-0.1,-0.15,-0.2\}$, $\hat\chi_2=-\hat\chi_1/2$, $\hat\kappa_1\in\{0.03,0.05,0.07\}$, $\hat\zeta_1\in\{0, 0.05\}$, $\hat\zeta_2\in\{0.1,0.2,0.3\}$ and $v_m\in\{0.01,0.02,0.03\}$, while keeping $\bar\rho=1.1$, $D_\rho=1/32$, $\alpha_2=\alpha_3=-\nu=0.05$, $\hat\kappa_2=0$, $\hat\lambda_1=-\hat\lambda_2=0.5$ and $D_m=0.2$ fixed. The bilayer profiles were extracted by taking the perpendicular cross-section of each edge of the network for every parameter set. For Fig.~\ref{fig:profiles}g, we extracted profiles for $\hat\chi_1=-0.15$, $\hat\chi=0.075$, $\hat\kappa_1=0.03$, $\hat\zeta_1=0$, $\hat\zeta_2=0.2$.

For the measurement of the micelle profiles (bottom row of Fig.~\ref{fig:profiles}), we ran simulations of the phenomenological model in a geometry of size $80\times80$, for run times $t=600$. The initial perturbation amplitudes were chosen as $\delta\rho=0.05$, $\delta p_{x,y} = 0.01$ and $\delta m = 0.05$. We varied the parameters $\hat\chi_1\in\{-0.15,-0.125,-0.1,-0.075\}$, $\hat\chi_2\in\{0,0.025,0.05\}$, $\hat\kappa_1\in\{0.05,0.075,0.1\}$, $\hat\lambda_2\in\{-0.2,-0.4,-0.6\}$, while keeping $\bar\rho=1.2$, $D_\rho=1/32$, $\alpha_2=\alpha_3=-\nu=0.05$, $\hat\kappa_2=\hat\zeta_1=0$, $\hat\lambda_1=-1.0$, $v_m=0.02$ and $D_m=0.2$ fixed. The micelle profiles were extracted by taking cross-sections passing through the center of every micelle that appeared at the end of the simulations.

For the inspection of the micelle instabilities in section \ref{sec:mic_instab}, we prepared a stable micelle in a $40\times 40$ geometry at the parameter values $\bar\rho=1.1$, $D_\rho=1/32$, $\alpha_2=\alpha_3=-\nu=0.05$, $\hat\chi_1=-0.15$, $\hat\chi_2=\hat\kappa_2=\hat\zeta_1=0$, $\hat\kappa_1=0.03$, $\hat\lambda_1=-1.0$, $\hat\lambda_2=0$, $\hat\zeta_2=-0.1$, $v_m=0.02$ and $D_m=0.2$, waiting for it to reach steady state. Then, to activate the branching instability, we started from that micelle as the initial condition and set $\hat\lambda_1=-0.5$, $\hat\zeta_2=-0.8$, $\hat\lambda_2=-1.1$ and $\hat\kappa_1$ to the values indicated in Fig.~\ref{fig:BrInst}, resuming the simulation. For the fingering instability, we started from the stable micelle as the initial condition, set $\hat\lambda_1=-0.7$ and $\hat\zeta_2$ to the values indicated in Fig.~\ref{fig:ShInst} and resumed the simulation.

\providecommand{\noopsort}[1]{}\providecommand{\singleletter}[1]{#1}%


\begin{thebibliography}{84}
\bibitem [{del()}]{delp}%
  \BibitemOpen
  \href@noop {} {}\bibinfo {note} {Writing $\polv =p\nv $, to lowest order, we
  find $\delta \polv =\delta p\vcu {e}_x + p_0\delta \nv $. Multiplying this
  equation with $\vcu {e}_{x,y}$, using $\delta \nv \vdot \vcu {e}_x=0$ (due to
  the unit vector condition), and comparing with the expression for $\polv $ in
  the main text, we obtain $\delta p = \delta p_\parallel e^{i\kv \vdot \rv }$
  and $\vcu {e}_y\vdot \delta \nv = \delta p_\perp e^{i\kv \vdot \rv
  }/p_0$.}\BibitemShut {Stop}%
\end{thebibliography}
\end{document}